\documentclass[a4paper,10pt]{article}
\usepackage{ucs}
\usepackage[utf8]{inputenc}
\usepackage{amsmath}
\usepackage{amsfonts}
\usepackage{amssymb}
\usepackage{amsthm}
\usepackage{rotating}
\usepackage{xcolor}
\usepackage[english]{babel}
\usepackage[T1]{fontenc}
\usepackage{graphicx}
\usepackage{float}
\usepackage{hyperref}
\usepackage{cuted}
\usepackage{verbatim}
\usepackage{subfigure}
\usepackage{multirow}
\usepackage{bbold}
\usepackage{slashed}
\usepackage{indentfirst}
\usepackage{tcolorbox}
\usepackage{xcolor}
\usepackage{placeins}
\usepackage[a4paper, total={7in, 10in}]{geometry}
\DeclareMathOperator{\sign}{sign}
\usepackage{placeins}
\usepackage{multicol}
\usepackage{fmtcount}
\usepackage{diagbox}

\newcommand{\bpsi}{\bar{\psi}}
\newcommand{\xlam}{\lambda}

\newcounter{eq}

\title{ \bf 
Classical   Yang Mills equations with sources: 
consequences of specific scalar potentials}

\author{\bf  Igor de M. Froldi,
 Fabio L. Braghin\\
Instituto de F\'\i sica, Federal University of Goias,
Av. Esperan\c ca, s/n,
 74690-900, Goi\^ania, GO, Brazil 
}

\date{\today}

\begin{document}

\maketitle

\begin{abstract}
Some well known gauge scalar potential very often considered or used 
 in the literature are 
investigated by means of the
 classical Yang Mills  equations  for the $SU(2)$  subgroups of $N_c=3$.
By fixing a particular shape for the scalar potential, the resulting 
vector potentials 
and the corresponding color-charges sources are found.
By adopting the spherical coordinate system, it is shown that 
spherically symmetric solutions, only dependent on the radial coordinate, are only possible 
for the Abelian limit, otherwise, there must have angle-dependent component(s).
The following solutions for the scalar potential are   investigated:
the Coulomb potential and a non-spherically symmetric   generalization,
a linear potential $A_0 (\vec{r}) \sim (\kappa r)$,
a Yukawa-type potential $A_0 (\vec{r}) \sim (C e^{-r/r_0}/r)$
and  finite spatial regions in which the scalar potential assumes constant values.
The corresponding   chromo-electric and chromo-magnetic fields, as well as the 
 color-charge densities, are found to 
have strong deviations from the spherical symmetric configurations.
We speculate these types of non-spherically symmetric configurations may
contribute (or favor) for the (anisotropic) confinement mechanism since they
 should favor color charge-anti-charge 
(or three-color-charge)
bound states that are intrinsically non spherically symmetric
with  (asymmetric) confinement of fluxes.
Specific conditions and relations between the parameters of the solutions are also presented.
\end{abstract}

\section{Introduction}

The Classical Yang-Mills Lagrangian \cite{YM} 
corresponds to the leading lower dimension local gauge-invariant
terms for 
a Lorentz covariant description of non-Abelian gauge-potentials.
It provided an important framework for the development 
of  the Standard Model (SM) in which non-Abelian gauge fields 
are at the basis of both Electroweak Model  and Quantum Chromodynamics
(QCD)  for the strong interactions \cite{SM,QCD,gaugetheory}.
The strong interactions sector 
 is formulated with  the  color group
$SU(N_c=3)$ with  eight four-vector potentials $A_\mu^a$
($a=1,...N_c^2-1$), 
for three color charges of the fundamental representation,
being, therefore, the number of degrees of freedom considerably large.
The complicated non-Abelian structure of the Yang-Mills fields makes it 
 difficult to be solved already at 
 the classical level and still more difficult at the quantum level.
Despite the need and  importance of the quantum field  description of the 
strong interactions,
classical configurations in Yang-Mills (YM) theories have been
proved to be of  interest for
both, QCD and  the 
SM,  also with Landau-Ginzburg/Abelian-Higgs models. 
Standard monopolar or dipolar configurations of the chromo-electromagnetic fields
have been found either in Minkowski or  Euclidean space-times, for example in Refs. 
\cite{monopole-rossi,actor-review,wu-yang-rosen}.
Besides their eventual roles  in particular physical situations, as discussed below, 
some of these classical  solutions, in Euclidianized spacetime,
 have been shown to be relevant,  for  example, in 
lattice QCD simulations, that is the case of  instantons  
\cite{instantons1,instantons-meron,bali}.
At least, one might expect that  solutions for the classical YM theory 
might help to gain some insights for 
the more complete quantum system.
Very often scalar gauge potentials, sometimes obtained from lattice QCD, 
are plugged into quark-bound state equations, 
either Dirac equation or  Schrodinger equation for heavy hadrons,
 to provide predictions for hadron spectroscopy, for example
\cite{review1,griffiths}.
It turns out to be interesting to verify if the use of these scalar gauge potentials
would bring further consequences or requirements when analyzed from the 
point of view of the YM theory, such as the classical YM equations.
The first step of this program is carried out in this work.

Besides an initial interest in finding solutions of pure Yang-Mills theory,
their coupling to fermion sources,  quarks, 
are also needed.
The investigation of the equations of motion 
have been already considered as a sort of first framework for these problems.
Different approaches for solving the equations of motion 
have been developed 
\cite{sikivie-weiss-prd,sikivie-weiss-prl,mandula,jackiw,passarino,gausslaw-greek}
 since the first investigations \cite{YM,ikeda-miyachi-1962}.
 The usual Coulomb solution, an Abelian solution,
 is recovered in the 
non-Abelian limit of the full theory as a particular limit.
Many solutions for the YM equations with sources that have smaller energies than the Coulomb
potential have also been found, for example in  \cite{sikivie-weiss-prd,jackiw}
and fluctuations around the Coulomb solution with conditions for
their stability have been found earlier
for example in \cite{mandula}.
The effects of different color-charge distributions with particular symmetries 
have   been analyzed within different specific conditions 
 such as for cylinder color-distribution shape \cite{cilinder},
 plane color distribution \cite{plane-color} and dynamical color-current as sources 
\cite{current-sources}.
Some finite energy conditions  for the Yang-Mills equations
 were formulated by
S.  Coleman   and others
by assuming that
 gauge fields go
asymptotically to constant values \cite{actor-review}.
Constant gauge potential configurations   have been
envisaged both in classical and quantum calculations
 \cite{constant-gauge,huang-levi} with resulting
instabilities already known from the Abelian case.
Although some celebrated solutions have constant
chromomagnetic fields \cite{constant-Bc1},
a realistic vacuum however cannot possess either constant gauge potential
 nor constant $F_{\mu\nu}$
and, in general, the QCD ground state is believed to have neither $E_c$ nor $B_c$,
chromo-electric and chromo-magnetic fields, equal to zero
 \cite{wetterich-reuter-1994}.
Given the large number of degrees of freedom,
several different simplifying (symmetric) limits  and situations have been considered
and some general properties have been determined at the classical level.
Among the several earlier  works,
the papers by
by  Sikivie and Weiss \cite{sikivie-weiss-prd,sikivie-weiss-prl}
were considered as a partial guide for the present work,
with 
different configurations for monopole and dipole-type color-charge.
To what extent classical configurations are stable or survive in the quantum description is of course 
extremely important, although it may be interesting by itself
 to trace back each effect of the realistic exact solution to eventual classical or quantum origin(s).

Despite the more general  interest in determining
the gauge field dynamics by starting from the classical description,
there might have 
different situations in which classical gauge configurations have been argued or shown to 
play specific or important roles.
Possible effects of  (semi)classical configurations of the gluon field
on  hadrons structure have been investigated
for heavy quarkonia in 
\cite{coimbra-quarkonia}.
Classical gluon fields are also expected to contribute in the 
initial conditions of 
 relativistic heavy-ion collisions when  gluons should be copiously produced
being the occupation number so high that they may be treated classically,
for example in Ref. \cite{initial-cond1}.
The stability of fast varying classical configurations becomes important
\cite{stability}.
Although static solutions may be considered as initial conditions for such time evolution,
the  time dependence  is extremely important
\cite{rhic3}
and it usually involves
non-equilibrium situations 
\cite{cgc-class-YM}.
The corresponding (classical) shear viscosity was calculated in \cite{shear-YM}.
Confinement is one of the most intriguing problems in Physics
and its description in the quantum regime still presents several difficulties.
One can try to understand, nevertheless, to what extent classical gauge field configurations
can mimic, favor, or be associated to the full confining gauge field solutions.
The pure Yang-Mills theory, in three dimensions, has been shown to provide 
confinement classically \cite{3dim-YM}.
A linear increase of the gauge scalar potential is the basis of one of the 
currently investigated mechanisms of 
confinement in four space-time dimensions,
the
dual superconductor confinement mechanism
 \cite{greensite}.
Quantum calculations in lattice QCD lead to a
linear potential 
as well as a (screened) Coulomb-type solution.
These well known solutions, sometimes considered as simple Abelian limiting cases,
are very often considered in the literature to be plugged into dynamical or bound state equations
to provide hadron spectroscopy.
In the present work we intend to provide a more dynamical framework which may
help to extend such  phenomenological potentials.
When they are considered,
they can be expected to have  corresponding vector potentials and color-charge  distributions
at the classical Yang-Mills equations level.
since color degrees of freedom do not show up outside hadrons.
Besides that, it is interesting to note that in confined systems, i.e.
mesons and baryons  \cite{fluxtube-qqb,fluxtube-qqb2,fluxtube-baryon},
these chromo-electromagnetic fluxes must 
not  be spherically symmetric unless some sort of dynamical rotation
restores spherical symmetry in the system, being, therefore, time-dependent.

In the present work new time-independent solutions for 
the classical YM equations in the presence of
punctual and extended sources are presented.
Not only the Euler-Lagrange equations are considered, but also 
the constraint-like Jacobi equations.
Although the problem is formulated for $SU(N_c=3)$
the  equations  for the $SU(2)$ subgroups will be solved.
After the proper alignment of color charges, as 
presented in Ref. \cite{sikivie-weiss-prd}
 we find that there must have no fully spherically  symmetric 
non-Abelian solutions simultaneously for the scalar and vector potentials
as defined below, i.e. $A_0 (\vec{r})=A_0(r)$ and 
$\vec{A}(\vec{r}) = \vec{A}(r)$.
 Given a necessary  anisotropy, 
the logics of the work is basically 
 to require a particular known  shape for  (spherically symmetric) scalar potential 
 and to verify how
the color charge distribution and vector potential must behave to ensure those 
specific scalar potential solutions.
This type of procedure can be also thought as a (classical) verification of 
situations in which Coulomb potential makes sense.
For that, the following scalar potentials will be considered:
 the Coulomb potential and 
an asymmetric generalization, a  linear potential that may  be expected to partially 
mimics a confinement potential, and an  Yukawa potential. 
Besides that, finite spatial regions with constant scalar potential are also considered.
The resulting corresponding vector potentials 
and the color-charge distribution 
contain strong spatial anisotropies. 
Also, phenomenological interpretation  of  some solutions 
are eventually provided by means of classical models of interest for general aspects of hadrons.
Many cases of  instabilities of a uniform electric field in classical Abelian and non-Abelian YM 
were found 
such as in Refs.
\cite{instabilities,instability-sikivie,constant-gauge}. 
Therefore, in general, we chose the boundary conditions
for  the chromo-electromagnetic fields 
to go asymptotically to zero.
The work is organized as follows.
In the next section the equations for the $SU(N_c=3)$ case, and the 
equations for the $SU(N_c=2)$ subgroups, are presented 
and particular limits for which the equations reduce to the Abelian equations are identified.
In section (\ref{sec:deviationsspherical}) it is shown 
that  
 there cannot have a 
fully spherically symmetric solution for both scalar and vector potentials.   
In section (\ref{SectionCoulomb})
the scalar potential is fixed to be a Coulomb potential and the
resulting vector potential, with a corresponding  color-charge density, that allows such configuration
are found.
Next, in section (\ref{sec:nonsphericalcoulomb}) a 
particular non-spherical generalization of the Coulomb potential
 will be  considered and the corresponding vector potentials and charge distributions 
found also analytically.
In section (\ref{sec:linearpot})
the scalar gauge potential is fixed to be  
a linear  one, $\kappa r$.
The corresponding vector potential, chromo-electric and chromo-magnetic fields
and needed color charge distributions
 are found analytically  from the equations.
The scalar potential is considered to be a Yukawa type potential
in section  (\ref{sec:yukawa}), and
the corresponding vector potentials are found numerically.
The typical length of the Yukawa potentials is traced back to a 
color-charge distribution size,  out of which gauge vector fields disappear.
Finally, a specific case of constant gauge field
  solution is proposed to be valid within a very specific
finite spatial regions  in section (\ref{sec:Vconstant}).
The constant scalar potentials is kept  inside two  semi-spheres of radius $R$
in a  dipole-type configuration (positive and negative).
The corresponding color charge-anticharge sources are distributed inside a sphere
that might be modeled by 
 two quarks of color charge and two anti-quarks of anti-color charge,
and it  is considered as  a classical (relativistic) model to 
describe masses and radii of heavy tetraquarks.
 The corresponding
vector potentials and color-charge distribution are found analytically.
 Finally, in the last section, there is a summary with conclusions.

\section{ Classical Yang Mills  equations with sources - $N_c=3$}

The Yang-Mills Lagrangian corresponds to the leading (lower dimension)
  gauge-invariant terms for gauge fields, $A_\mu^a (x)$,
displayed according to  the compact Lie group SU($N_c$).
 Classical Chromodynamics (CCD) is obtained by adding 
a  gauge invariant coupling with
quark sources and the corresponding free terms.
It yields:
\begin{eqnarray}
{\cal L} &=&
\bar{\psi} \left( i D \cdot \gamma - m \right) \psi
 - \frac{1}{4 } F_{\mu\nu}^a F^{\mu\nu}_a , 
\end{eqnarray}
where  $\psi$ is the quark field and its mass $m$
 can be taken to be an unique flavor state 
of the SU($N_f$) fundamental representation for $N_f$ flavors,
and the three states for SU($N_c=3$) in the fundamental representation.
$a=1,...,(N_c^2-1)$ are indices for color in the
adjoint representation,
The (gauge invariant) covariant derivative, 
that can be considered a basic way to build up
gauge invariant quantities,  is given by:
\begin{eqnarray} \label{covderiv}
D_\mu &=& \partial_\mu - i A_\mu g  ,
\end{eqnarray}
where $A_\mu = A_\mu^a \cdot \lambda^a$,
being  $\lambda_a$ the
GellMann matrices  as the generators of the algebra,
and the (gauge invariant)  non-Abelian stress tensor has been
defined as:
\begin{eqnarray} \label{Fmunu}
F_{\mu\nu} &=& \partial_\mu A_\nu - \partial_\nu A_\mu - i
g [ A_\mu , A_\nu] .
\end{eqnarray}


A finite gauge transformation
for this gauge field can be written as:
\begin{eqnarray}
A_\mu (x) \to A_\mu^U (x)= \partial_\mu U (x) U^{-1} (x) + U(x) A_\mu U^{-1} (x),
\end{eqnarray}
where the algebra element can be written in terms of a closed path C in a path 
ordering operator:
\begin{eqnarray}
U(C ; A ) = P \; exp \left( \int_C A(x) \cdot d x \right).
\end{eqnarray}
The (color)  Noether current is given by
\begin{eqnarray} \label{noethercurr}
J^\mu = j^\mu - i g [ A_\nu , F^{\nu\mu} ] 
= \partial_\nu F^{\nu\mu} ,
\end{eqnarray}
where 
the quark  currents can  be written shortly, for $j_a^\mu \cdot \lambda_a = j^\mu$,  as:
\begin{eqnarray} \label{currents}
j_a^\mu = \bar{\psi} \gamma^\mu \lambda_a \psi.
\end{eqnarray}
being that
\begin{eqnarray}
D_\mu J^\mu = \partial_\mu j^\mu + i  g [ A_\mu , j^\mu ] =0.
\end{eqnarray}

\subsection{ Dynamical  equations }

The Euler-Lagrange equations for the gluon field can be written as:
\begin{eqnarray} \label{eom1}
D_\mu F^{\mu\nu} = \partial_\mu F^{\mu\nu} - i g [A_\mu , F^{\mu\nu} ] = j^\nu.
\end{eqnarray}
These  equations can be reduced to the non-Abelian
 versions of electric Gauss's law and the Ampère-Maxwell equations.
Also, to present a set of equations with clear correspondence  with the
Maxwell equations of  electromagnetism,  the Jacobi identity 
must also be considered.
It can be written as:
\begin{eqnarray} \label{jacobi1}
D_\rho F^{\mu \nu} + D_\mu F^{\nu \rho} + D_\nu F^{\rho \mu} = 0.
\end{eqnarray}
Due to the anti-symmetry of $F_{\mu\nu}$, these
 equations can be written
as two  equations that correspond to non-Abelian generalizations of the 
magnetic Gauss's law and the Faraday equation, both discussed below.


For the  time independent gauge potentials, i.e. 
$\partial_t  A_\mu = 0$, the Euler Lagrange equations   (\ref{eom1})  and
the equations obtained from the Jacobi identity   (\ref{jacobi1}) 
can be written as:
\begin{eqnarray}\label{gaussEc}
 - J^0_a &=& \nabla^2 A^0_a -  g f_{abc} \left( \vec{\nabla} \cdot \vec{A}_b \right) A^0_c  
+ g^2 f_{abc} f_{cde} \vec{A}_b \cdot \vec{A_d} A^0_e;
\\ 
\label{magnetic}
 0 &=& g f_{abc} \left( \vec{\nabla} \times \vec{A}_b \cdot \vec{A}_c - 3 \vec{A}_b \cdot \vec{\nabla} \times \vec{A}_c - g f_{cde} \vec{A}_b \cdot \left( \vec{A}_d  \times \vec{A}_e \right) \right);
\\
  \label{faraday}
0 &=& 
g f_{abc} \left( \vec{\nabla} \times \left( \vec{A}_b A^0_c \right) + \vec{A}_b \times \vec{\nabla} A^0_c - A^0_c \vec{\nabla} \times \vec{A}_c  - g f_{cde} \left( \vec{A}_b \times \vec{A}_d A^0_e +
 \frac{ A^0_b \vec{A}_d \times \vec{A}_e}{2} \right) \right) ;
\\  
\label{ampere}
\vec{J}_a &=& 
\vec{\nabla} \times \vec{\nabla} \times \vec{A}_a +  g f_{abc} \left( 
\frac{\vec{\nabla}
 \times \left( \vec{A}_b \times \vec{A}_c \right)}{2}
 - \vec{A}_b \times \vec{\nabla} \times \vec{A}_c + A^0_b \vec{\nabla}A^0_c \right) 
\nonumber
\\
&-& g^2 f_{abc}f_{cde} 
 \left( A^0_b \vec{A}_d A^0_e -
 \frac{\vec{A}_b \times \vec{A}_d \times \vec{A}_e}{2} \right) .
\end{eqnarray}
These equations are also written in terms of the corresponding
(non-Abelian) chromo-electric and chromo-magnetic fields
defined analogously to the electromagnetic ones in the Appendix (\ref{appEcBc}).
In the most general non-Abelian case, there are 8 scalar potentials and 8 vector potentials
whose equations are coupled, although it is possible 
to find particular solutions for a restricted number of gauge fields in which equations might decouple.
This reduction makes possible analytical developments.
Gauge transformations of a restricted solution can help to extend their validity 
although they still will not correspond to the most general possible solution.

\subsection{ Alignment in internal-space }
\label{secIVU}

For the color charge currents in Eq. (\ref{currents}),
the color components, 
red, green and blue 
\cite{greenberg}, can be written in terms of the 
adjoint representation for the gluon field components as:
\begin{eqnarray} \label{bgr}
b &=& - \frac{1}{\sqrt{3}} \delta_{a8} ,
\nonumber
\\
g &=& \frac{1}{\sqrt{3}} ( \frac{1}{2} \delta_{a8} 
- \frac{\sqrt{3}}{2} \delta_{a3} ) ,
\nonumber
\\
r &=& \frac{1}{\sqrt{3}} ( \frac{1}{2} \delta_{a8} 
+ \frac{\sqrt{3}}{2} \delta_{a3} ).
\end{eqnarray}
Because of these compositions and the structure of Gauss's law, it becomes interesting, and to some 
extent natural, to
make some choices among all the possibilities for the gauge fields.
Usually, to keep contact to the known Abelian case, the scalar potentials can be mostly associated to 
chromo-electric (c-electric) fields 
and the vector potentials mostly  to the chromo-magnetic  (c-magnetic) fields,
although non-Abelian contributions can modify them.
Therefore, for static potentials configurations,
 the following sector of the gauge fields will be considered:
\begin{align} \label{aligment1}
A_a^0 (x)  &= \left( \delta_{a3} A_3^0 (\vec{r}) + \delta_{a8}
A_8^0 (\vec{r})  \right),
\\ \label{alignment2}
\vec{A}_a (x) &= \sum_{j \neq 3,8} \delta_{a,j} \vec{A}_{j} (\vec{r}),
\end{align}
\textit{i.e.} the temporal component of the gauge field is given by the elements corresponding 
to the  diagonal 
generators of the algebra  and the vector potential to the others.

\subsection{ Sub-groups}

The $SU(3)_c$ group has 
three  $SU(2)$ subgroups, namely the so-called I-spin, V-spin, and U- spin groups.
 They can be defined by the generators of $SU(3)$ as follows:
 the I-spin ($a=1,2$),
V-spin ($a=4,5$) and U-spin ($a=6,7$) and the 
complete sets of equations for each case  are displayed in the Appendix (\ref{EquaçõesSU2}).
It is possible to write the sub group equations in an unified way by considering some specific combinations of 
the gauge potentials, such as (\ref{aligment1}) and (\ref{alignment2}),
 corresponding to alignments in color space.
For the choices presented in Table (\ref{tabtabela1}) 
 the following equations are obtained ($(c,d)=(1,2)$ or $(4,5)$ or $(6,7)$):
\begin{align}\label{EquaçãoGeneralizada1}
\nabla^2\sigma &= -j_\sigma,\\
 \label{EquaçãoGeneralizada2}
\nabla^2 \varphi - g^2 \left( A_{c}^2 + A_{d}^2 \right) \varphi &= -\rho,
\\ \label{EquaçãoGeneralizada3}
\vec{\nabla}\left( \vec{\nabla} \cdot \vec{A}_c \right) - \vec{\nabla}^2 \vec{A}_c - g^2 \varphi^2 \vec{A}_c &=0,\\
g \varphi \vec{\nabla} \cdot \vec{A}_c &= j^0_c,
\\ \label{EquaçãoGeneralizada4}
g \left( \vec{\nabla} \times \left( \vec{A}_{c} \times \vec{A}_{d} \right) - \vec{A}_d \times \vec{\nabla} \times \vec{A}_c + \vec{A}_c \times \vec{\nabla} \times \vec{A}_d \right) &= 0.
\end{align}
 
 From now on, we refer
 to the letters $(c,d)$ to the corresponding values in  Table (\ref{tabtabela1}):
\begin{table}[H]
\centering
\begin{tabular}{|c|c|c|c|}
\hline
\diagbox{Variable}{$(c,d)$}   & $(1,2)$ & $(4,5)$ & $(6,7)$  \\
\hline\hline
$\sigma$ & $A^0_8$ & $(A^0_8 - \sqrt{3} A^0_3)/2$ & $(A^0_3 + \sqrt{3} A^0_8)/2$ 
\\
\hline
$\varphi$ & $A^0_3$ & $(A^0_3 + \sqrt{3}A^0_8)/2$ & $(A^0_3 - \sqrt{3} A^0_8)/2$
 \\
\hline
$j_\sigma$ & $j^0_8$ & $(j^0_8 - \sqrt{3} j^0_3)/2$ & $(j^0_3 + \sqrt{3} j^0_8)/2$  \\
\hline
$\rho$ & $j^0_3$ & $(j^0_3 + \sqrt{3}j^0_8)/2$ & $(j^0_3 - \sqrt{3} j^0_8)/2$\\
\hline
$\vec{\varphi}$ & $\vec{A}_3$ & $(\vec{A}_3 + \sqrt{3}\vec{A}_8)/2$ & $(\vec{A}_3 - \sqrt{3} \vec{A}_8)/2$ \\
\hline
$A^0$ & $A^0_{1,2}$ & $A^0_{4,5}$ & $A^0_{6,7}$\\
\hline  
\end{tabular}
\caption{The corresponding quantities for different subgroups for the set of eqs.  
(\ref{EquaçãoGeneralizada1} - \ref{EquaçãoGeneralizada4}):
I-spin (1,2), V-spin (4,5) and U-spin (6,7).
 }
 \label{tabtabela1}
\end{table} 
 
If we are interested in tackling the I-Spin case, for instance, we'd choose $(c,d)=(1,2)$, so all the other vector potentials 
(i.e., the 4,5,6,7) are zero. In this case, the color-charge associated with the Abelian 
direction $a=8$ is given by Poisson's equation (electrostatic solution), while the charge in the 
Abelian direction $a=3$ receives non-Abelian contributions
as seen in Eq. (\ref{EquaçãoGeneralizada2})  for the I-spin.
It's interesting to notice that, equations above are also invariant 
under 
 $\vec{A}_c \to \pm \vec{A}_d$
together with $j^0_c \to \pm j^0_d$.
Correspondingly:
$\left(\vec{E}_c, \vec{B}_c \right) \rightarrow \pm \left( \vec{E}_c, \vec{B}_c \right)$.

It is also somewhat convenient, as seen in equations above, to adopt the Coulomb gauge:
\begin{eqnarray}\label{CoulombGauge}
\vec{\nabla} \cdot \vec{A}_c = 0,
\end{eqnarray}
 in which we necessarily must have $j^0_c = j^0_d=0$.
 With that, the non-Abelian directions $(c,d)$ are completely 
equivalent, so we choose to make, in each case, for the non-vanishing vector potentials:
\begin{align}
\vec{A}_c &= \vec{A}_d \equiv \vec{A}.
\end{align}
A completely equivalent choice would be $\vec{A}_c = -  \vec{A}_d \equiv \vec{A}$.

Also, the c-eletric and c-magnetic fields, from Appenidx \ref{appEcBc}, become:
\begin{align} \label{EaUspin}
\vec{E}_a =& \delta_{a3} \left( - \vec{\nabla} A^0_3  \right) + \delta_{a8} \left(- \vec{\nabla} A^0_8  \right) + \delta_{ac} \left(g \varphi  \vec{A} \right) +\delta_{ad} \left( - g \varphi \vec{A} \right);\\  \label{BaUspin}
\vec{B}_a =& \delta_{ac} \left( \vec{\nabla} \times \vec{A} \right) + \delta_{ad} \left( \vec{\nabla} \times \vec{A} \right) .
\end{align}
Therefore the non-Abelian effects will appear firstly in the c-clectric sector and secondly, as an indirect effect,
in the c-magnetic field.

 In the next sections, some solutions for $\varphi$ will be presented, 
but one has to keep in mind that $\varphi$, for U-spin and V-spin,
is a combination of the scalar potentials.
To obtain a complete solution for $A^0_3$ and $A^0_8$  
a electromagnetic-type solution for the Poisson equation ($\sigma$) 
that must be added or subtracted.
Since for some of the cases discussed in this work,
the color-charge distribution is not 
 simply a single punctual charge,
 we also can impose the following  restriction: 
\begin{eqnarray} \label{sigma0}
\sigma=0.
\end{eqnarray}
This means that we restrict solutions to the following cases
$A_8^0 = c A_3^0$ where
$c = \sqrt{3}$ ($- 1/\sqrt{3}$)
for the V-spin (U-spin).
Therefore the $SU(3)_c$ problem is reduced to 
 one of the $SU(2)$ subgroups and the equations to be solved are the
 ones in Appendix (\ref{EquaçõesSU2})  or
 the eqs. above (\ref{EquaçãoGeneralizada1} - \ref{EquaçãoGeneralizada4}).
Another possibility we will envisage is that the color source for each of the equations 
for the SU(2) subgroups
 $q_\eta$ can be written as:
\begin{align} \label{qeta}
q_I &= r-g;\\
q_V &= r-b;\\
q_U &= b-g.
\end{align}
Therefore, to address the   punctual color charge  one might consider either
the fundamental or the adjoint representation.
In the latter case, one might deal with two color-charges and two anti-color-charges
(of the fundamental representation)
as discussed in the section (\ref{sec:classicalmeson}).

\subsection{{\it Effective} Abelian Limits }

The Abelian case is
 the trivial one in the sense that it reduces completely to the eletromagnetic theory.
Even if one is interested in the non-Abelian effects, 
it might be 
useful to have an approach that only partially meets the criteria in table (\ref{tabtabela1}), so 
that
 a solvable set of equations can be handled analytically.
 This process is associated with the alignment in internal space, and its discussed in \ref{secIVU}.

The usual trivial Abelian limit of the four equations above (\ref{gaussEc} - \ref{ampere})
 corresponds to the Maxwell electromagnetism and it is
obtained with $g=0$.
There are, however, other gauge field configurations that 
present the same limit of the equations above without taking $g=0$.
Therefore they {\it effectively} behave as Abelian fields. 
Different Abelian limits can be extracted in certain gauges.
For example, in the  Coulomb Gauge
 $f_{abc} A_c^0 \nabla \cdot \vec{A}_b=0$.
The complete {\it Abelian-limit} can be reached by imposing the following
\begin{eqnarray} \label{abelianlimit}
f_{abc} f_{cde} \vec{A}_b \cdot \vec{A_d} A^0_e =0.
\end{eqnarray}

The different possible {\it effective} Abelian limits of the  SU(2)  subgroups are summarized in 
Table (\ref{tabtabela2}).
\begin{table}[ht]
\centering
\begin{tabular}{|c|c|}
\hline
Variable & Description \\
\hline
$\varphi=\vec{\varphi}=0$ & There are no charge distributions. $\vec{A}$ and $A^0$ are given by Laplace's equation.\\
\hline
$\varphi = A^0 = \vec{A} = 0$&  Two associated copies of electromagnetism.\\
\hline
$A^0 = \vec{A} = 0$& The $\varphi$ potential is given by Poisson's equation, while $\vec{\varphi}$ is given by Laplace's.\\
\hline
$A^0 = \vec{A} \times \vec{\varphi} = \vec{\varphi} \times \vec{\nabla} \times \vec{A} = 0$& No charges, while $\vec{A}$ and $\vec{\varphi}$ have the same (laplace) equation.\\
\hline
$\vec{\varphi} = \vec{A} = \varphi \vec{\nabla} A^0 - A^0 \vec{\nabla} \varphi = 0$& Poisson's equation for $\varphi$ and Laplace's for $A^0$.\\
\hline
\end{tabular}
\caption{The different Abelian cases obtained from eq. (\ref{abelianlimit}).}
\label{tabtabela2}
\end{table}

\subsection{ Finite energy condition for constant potentials}

A great concern when investigating gauge fields, classical or not, is 
the energy (and pressure) content associated with their configurations.
One usually needs to search  non only the lowest energy level configuration 
but also one may have  to guarantee finiteness of the total energy \cite{actor-review}
and references therein.
The energy-momentum tensor for the Yang-Mills fields is written explicitly in Appendix
 (\ref{appemtensor}).
The energy density, eq. (\ref{EMTensor00}),  can then be written as:
\begin{align}\label{energyatinfinity}
\theta^{00} &= \frac{1}{2}\sum_a \mid \vec{\nabla} A^0_a \mid ^2 + \frac{1}{2}\sum_a \mid \vec{\nabla} \times \vec{A}_a \mid ^2 + g^2 \left( \mid \vec{\varphi} \mid^2 (A^0)^2 + \varphi^2 A^2 - \varphi A^0 \vec{\varphi} \cdot \vec{A} + \mid \vec{\varphi} \times \vec{A} \mid^2 \right),
\end{align}
where the $a$ index covers the $SU(2)$ indices.
By considering that either  inside a  spatial region  or in asymptotically large distances
 the four-potential assume nearly constant values and don't vanish, 
we have a condition for finite energy 
given by:
\begin{align}\label{cosPot}
\cos(\zeta) |_{r\to \infty} &=  \frac{\vec{\varphi} \cdot \vec{A}}{\mid \vec{\varphi} \mid \mid \vec{A} \mid}
=
 -\frac{\varphi A^0}{2\mid \vec{\varphi} \mid \mid \vec{A} \mid} \pm \sqrt{\left( \frac{\varphi A^0}{2\mid \vec{\varphi} \mid \mid \vec{A} \mid} \right)^2 + \left( \frac{\varphi}{\mid \vec{\varphi} \mid} \right)^2 + \left( \frac{A^0}{ \mid \vec{A} \mid} \right)^2 + 1}.
\end{align}
These functions are valued inside a finite spatial region.
 It's important to notice that, from (\ref{energyatinfinity}), we see that if any of the potentials is zero at $r\rightarrow\infty$, all of them must also be for finite energy (since the negative term vanishes). 
Except for this latter situation,
 equation \ref{cosPot} gives a constraint between the different potentials, depending on the angle between the vector potentials in the Abelian ($\vec{\varphi}$) and non-Abelian ($\vec{A}$) directions. The simplest cases ($\zeta=0,\pi/2,\pi$) give the following constraints:

$\bullet$ $\zeta = 0$:
\begin{align}
\varphi A^0  = \varphi^2 \frac{|\vec{A}|}{|\vec{\varphi}|} + (A^0)^2 \frac{|\vec{\varphi}|}{|\vec{A}|}.
\end{align}

$\bullet$ $\zeta = \frac{\pi}{2}$:
\begin{align}
\varphi = A^0 = 0.
\end{align}

$\bullet$ $\zeta = \pi$:
\begin{align}
\varphi A^0 = - \varphi^2  \frac{|\vec{A}| }{|\vec{\varphi}|} - (A^0)^2 \frac{|\vec{\varphi}| }{|\vec{A}|}.
\end{align}
The special case for \ref{cosPot} that minimizes \ref{energyatinfinity} reads:
\begin{align}
\cos(\zeta) = -\frac{\varphi A^0}{2\mid \vec{\varphi} \mid \mid \vec{A} \mid}.
\end{align}
This corresponds to the conditions for $\zeta=\pi/2$
which, again, must mean that all the potentials are zero in this region). With that in mind,
by using  the alignment used in 
\ref{aligment1} and \ref{alignment2}, 
the energy for the trivial case $\zeta = \frac{\pi}{2}$
 is lower than any configuration in the full classical $SU(2)$ picture
 with constant potentials at infinity or as from a finite surface.
 This emphasizes the need for the boundary condition that returns 
$A_\mu=0$ in such a region (or as from it).

\section{ Deviations from spherically symmetric solutions}
 \label{sec:deviationsspherical}

Let us consider
the equations for $\varphi$ and $\vec{A}$, Eqs. 
(\ref{EquaçãoGeneralizada2})
 and (\ref{EquaçãoGeneralizada3}),
 in the spherical coordinate system,  wtih  the definitions in Table (\ref{tabtabela1}). 
For an unique  single vector potential,  we write:
\begin{eqnarray} \label{Aspher}
\vec{A} (\vec{r}) = 
A_r (r,\theta,\phi) \hat{r} + A_\theta (r,\theta,\phi) \hat{\theta}
+ A_\phi (r,\theta,\phi) \hat{\phi}.
\end{eqnarray}
The equations for $\varphi$ and $\vec{A}$ - (\ref{EquaçãoGeneralizada2}) and
 (\ref{EquaçãoGeneralizada3})
- in each direction ($\hat{r}, \hat{\theta}, \hat{\phi}$) \cite{laplacianoMoon}, 
can be written  
as :
\begin{align} \label{varphiesf}
\nabla^2 \varphi - 2g^2A^2 \varphi &= -\rho; \\\label{fesfr}
\nabla^2 A_r - \frac{2 A_r}{r^2} - \frac{2}{r^2} \frac{\partial A_\theta}{\partial \theta} - \frac{2\cot(\theta)}{r^2} A_\theta - \frac{2}{r^2 \sin (\theta)} \frac{\partial}{\partial \phi} A_\phi + g^2 \varphi^2 A_r &= 0;
\\ \label{fesftheta}
\nabla^2 A_\theta  + \frac{2}{r^2} \frac{\partial A_r}{\partial \theta} - \frac{1}{r^2 \sin^2(\theta)} A_\theta  - \frac{2 \cos (\theta)}{r^2 \sin( \theta)} \frac{\partial A_\phi}{\partial \phi} + g^2 \varphi^2 {A}_\theta &= 0;
\\ \label{fesfphi}
\nabla^2 A_\phi - \frac{1}{r^2 \sin^2 (\theta)}A_\phi + \frac{2}{r^2 \sin (\theta)} \frac{\partial A_r}{\partial \phi} + \frac{2 \cos(\theta)}{r^2 \sin^2(\theta)}\frac{\partial A_\theta}{\partial \phi}+ g^2 \varphi^2 {A}_\phi  &= 0.
\end{align}
A few solutions for these equations will be worked out below.
Before doing that, the possibility of complete spherically symmetric solutions for these equations will be
discussed.

Let us assume there is a spherically symmetric solution for
the vector potential
 that can lead to circular c-magnetic field configurations, being independent of $\phi$.
By imposing that:
\begin{align} \label{spher}
\vec{A}(\vec{r}) &= \vec{A}(r);\\
\varphi (\vec{r}) &= \varphi (r),
\end{align}
equations \ref{fesftheta} and \ref{fesfphi} can be written respectively as:
\begin{align}
\frac{1}{r^2}\frac{d}{dr}\left( r^2 \frac{dA_\theta}{dr} \right) - \frac{1}{r^2 \sin^2(\theta)} A_\theta  + g^2 \varphi^2 {A}_\theta &= 0;
\\
\frac{1}{r^2}\frac{d}{dr}\left( r^2 \frac{dA_\phi}{dr} \right) - \frac{1}{r^2 \sin^2 (\theta)}A_\phi + g^2 \varphi^2 {A}_\phi  &= 0.
\end{align}
From this equations it is clear that:
\begin{align}
A_{\theta} = A_{\phi} &= 0.
\end{align}
This means only the radial component of the vector potential may
 be non-zero. From Coulomb's gauge condition, we obtain:
 \begin{align}
 A_r(\vec{r}) = \frac{C}{r^2}.
 \end{align}
Plugging this into Eq. \ref{fesfr}, it yields :
\begin{align}\label{trivialesf}
g^2 \varphi^2 A_r &=0.
\end{align}
From \ref{trivialesf} we have three possibilities:
\begin{enumerate}
\item $g=0$ (Abelian limit);
\item $\vec{A}_{c,d}=0$ (Effective electrostatic regime);
\item $\varphi = 0$ (no 
color charges).
\end{enumerate}
 With that, we conclude that for the separate  (I,U,V)-spin cases, given by \ref{EquaçãoGeneralizada1} - \ref{EquaçãoGeneralizada4}, 
there aren't non-trivial solutions with full spherical symmetry for the scalar and vector potentials.
Also, the completely symmetric scalar potential  necessarily implies: $\vec{B}_c = 0$. 
These conclusions   agree with
Refs. \cite{ikeda-miyachi-1962,Loos-1965}
according to which all spherically symmetric solutions for the YM equations
 can be  gauge transformed to the Coulomb potential being therefore
 Abelian solutions,
although we make use of a considerable simpler argument.
These conclusions extend
more restricted ansatzse
 for defining a spherically symmetric solution or configuration, for example in Ref.
\cite{hadicke-pohle-spher-sym}.

\subsection{ Equations for spherically symmetric scalar potential}
\label{secPartialSym}

Let us consider  a spherically symmetric scalar potential,
$\varphi (\vec{r}) = \varphi(r)$.
The corresponding vector potential will be considered to have an axial symmetry
being independent of the coordinate $\phi$ for all the cases addressed below.
I can be  therefore written as:
\begin{align}\label{SEMISYMADIRECTION}
\vec{A} (\vec{r}) &= A_\theta(r,\theta) \hat{\theta} + A_\phi(r,\theta) \hat{\phi}.
\end{align}
In this case, the Coulomb gauge condition is given by:
\begin{align}
 \frac{1}{r \sin(\theta)} \frac{\partial}{\partial \theta} \left( A_\theta \sin(\theta) \right) =0,
\end{align}
and therefore the following angular dependency is obtained:
\begin{align} \label{Theta-theta}
A_\theta(\vec{r}) = \frac{R_\theta(r)}{\sin(\theta)}.
\end{align}
It's interesting to notice that this is a general result for the prescription given in 
(\ref{SEMISYMADIRECTION}), in the Coulomb gauge.
As discussed above, to have non-Abelian solutions one must have an angular dependence.
The resulting equations, with a
color charge distribution $\rho^c(\vec{r})$, can be solved
by separation of variables, $A_\phi = R_\phi(r) \Theta_\phi(\theta)$, and they can be written as:
\begin{align} \label{SEMISYM1}
\frac{1}{r^2} \frac{d}{dr}\left( r^2 \frac{d \varphi}{d r} \right) - 2g^2 \left( \frac{R_\theta^2}{\sin^2(\theta)} + \left(R_\phi \Theta_\phi \right)^2 \right) \varphi &= - \rho^{c} (\vec{r}) ;
\\  \label{SEMISYM2}
r^2 \frac{d^2R_\theta}{dr^2} +2r \frac{dR_\theta}{dr} + (g \varphi r)^2 R_\theta &= 0;
\\  \label{SEMISYM3}
\frac{1}{R_\phi} \frac{\partial}{\partial r} \left( r^2 \frac{\partial R_\phi}{\partial r} \right) + \frac{1}{ \sin(\theta) \Theta_\phi} \frac{\partial}{\partial \theta} \left( \sin(\theta) \frac{\partial \Theta_\phi}{\partial \theta} \right) - \frac{1}{\sin^2 (\theta)} + \left(g\varphi r\right)^2   &= 0.
\end{align}
Note that the corresponding dependence on $\theta$, i.e.  $\Theta_\theta (\theta)$, was 
found in Eq. (\ref{Theta-theta}).
Some solutions for these equations are presented in the next sections.

\section{ Vector potential $\vec{A}_{c,d}$ associated to the Coulomb potential $\varphi$}
\label{SectionCoulomb}

The Coulomb potential is one of the very well known   solutions
for the Abelian and non-Abelian scalar potentials.
For the different SU(2) subgroups the different gauge directions 
in the fundamental representation (\ref{bgr}) will be labeled by
$\eta= I, V,U$ that may or not be a combination of $r, g, b$ 
according to Table (\ref{tabtabela1}), given in
Eq. (\ref{qeta}).
The Coulomb solution will be given by:
\begin{align} \label{coulomb}
\varphi (\vec{r}) = \varphi_\eta (\vec{r}) = \frac{q_\eta}{ 4 \pi r}.
\end{align}

Next, we  calculate the corresponding vector potential resulting from the equations
(\ref{SEMISYM1}-\ref{SEMISYM3}) by requiring that part of the
 non-Abelian scalar potential to be given by (\ref{coulomb}).
Due to the non-Abelian interactions, the chromo-electric field of course will receive 
further corrections and present deviation from the Coulomb field.
Let us consider a punctual color-charge at the origin and an additional unknown color-charge
distribution both contributions encoded in $\rho(\vec{r})$.
By denoting $\cos(\theta)\equiv x$, for the  variables $R_\theta$ and $R_\phi$ defined above,
 the following equations are obtained from 
the non-Abelian generalizations of 
the Gauss's and Amp\`ere's Law \ref{SEMISYM1} - \ref{SEMISYM3}, respectively:
\begin{align} \label{coulomb1}
2g^2 \left( \left( R_\theta \Theta_\theta \right)^2 + \left( R_\phi \Theta_\phi \right)^2 \right) \frac{q}{4\pi r} &= \rho (\vec{r}) - q \delta(\vec{r}),
\\ \label{coulomb2}
r^2 \frac{d^2R_\theta}{dr^2} +2r \frac{dR_\theta}{dr} + \left( \frac{qg}{4\pi} \right)^2 R_\theta &= 0,
\\   \label{coulomb3}
r^2 \frac{d^2R_\phi}{dr^2} + 2r \frac{dR_\phi}{dr} - \beta R_\phi &=0,
\\ \label{coulomb4}
(1-x^2) \frac{d^2\Theta_\phi}{dx^2} -2x \frac{d\Theta}{dx} + \left( l(l+1)- \frac{1}{1-x^2} \right) \Theta_\phi &=0,
\end{align}
where   eq. (\ref{coulomb1}) 
  can be seen as a sort of {\it constraint} that will in fact determine
 the corresponding charge distribution that produces the Coulomb potential for $\varphi$.
The relation between the integration constants $l$ and $\beta$ in the equations above is given by:
\begin{align} \label{beta}
\beta &= l(l+1) - \left( \frac{q_\eta g}{4\pi} \right)^2 \equiv l(l+1) - \alpha^2.
\end{align}
Note that, in spite of the label $\alpha$, the constant $q g/(4 \pi)$ is not exactly the QCD 
fine structure constant $\alpha_s (M^2)$ \cite{alpha-qcd},
whose definition is usually quantum.
To solve the equations for the vector potential  there is no  
need to know the charge density $\rho(\vec{r})$.

Equations \ref{coulomb2} and \ref{coulomb3} are Euler's differential equations \cite{butkov}, 
with  the  following solutions:
\begin{align}
R_\theta(r) &= a_\alpha r^{\frac{-1 + \sqrt{1-4 \alpha^2}}{2}} + b_\alpha r^{\frac{-1 - \sqrt{1-4 \alpha^2}}{2}};\\
R_\phi(r) &= c_\beta r^{\frac{-1 + \sqrt{1+4 \beta}}{2}} + d_\beta r^{\frac{-1 - \sqrt{1+4 \beta}}{2}}.
\end{align}
Equation \ref{coulomb4}
is Legendre's associated equation with degree $\pm 1$ and order $l$, so:
\begin{align}
\Theta_\phi(\theta) = p_l P_l^1(\cos(\theta)) + q_l Q_l^1(\cos(\theta)).
\end{align}
with $p_l$ and $r_l$ being the integration constants, while $P_l^m(x)$ and $Q_l^m(x)$ are the associated Legendre's functions of order $m$ and degree $l$, of the first and second kind, respectively.

By considering (\ref{beta}), one can write these solutions in an uniform way only in terms of $\alpha$ 
and $l$ as:
\begin{align}
\vec{A} \left( \vec{r},\alpha \right) &= \frac{1}{\sin(\theta)}\left( a_\alpha r^{\frac{-1 + \sqrt{1-4 \alpha^2}}{2}} + b_\alpha r^{\frac{-1 - \sqrt{1-4 \alpha^2}}{2}} \right) \hat{\theta} \\
&+ \sum_{l=1}^{\infty}\left( c_l r^{\frac{-1 + \sqrt{1+4 l(l+1)-4\alpha^2}}{2}} + d_l r^{\frac{-1 - \sqrt{1+4 l(l+1)-4\alpha^2}}{2}} \right) \left( p_l P_l^1(\cos(\theta)) + q_l Q_l^1(\cos(\theta)) \right) \hat{\phi}.
\end{align}
Note that, to assure stability of the (non-imaginary) solution,
 there is an upper and lower bound for $\alpha$:
\begin{align} \label{cond-qg-coulomb}
0 \leq \alpha = \mid \frac{q g}{4\pi} \mid = \sqrt{l(l+1) - \beta} \leq \frac{1}{2}.
\end{align}
If one considers results from lattice QCD  for the running coupling constant
at low energies \cite{lattice-g}
$g \sim 1$ and with $q$ an integer number,  this bound is satisfied even in the quantum case.
Equivalent conditions for the stability of the solutions 
have been found in other works that investigated time-dependent configurations
 \cite{mandula}. 
Although
usually the stability is analyzed in time-evolving systems,
give that Gauss' law is satisfied at an initial condition, that could be the above solution,
the time evolution should preserve it \cite{CYM-book}.
This remark is valid for the solutions analyzed in the following sections.

So we see that the coupling $\alpha$ gives the regime of interaction,
 in the region we can approximate the scalar potential by the Coulomb one. 
Some of the  simplest non-trivial cases (that can be  taken together)
 may be seen as:
\begin{itemize}
\item $a_\alpha = b_\alpha =  \rightarrow$ no divergence for $\vec{A}$ at $\theta = 0,\pi$ and
\item $l=1 \rightarrow$ lowest order of angular dependence.
\end{itemize}
All of these, together with the condition of $\vec{A}(r\rightarrow \infty) = 0$,
 result in:
\begin{align}
\vec{A}(\vec{r}) = d_1 r^{\frac{-1-\sqrt{9-4\alpha^2}}{2}} \cos(\theta) \hat{\phi}.
\end{align} 
Note that the dimension of the constant of integration depends on $\alpha$.
In particular, for the Abelian case, $\alpha=0$ and   $d_1=0$.
The non-Abelian correction induces a redefinition of $d_1$ and a slight deviation 
from the dipolar solution since $\alpha < 1/2$.
This will be made explicit in the following.
 $d_1$ can be defined  by a new length parameter $r_0$ as:
\begin{align}
d_1 =  r_0^{\frac{-1+\sqrt{9-4\alpha^2}}{2}},
\end{align}
and the vector potential is written as:
\begin{align}
\vec{A}(\vec{r}) = \frac{1}{r_0} \left( \frac{r_0}{r} \right)^{\frac{1}{2}
+\sqrt{\frac{9}{4}-\alpha^2}} \cos(\theta) \hat{\phi}.
\end{align}
Notice that, in order to regain the Abelian solution as
 $\alpha \rightarrow 0$, $r_0$ must be a function of $\alpha$ such that: 
\begin{align}
\lim_{\alpha\rightarrow 0} r_0(\alpha) = 0 .
\end{align}

The resulting c-electric and c-magnetic fields,  by adding the contribution of the
Coulomb potential  for each subgroup respectively:
\begin{align}\nonumber
\mbox{\bf{I-spin}}\\ \label{CoulombEIspin}
\vec{E}_a(\vec{r}) &= \frac{q_\eta}{4\pi} \frac{\vec{r}}{r^3} \delta_{a3} + \frac{gq}{4\pi}\frac{1}{r_0^2} \left( \frac{r_0}{r} \right)^{ T + 1}  \cos(\theta) \hat{\phi} \left( \delta_{a1} - \delta_{a2} \right),
\\ \label{CoulombBIspin}
\vec{B}_a(\vec{r}) &= \left( \frac{\cos(2\theta)}{\sin (\theta)}  \hat{r} - \cos(\theta) \left( T-1  \right)   \hat{\theta} \right) \frac{1}{r_0^2} \left( \frac{r_0}{r} \right)^{
T+1} \left( \delta_{a1} + \delta_{a2} \right).
\\ \nonumber
\mbox{\bf{V-spin}}
\\
\vec{E}_a(\vec{r}) &= \frac{q_\eta}{8\pi} \frac{\vec{r}}{r^3} \left( \frac{\delta_{a3} + \sqrt{3}\delta_{a8}}{2} \right) + \frac{gq}{4\pi}\frac{1}{r_0^2} \left( \frac{r_0}{r} \right)^{T+1}  
\cos(\theta) \hat{\phi} \left( \delta_{a4} - \delta_{a5} \right),
\\
\vec{B}_a(\vec{r}) &= \left( \frac{\cos(2\theta)}{\sin (\theta)}  \hat{r} - \cos(\theta) \left( T-1 \right)   \hat{\theta} \right) \frac{1}{r_0^2} \left( \frac{r_0}{r} \right)^{
T+1} \left( \delta_{a4} + \delta_{a5} \right).
\\ 
\nonumber
\mbox{\bf{U-spin}}
\\ \label{coulombEUspin}
\vec{E}_a(\vec{r}) &= \frac{q_\eta}{4\pi} \frac{\vec{r}}{r^3} \left(\delta_{a3}- \frac{\sqrt{3}}{3}\delta_{a8} \right) + \frac{gq}{4\pi}\frac{1}{r_0^2} \left( \frac{r_0}{r} \right)^{T+1} 
 \cos(\theta) \hat{\phi} \left( \delta_{a6} - \delta_{a7} \right),
\\
\label{coulombBUspin}
\vec{B}_a(\vec{r}) &= \left( \frac{\cos(2\theta)}{\sin (\theta)}  \hat{r} - 
\cos(\theta) \left( T-1 \right)   \hat{\theta} \right) \frac{1}{r_0^2} \left( \frac{r_0}{r} \right)^{
T+1} \left( \delta_{a6} + \delta_{a7} \right),
\end{align}
where the punctual  color charge $q$ stands for a particular subgroup given in 
Eqs. (\ref{qeta}), and
where $T=\frac{1}{2}+\sqrt{\frac{9}{4}-\alpha^2}$.
In this expression, to assure real solutions, $\alpha$ is bound as shown above.
In the Abelian limit only the Coulomb potential remains non zero.
The two components of the c-magnetic field have different angular dependence,
$\cos(2 \theta)/\sin(\theta)$ and $\cos(\theta)$.
The c-magnetic field has analytical cuts  along the direction $\hat{r}$ for $\theta=0$ and $\theta=\pi$,
besides the singularity at $r=0$.
It may be seen as  a (static)  non-Abelian
effect that mimics a color-current  
that yield the magnetic field.

\begin{figure}[ht!]
    \centering
    \begin{minipage}{0.5\textwidth}
        \centering
        \includegraphics[width=1.\textwidth]{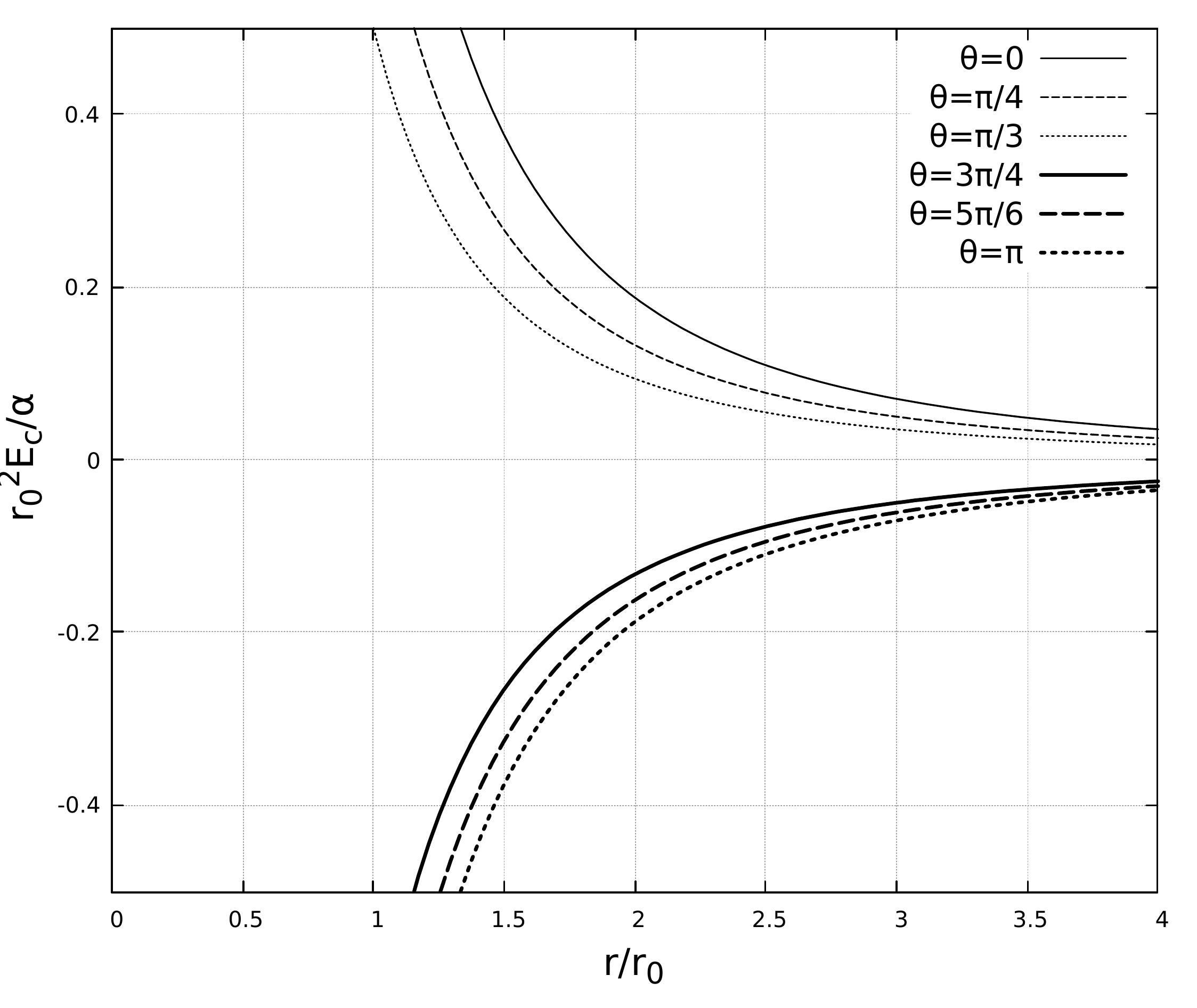}
        \caption{$\vec{E}_c \cdot \hat{\phi} \equiv E_c$, Eq. (\ref{CoulombEIspin}), as a function of $r/r_0$, for $\alpha=1/2$ for different angles $\theta$.}
        \label{figfig1Ec}
    \end{minipage}
\end{figure}

\begin{figure}[H]
    \centering
    \begin{minipage}{0.49\textwidth}
        \centering
        \includegraphics[width=1.\textwidth]{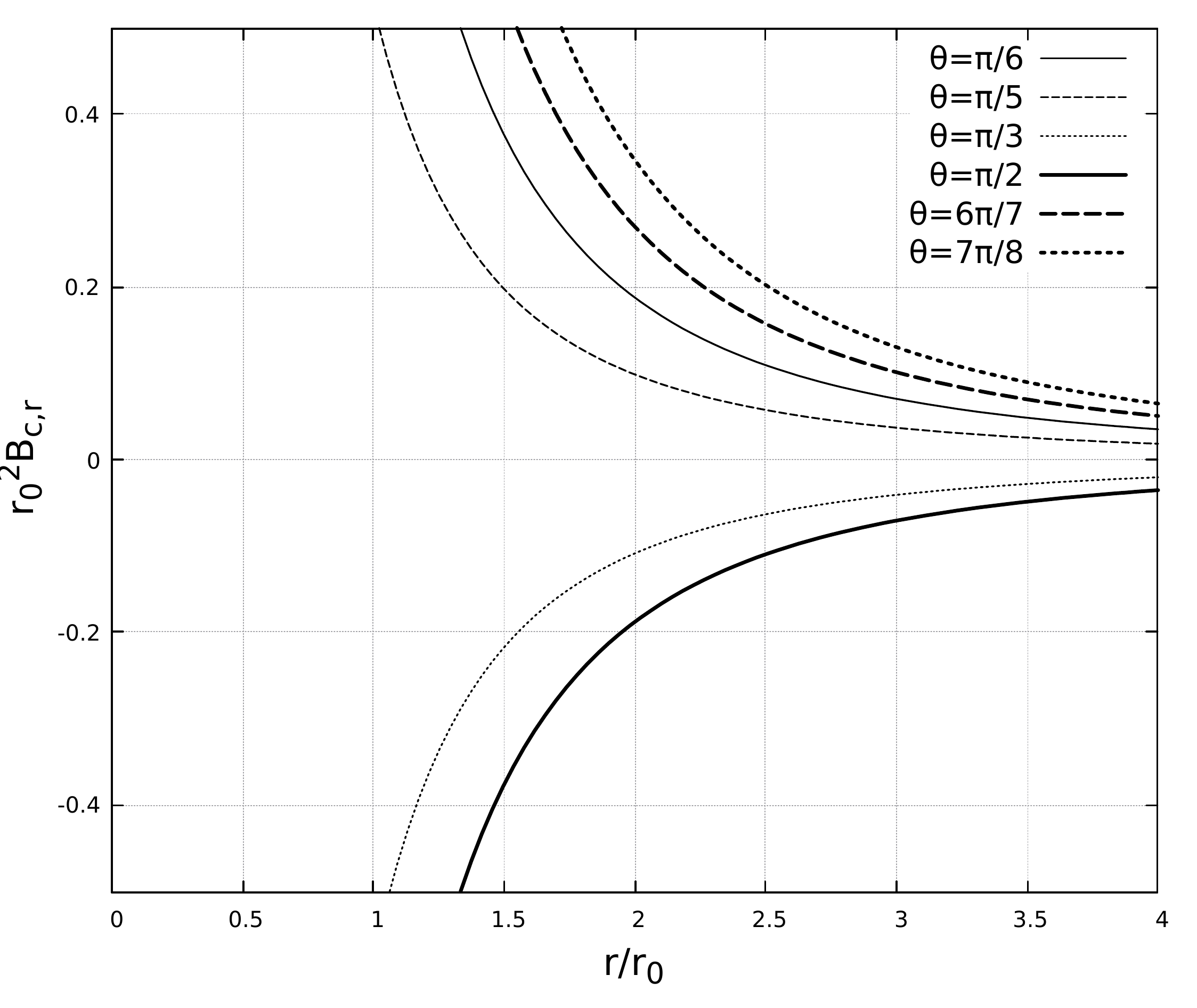}
\caption{$\vec{B}_c \cdot \hat{r} \equiv B_{c,r}$, Eq. (\ref{CoulombBIspin}),
 as a function of $r/r_0$, for $\alpha=1/2$ and different angles $\theta$.}
        \label{figfig2Bcr}
    \end{minipage}\hfill
    \begin{minipage}{0.49\textwidth}
        \centering
        \includegraphics[width=1.\textwidth]{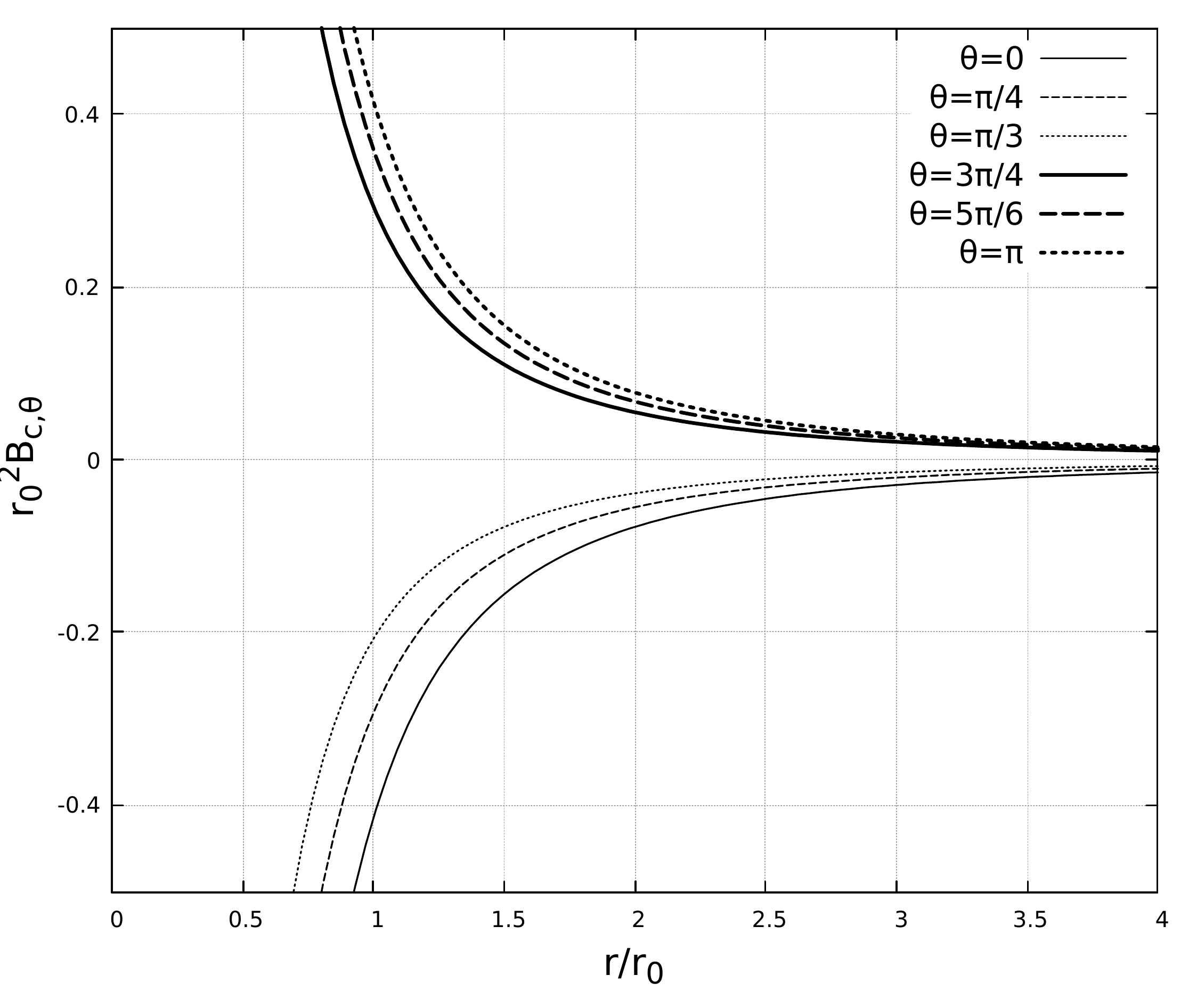}
\caption{$\vec{B}_c \cdot \hat{\theta} \equiv B_{c,\theta}$, Eq. (\ref{CoulombBIspin}),
 as a function of $r/r_0$, for $\alpha=1/2$
 and different angles $\theta$.}
        \label{figfig2Btheta}
    \end{minipage}
\end{figure}

\begin{figure}[H]
    \centering
        \includegraphics[width=0.49\textwidth]{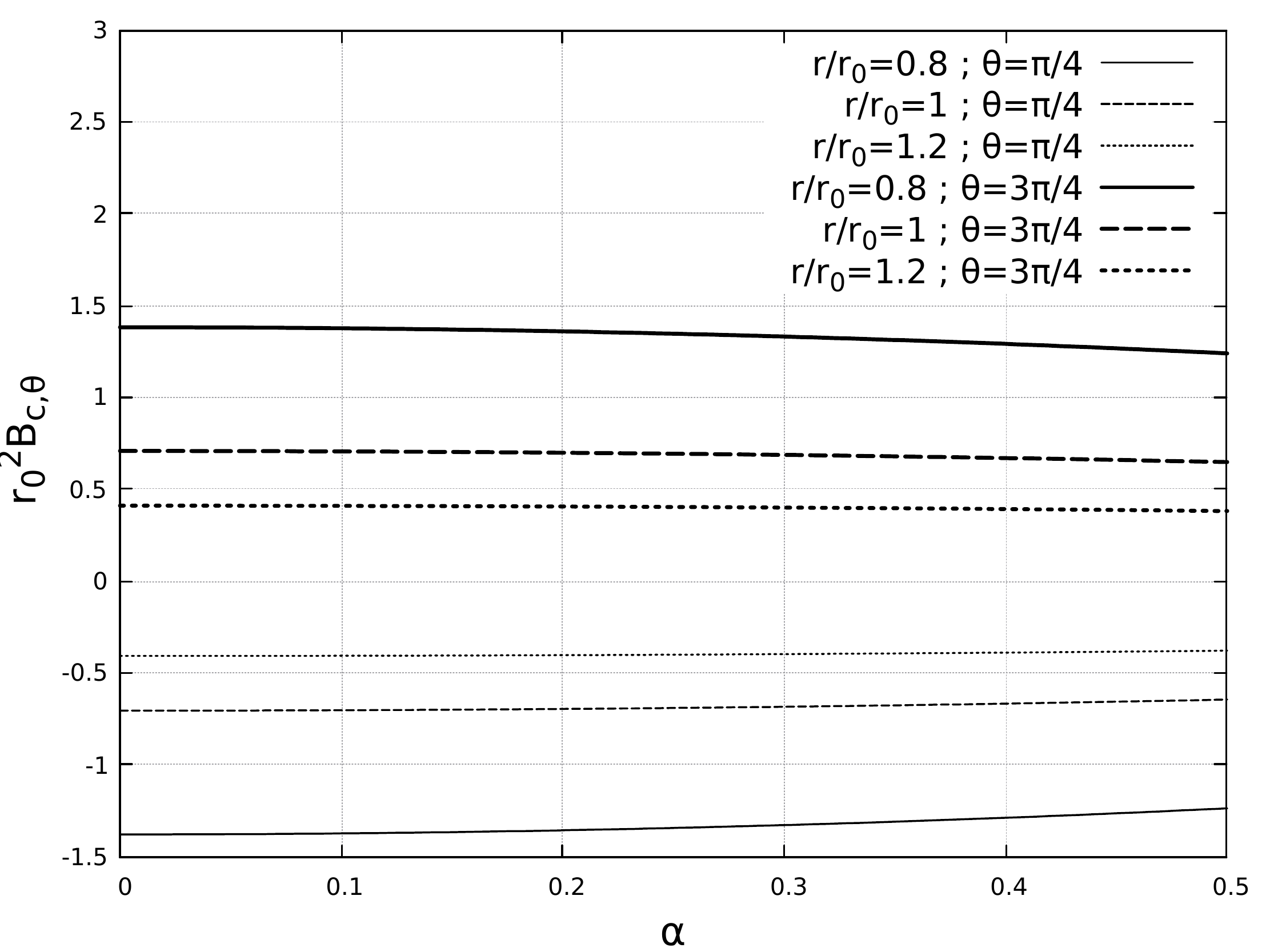}
\caption{$\vec{B}_c \cdot \hat{\theta} \equiv B_{c,\theta}$,  Eq. (\ref{CoulombBIspin}), as a function of $\alpha$
for different radial coordinates $r$ for different angles $\theta$.}
        \label{figfig3Bcalpha}
\end{figure}

The c-electric field now has a (non-Abelian) $\phi-$ component that decreases quite fast
and its normalized value is shown in Fig. (\ref{figfig1Ec}) as a function of the radial coordinate $r/r_0$
for several values of the angle $\theta$.
Its strength depends on $cos (\theta)$ with a quite complicated shape, and it does not contribute for 
the total flux across a spherical surface,   which is however anisotropic.
For this figure, and the next ones. we chose  $\alpha =1/2$ because the 
difference between curves for the different angles is more evident.
The c-magnetic field has two (non-Abelian) components with a quite complicated form and 
these components are  presented in 
Figs. (\ref{figfig2Bcr})  and (\ref{figfig2Btheta}) for different angles $\theta$.
As a purely non-Abelian effect both the Ec and the Bc fields change signs
depending on the quadrant or octant they are displayed, although the 
non-Abelian color-charge distribution has a unique sign.
The sign in each octant of the c-electric field is basically the same of the 
component $B_{c,\theta}$ and quite different from the component
$B_{c,r}$.
In Fig. (\ref{figfig3Bcalpha}) the dependence of $B_c^{\theta}$ on the
integration parameter $\alpha$ is presented to show that the choice $\alpha=1/2$ does 
not make much difference.
Of course, these highly anisotropic field configurations lead to anisotropic fluxes.
These anisotropies may be related to further consequences that are expected in
the full quantum problem,
since the c-electric and c-magnetic fields lines, and their fluxes, cannot be isotropic
inside mesons and baryons.
Therefore one might expect a sort of anisotropic confinement of fluxes
and that may  favor anisotropic quark-antiquark (color -anticolor) 
or three-color configurations. 
Although the quantum dynamics is missing, 
these effects may have  consequences for the full quantum system.
The total non-Abelian c-electric (c-magnetic) field flux   goes to the total charge
(zero)  across a closed sphere although
they are highly anisotropic as it can be noted from their expressions 
(\ref{CoulombEIspin}-\ref{coulombBUspin}).
The c-magnetic flux in different directions 
decreases with $\sim 1/r^{T}$.

\begin{figure}[H]
    \centering
    \begin{minipage}{0.5\textwidth}
        \centering
        \includegraphics[width=1\textwidth]{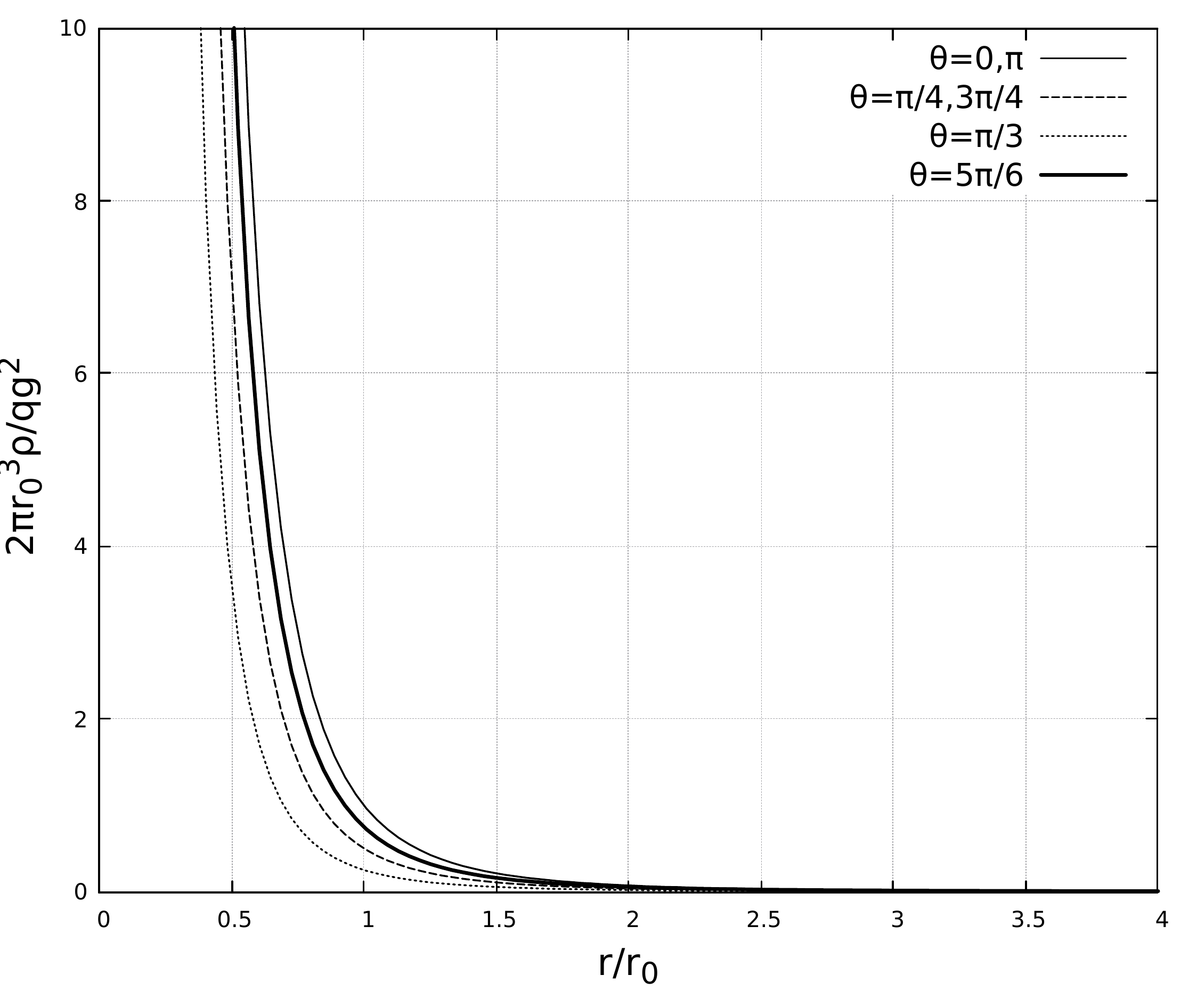} 
        \caption{$\rho (r)$ as a function of $r/r_0$, Eq. (\ref{CoulombRhoEff}),
 for $\alpha=1/2$ and different angles $\theta$.}
        \label{figfig4rho}
    \end{minipage}\hfill
\end{figure}

The corresponding non-Abelian charge density is given by:
\begin{align}\label{CoulombRhoEff}
\rho' (\vec{r}) = \rho (\vec{r})
 - q\delta(\vec{r}) &=
\frac{g^2q}{2\pi} \frac{1}{r_0^3} \left( \frac{r_0}{r} \right)^{2T+1} \cos^2(\theta) .
\end{align}
Being that $ \frac{1 + \sqrt{2}}{2} \leq T \leq 2 $.
Note that, although the scalar potential $\varphi$ is a Coulomb potential, the 
corresponding charge distribution is not a single punctual charge but it also contains
a non-spherically symmetric distribution.
This is a purely non-Abelian effect.
In Fig. (\ref{figfig4rho}) the (normalized)
 charge density profile is shown  as a function of $r/r_0$ 
for few different angles, except 
$\theta^0 =  \pi/2$  in which direction $\rho(r, \theta^0) = q\delta(\vec{r})$. 
Note that, in this direction, the solution reduces 
to the Coulombic solution according to the solutions for instance in \ref{CoulombEIspin} and \ref{CoulombBIspin}.
Non-Abelian effects, therefore, are related  to the fact that 
anisotropic  charge distribution leads to a simple Coulomb potential for $A_0$
at the expense of the non-existence of  chromo-electric/magnetic field lines in 
 part of the 
space, i.e. there appear anisotropic Ec and Bc lines due to the vector potential.
It is also associated with the  fact that these fields lines remain 
 to more restricted region(s), strictly where there is non-zero charge distribution.
 The total color- charge $Q$ contained inside a sphere of radius $R_0$ is given by:
\begin{align} \label{Q-coulomb}
Q = q \left( 1 + \frac{g^2}{T-1} \left( \frac{r_0}{R_0} \right)^{2(T-1)} \right).
\end{align}
Since $T > 1$, for a fixed $R_0$ the quark color-charge is increased due to the 
gauge potentials.
For increasing $R$ this anti-screening effect is reduced, although 
the  gauge potentials are restricted to the region in which there is color charge distribution 
($r< R_0$)
as discussed above.
A natural choice should be $R_0=r_0$ for which 
$Q = q (1 + g^2/(T-1)) > q$.

\FloatBarrier

\FloatBarrier

\FloatBarrier

\FloatBarrier

The non-Abelian contribution to the energy density,
($u_{Ec,Bc}$), from Eq. \ref{EMTensor00}, is given by:
\begin{align}
u_{Ec,Bc}(\alpha) =& \frac{1}{2}\left( \frac{\cos^2(2\theta)}{\sin^2(\theta)} + \cos^2(\theta) \left( \frac{5}{2} - \sqrt{9-4\alpha^2} \right) \right)\frac{1}{r_0^4} \left( \frac{r_0}{r} \right)^{3+\sqrt{9-4\alpha^2}} .
\end{align}
The total  energy will have a divergence that would need 
a short distance cutoff, similarly to the classical electron radius problem.
However, 
besides an  (expected) divergence of the charge in the origin $1/r^{(3+\sqrt{9-4\alpha^2})}$,
 analytical cuts  appear
in the angular extrema directions, $\theta = 0, \pi$.

\section{ Anisotropic Coulomb    potential: class of solutions}
\label{sec:nonsphericalcoulomb}

Before investigating  other different well known solutions for the scalar potential,
now, by using the \textit{ansatz} \ref{SEMISYMADIRECTION},
let us consider an anisotropic modification of the Coulomb potential  by means of  the  following prescription:
\begin{align}\label{coulomblike}
\varphi (r,\theta) = \frac{f(\theta)}{r}.
\end{align}
As a source we consider a punctual charge in the origin and keep an additional arbitrary 
color-charge density $\rho(\vec{r})$ to be determined later.
 By considering  more  general equations,  Eqs. (\ref{EquaçãoGeneralizada2})-(\ref{EquaçãoGeneralizada4}),
than those  of the previous section,
for an anisotropic potential
 we obtain the following equations:
\begin{align}
\frac{1}{r} \nabla^2 f(\theta)-4\pi\delta(\vec{r}) f(\theta)  - 2g^2 A_\phi^2 \frac{f(\theta)}{r} &= - \rho;\\ \label{COULOMBLIKE2}
R_\theta  &= 0; \\ \label{COULOMBLIKE3}
r^2R_\phi'' + 2rR_\phi' -l(l+1) R_\phi &= 0;\\ \label{COULOMBLIKE4}
(1-x^2) \frac{d^2\Theta_\phi}{dx^2} - 2x \frac{d\Theta_\phi}{dx} + \left( g^2 f^2(\theta) 
- \frac{1}{1-x^2} + l(l+1) \right) \Theta_\phi &=0.
\end{align}

Equation \ref{COULOMBLIKE3} is the Euler-Cauchy equation \cite{butkov}, with the following solutions:
\begin{align}
R_\phi(r) = a_l r^{\frac{-1-\sqrt{1+4l(l+1)}}{2}} + b_l r^{\frac{-1+\sqrt{1+4l(l+1)}}{2}}.
\end{align}
The prescription \ref{coulomblike} can be solved for different cases of angular dependency, given by 
$f(\theta) = f(\cos(\theta)) = f(x)$. 
 As an example, consider the case:
\begin{align}
f(x)= \frac{\lambda}{4\pi\sqrt{1-x^2}},
\end{align}
 where $\lambda$ is a parameter that carries
color charge which initially might be a sort of punctual  although the rotational symmetry breaking
must be a non-Abelian effect.
With this prescription, equation \ref{COULOMBLIKE4} becomes:
\begin{align}
(1-x^2) \frac{d^2\Theta_\phi}{dx^2} - 2x \frac{d\Theta_\phi}{dx} + \left( l(l+1) - \frac{1-(\lambda g/4\pi)^2}{1-x^2} \right) \Theta_\phi &=0,
\end{align}
which is the Legendre's associated equation of order $\sqrt{1-(Cg)^2}$,
being  $C=\lambda/(4\pi)$, \cite{butkov}. The solutions can be written as:
\begin{align}
\Theta_\phi = c_l P_l^{\sqrt{1-(Cg)^2}}(\cos(\theta)) + d_l Q_l^{\sqrt{1-(Cg)^2}}(\cos(\theta)).
\end{align}

In order for these solutions to be real for all $\cos(\theta)$, 
we must have an integer for the order of the function.
 The only possibility is therefore given by:
\begin{align}\label{gelambda}
| C g | = \left| \frac{\lambda g}{4\pi} \right|  = 1.
\end{align}
This condition provides a large value for $|\lambda g/(4\pi)|$ than the 
corresponding solution of the (spherically symmetric) 
 Coulomb potential of the last section - in Eq. 
(\ref{cond-qg-coulomb}).
By considering the largest value $gC = 1$, and with 
a  condition that $\vec{A} (r\rightarrow\infty) \to 0$ , the solution for the vector potential is:
\begin{align}
\vec{A}(\vec{r}) = \sum_{l=0}^\infty  a_l r^{\frac{-1-\sqrt{1+4l(l+1)}}{2}} \left( c_l P_l(\cos(\theta)) + d_l Q_l(\cos(\theta)) \right) \hat{\phi}.
\end{align}
Since 
  no boundary conditions
will be applied  to this type of problem,
let us analyze the two different lowest energies cases of $l$.

\subsection{$l=0$}
\label{sec:asymcoulombl0}

\begin{figure}[H]
    \centering
    \begin{minipage}{0.49\textwidth}
        \centering
        \includegraphics[width=0.9\textwidth]{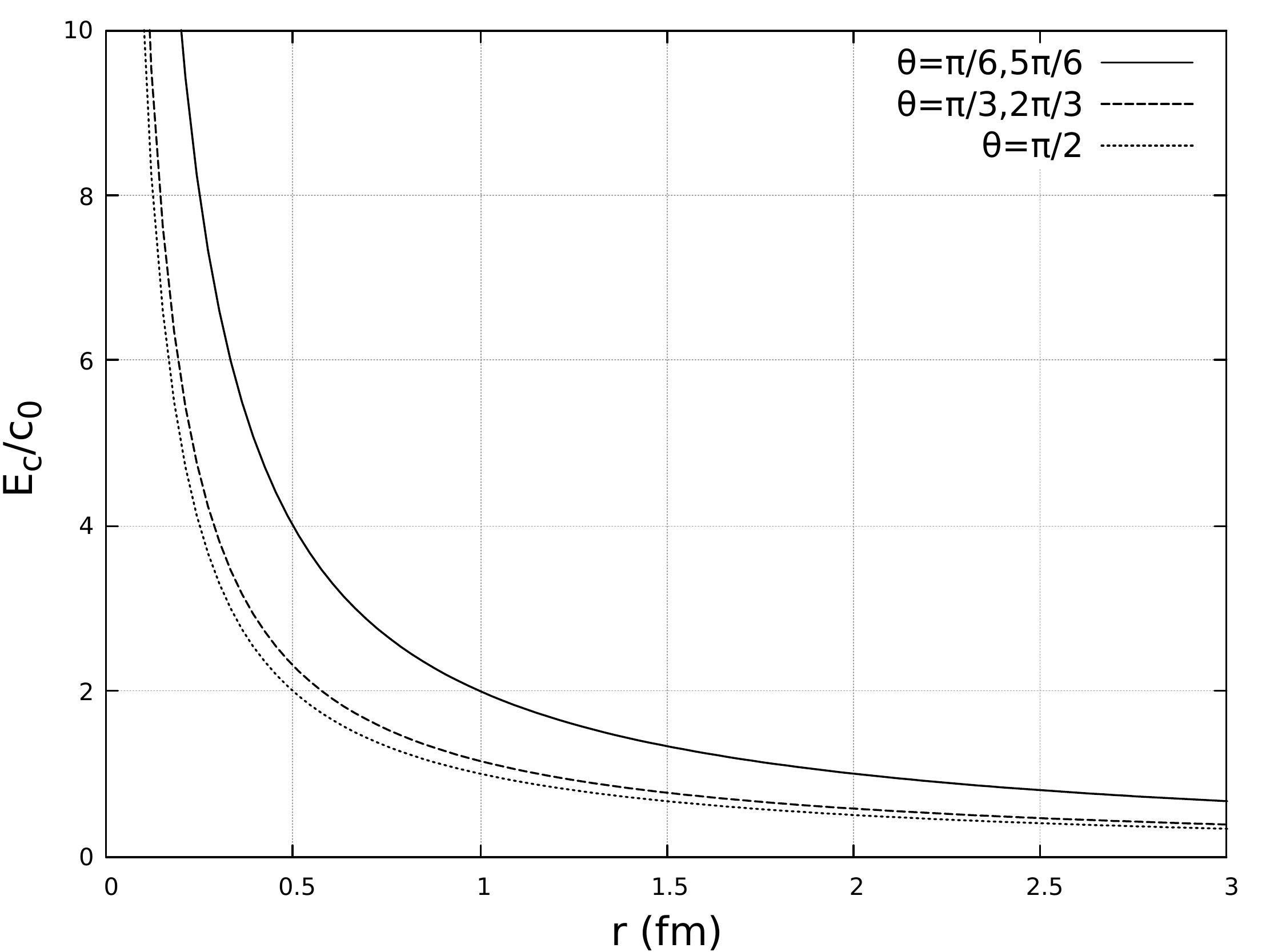}
\caption{$E_c^{\phi}$, from \ref{CoulLikeEl0}, for $d_0=0$.}
        \label{fig:Ecphic0}
    \end{minipage}\hfill
    \begin{minipage}{0.49\textwidth}
        \centering
        \includegraphics[width=0.9\textwidth]{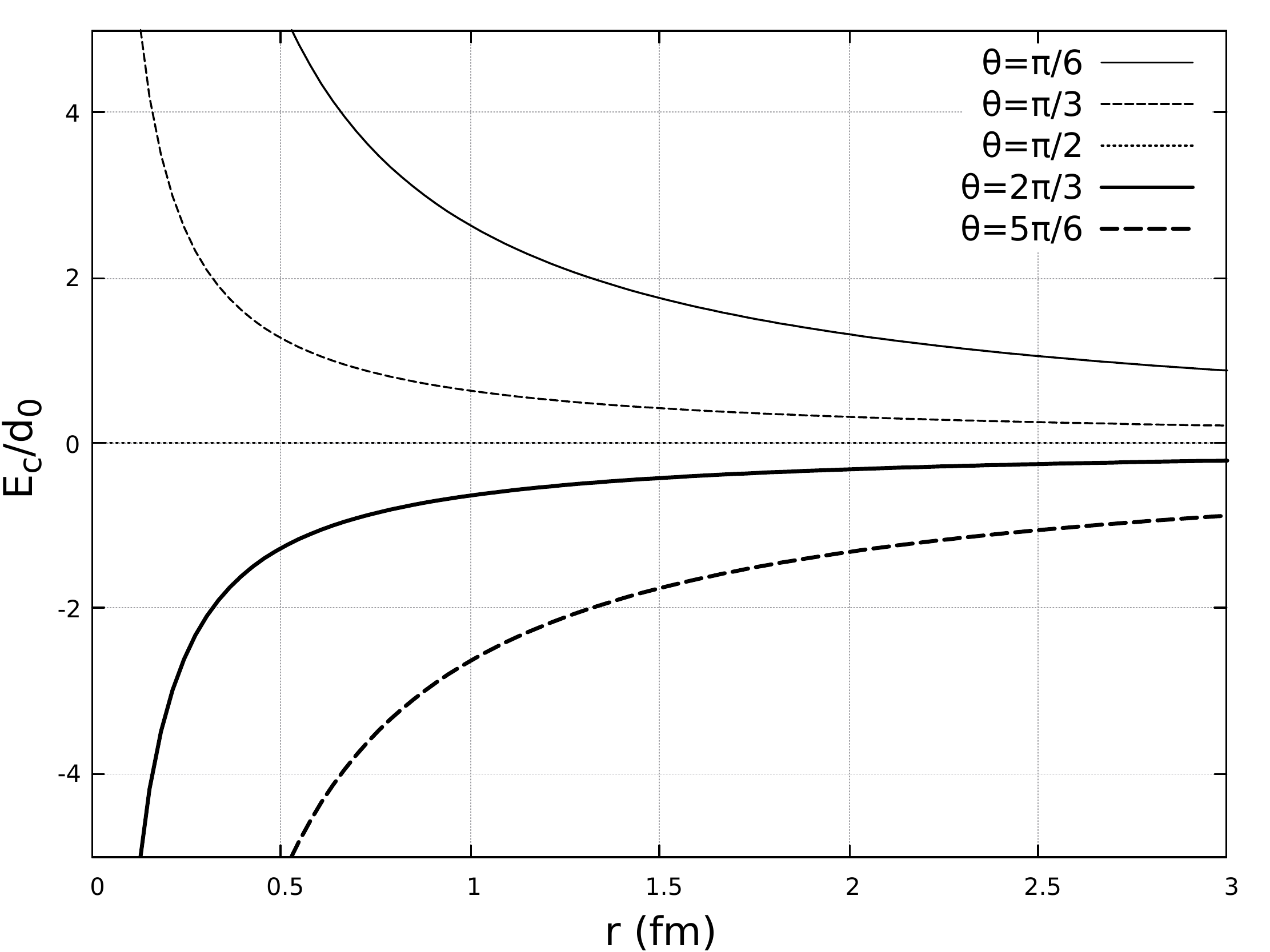}
\caption{$E_c^{\phi}$, from \ref{CoulLikeEl0}, for $c_0=0$.}
        \label{fig:Ecphid0}
    \end{minipage}
\end{figure}

\begin{figure}[H]
    \centering
    \begin{minipage}{0.49\textwidth}
        \centering
        \includegraphics[width=0.9\textwidth]{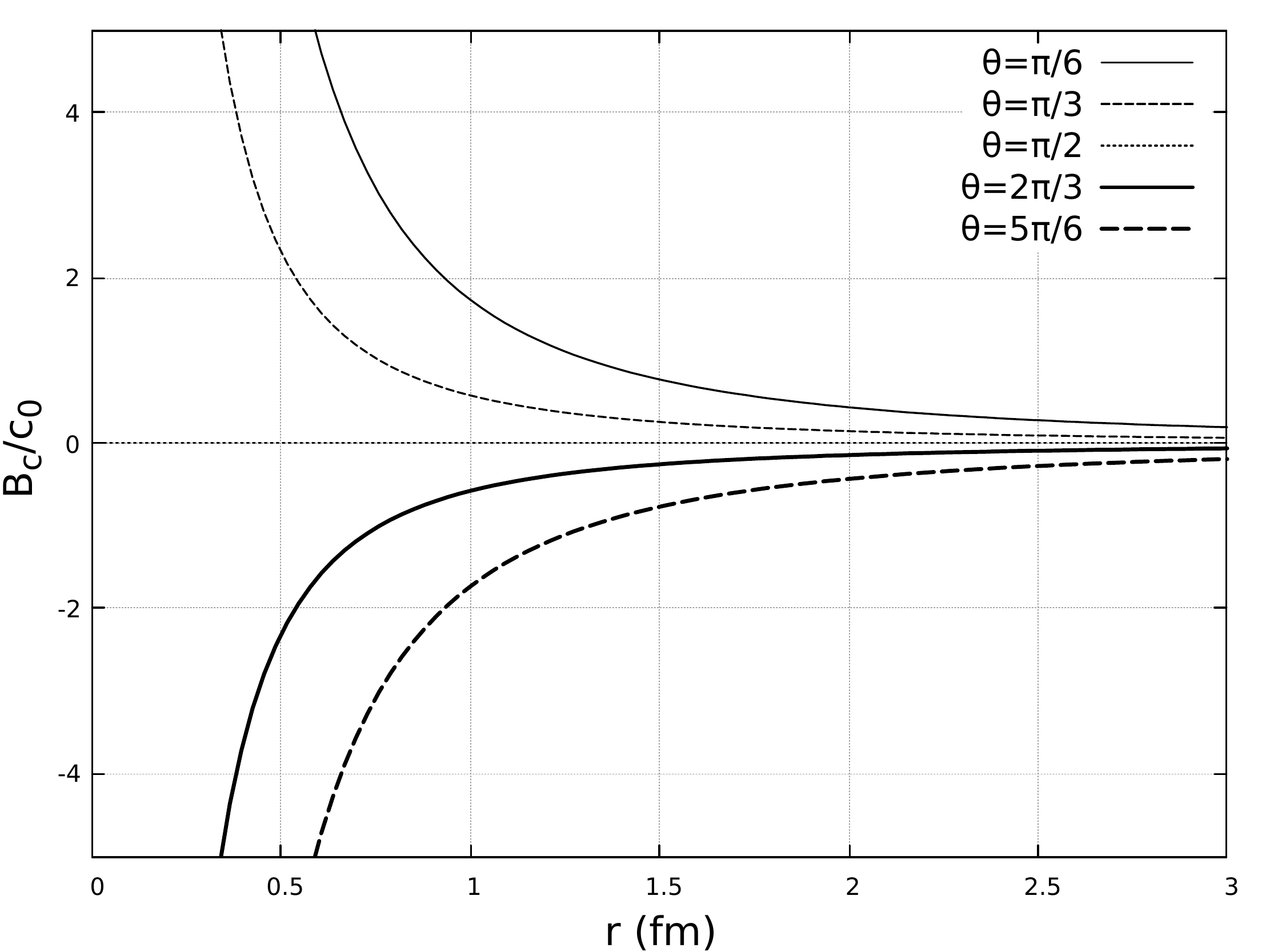}
\caption{$B_c^r$, from \ref{CoulLikeBl0}, for $d_0=0$.}
        \label{fig:Bcphic0}
    \end{minipage}\hfill
    \begin{minipage}{0.49\textwidth}
        \centering
        \includegraphics[width=0.9\textwidth]{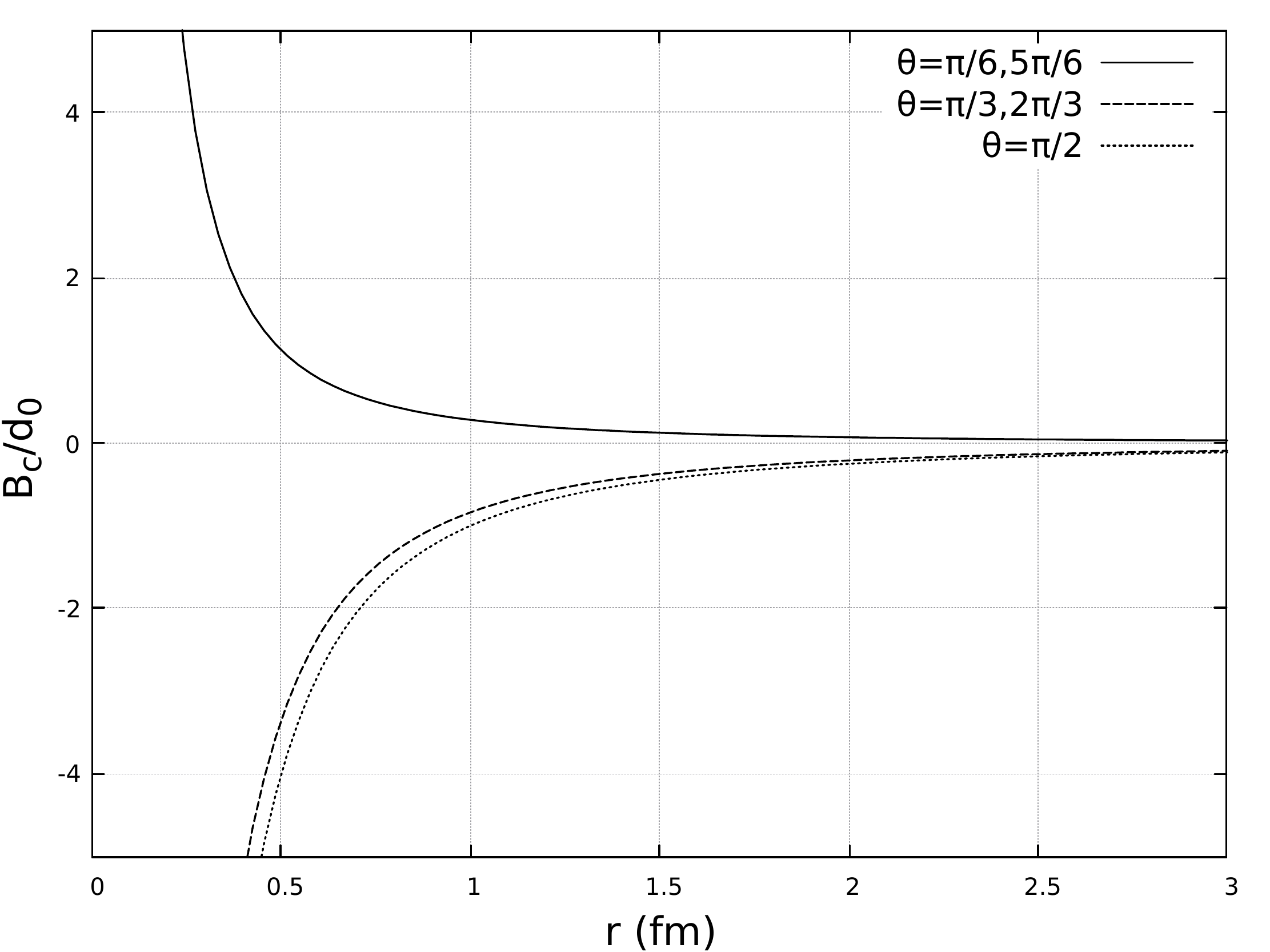}
\caption{$B_c^r$, from \ref{CoulLikeBl0}, for $c_0=0$.}
        \label{fig:Bcphid0}
    \end{minipage}
\end{figure}

For this case, $P_0=1$, so we only are left with a non-trivial case if $d_0 \neq 0$. 
Therefore, the solution is then:
\begin{align}
\vec{A}(\vec{r}) =  \frac{1}{r} \left[c_0 + d_0Q_0(\cos(\theta)) \right] \hat{\phi},
 \mbox{ } Q_0(x) = \frac{1}{2} \ln \left(\frac{1+x}{1-x} \right).
\end{align}
As we see, $c_0$ and $d_0$ are dimensionless constants. They also have to be functions of $g$, 
such that in the 
 Abelian limit ($g\rightarrow0$), we recover the solution $\vec{A}=0$:
\begin{align}\label{NonSymCoulEMLimit}
\lim_{g\rightarrow 0} c_0(g) = \lambda , 
\;\;\;\;\;  \lim_{g\rightarrow 0} d_0(g) = 0.
\end{align}
The corresponding c-electric and c-magnetic fields are given by
by considering Eq. (\ref{sigma0}):
\begin{align}\nonumber
\mbox{\textbf{I-spin}:}\\ \label{CoulLikeEl0}
\vec{E}_a(\vec{r}) &= \frac{\lambda}{4\pi\sin(\theta)r^2} \left( \hat{r} + \frac{\hat{\theta}}{\sin(\theta)} \right) \delta_{a3} + \frac{1}{\sin(\theta)r} \left( c_0 + \frac{d_0}{2}\ln \left( \frac{1+\cos(\theta)}{1-\cos(\theta)} \right) \right)\hat{\phi} (\delta_{a1}-\delta_{a2});
\\ \label{CoulLikeBl0}
\vec{B}_a(\vec{r}) &= \frac{\hat{r}}{r^2 \sin(\theta)} \left( c_0 \cos(\theta) + d_0 \left(\frac{\cos(\theta)}{2} \ln \left( \frac{1+\cos(\theta)}{1-\cos(\theta)}\right)  -1\right) \right) (\delta_{a1} + \delta_{a2}) ;\\ \nonumber
\mbox{\textbf{V-spin}:}\\
\vec{E}_a(\vec{r}) &= \frac{\lambda}{4\pi\sin(\theta)r^2} \left( \hat{r} + \frac{\hat{\theta}}{\sin(\theta)} \right) \left(  \frac{\delta_{a3}+\sqrt{3}\delta_{a8}}{2} \right) + \frac{1}{\sin(\theta)r} \left( c_0 + \frac{d_0}{2}\ln \left( \frac{1+\cos(\theta)}{1-\cos(\theta)} \right) \right)\hat{\phi} (\delta_{a4}-\delta_{a5});
\\ \nonumber
\vec{B}_a(\vec{r}) &=  \frac{\hat{r}}{r^2 \sin(\theta)} \left( c_0 \cos(\theta) + d_0 \left(\frac{\cos(\theta)}{2} \ln \left( \frac{1+\cos(\theta)}{1-\cos(\theta)}\right)  -1\right) \right) (\delta_{a4} + \delta_{a5});\\ \nonumber
\mbox{\textbf{U-spin}:}\\
\vec{E}_a(\vec{r}) &= \frac{\lambda}{4\pi\sin(\theta)r^2} \left( \hat{r} + \frac{\hat{\theta}}{\sin(\theta)} \right) \left( \delta_{a3} -  \frac{\sqrt{3}}{3} \delta_{a8}\right) + \frac{1}{\sin(\theta)r} \left( c_0 + \frac{d_0}{2}\ln \left( \frac{1+\cos(\theta)}{1-\cos(\theta)} \right) \right)\hat{\phi} (\delta_{a6}-\delta_{a7});
\\ \nonumber
\vec{B}_a(\vec{r}) &=  \frac{\hat{r}}{r^2 \sin(\theta)} \left( c_0 \cos(\theta) + d_0 \left(\frac{\cos(\theta)}{2} \ln \left( \frac{1+\cos(\theta)}{1-\cos(\theta)}\right)  -1\right) \right) (\delta_{a6} + \delta_{a7}).
\end{align}
The above solutions for the $\hat{\phi}$-component of the 
electric field and  the (radial) component of the magnetic field
 are shown in the next figures
 separately for the two components with $c_0$ or $d_0$.
Some profiles of the   $E^\phi_c$  for different angles (those that are non divergent)
are shown for $d_0=0$ ($c_0=0$) in  Fig. \ref{fig:Ecphic0} ((\ref{fig:Ecphid0})).
Although the Ec field decreases with $1/r$ the angular dependence is very strong.
The $c_0$ component has an unique sign for all directions, whereas
the $d_0$ changes sign for $ 3 \pi/ 2 > \theta > \pi/2$.
The c-magnetic field profiles  for different angles
 are shown in Figs. (\ref{fig:Bcphic0}) and (\ref{fig:Bcphic0})
respectively for $d_0=0$ and $c_0=0$.
The strong anisotropic behavior is also noted 
being that the change of sign of $B_c$
is present for the components  
 $d_0$ of the c-electric field,
in spite of being very different.
As a consequence,
the c-electric and c-magnetic fields, and also their fluxes, are strongly anisotropic.

The charge density, responsible for these configurations, is given by:
\begin{align}
\rho (\vec{r}) &= \frac{\lambda}{\sin(\theta)} \delta(\vec{r}) -
 \frac{\lambda}{4\pi r^3 \sin^3(\theta)} 
+ 2g\sign(\lambda) \frac{1}{r^3} \frac{\left[
c_0 + d_0Q_0(\cos(\theta) \right]^2}{\sin(\theta)}.
\end{align}
Note that the usual punctual charge has an anisotropic form with cuts along $\theta=0,\pi$
which are responsible for the scalar potential (\ref{coulomblike}), besides other
extended contributions all of them with cuts in the same directions
and the singularities in $r=0$.


\subsection{$l=1$}

In this case  the vector potential reads:
\begin{align}
\vec{A}(\vec{r}) =  \frac{1}{r^2} \left(c_1P_1(\cos(\theta) +  d_1Q_1(\cos(\theta)) \right) \hat{\phi}, \mbox{ } P_1(x)=x \mbox{ and } Q_1(x) = \frac{x}{2} \ln\left(\frac{1+x}{1-x} \right)-1.
\end{align}
As we expect $\vec{A}$ to vanish in the limit $g\rightarrow0$,
 $d_1$ must be some function of $g$ such that this condition is obtained, so we rename:
\begin{align}
c_1 &= \epsilon r_0(g);\\
d_1 &= \gamma r_0(g),
\end{align}
where
$r_0(g)$ is a function of g such that $r_0(0)=0$, representing our new integration constant that satisfies the eletromagnetic limit, and $\gamma$, $\epsilon$ are fitting dimensioneless constants. Therefore:
\begin{align}
\vec{A}(\vec{r}) = \frac{1}{r_0} 
 \left( \frac{r_0}{r} \right)^2 \left(\gamma \cos(\theta) + \epsilon \left[ \frac{\cos(\theta)}{2} 
\ln \left( \frac{1+\cos(\theta)}{1-\cos(\theta)} \right) 
- 1 \right] \right)\hat{\phi}.
\end{align}

The corresponding  c-electric and c-magnetic fields  for each of the SU(2) subgroups 
are given by (using \ref{sigma0}):
\begin{align}\nonumber
\mbox{\textbf{I-spin}:}\\ \label{ECoulombrho}
\vec{E}_a(\vec{r}) &= \frac{\lambda}{4\pi\sin(\theta)r^2} \left( \hat{r} + \frac{\hat{\theta}}{\sin(\theta)} \right) \delta_{a3} + \frac{\sign(\lambda)}{\sin(\theta)r_0^2}\left( \frac{r_0}{r} \right)^3 \left( \gamma P_1(\cos(\theta) + \epsilon Q_1(\cos(\theta) \right)\hat{\phi} (\delta_{a1}-\delta_{a2});
\\ \label{BCoulombrho}
\vec{B}_a(\vec{r}) &=  \frac{1}{r_0^2}  \left( \frac{r_0}{r} \right)^3 \left[ \frac{\hat{r}}{\sin(\theta)} \left( \gamma \cos(2\theta) - \epsilon \left( Q_0(\cos(\theta) + \frac{\cos(\theta)}{2} \right)  \right)+ \hat{\phi} \left( \gamma P_1(\cos(\theta) + \epsilon Q_1(\cos(\theta) \right) \right](\delta_{a1} + \delta_{a2});\\ \nonumber
\mbox{\textbf{V-spin}:}\\ \nonumber
\vec{E}_a(\vec{r}) &= \frac{\lambda}{4\pi\sin(\theta)r^2} \left( \hat{r} + \frac{\hat{\theta}}{\sin(\theta)} \right) \left(  \frac{\delta_{a3}+\sqrt{3}\delta_{a8}}{2} \right) + \frac{\sign(\lambda)}{\sin(\theta)r_0^2}\left( \frac{r_0}{r} \right)^3 \left( \gamma P_1(\cos(\theta) + \epsilon Q_1(\cos(\theta) \right)\hat{\phi} (\delta_{a4}-\delta_{a5});
\\ \nonumber
\vec{B}_a(\vec{r}) &=  \frac{1}{r_0^2}  \left( \frac{r_0}{r} \right)^3 \left[ \frac{\hat{r}}{\sin(\theta)} \left( \gamma \cos(2\theta) - \epsilon \left( Q_0(\cos(\theta) + \frac{\cos(\theta)}{2} \right)  \right)+ \hat{\phi} \left( \gamma P_1(\cos(\theta) + \epsilon Q_1(\cos(\theta) \right) \right](\delta_{a4} + \delta_{a5});\\ \nonumber
\mbox{\textbf{U-spin}:}\\ \nonumber
\vec{E}_a(\vec{r}) &= \frac{\lambda}{4\pi\sin(\theta)r^2} \left( \hat{r} + \frac{\hat{\theta}}{\sin(\theta)} \right) \left( \delta_{a3} -  \frac{\sqrt{3}}{3} \delta_{a8}\right) + \frac{\sign(\lambda)}{\sin(\theta)r_0^2}\left( \frac{r_0}{r} \right)^3 \left( \gamma P_1(\cos(\theta) + \epsilon Q_1(\cos(\theta) \right)\hat{\phi} (\delta_{a6}-\delta_{a7});
\\ \nonumber
\vec{B}_a(\vec{r}) &=  \frac{1}{r_0^2}  \left( \frac{r_0}{r} \right)^3 \left[ \frac{\hat{r}}{\sin(\theta)} \left( \gamma\cos(2\theta) - \epsilon \left( Q_0(\cos(\theta) + \frac{\cos(\theta)}{2} \right)  \right)+ \hat{\phi} \left( \gamma P_1(\cos(\theta) + \epsilon Q_1(\cos(\theta) \right) \right](\delta_{a6} + \delta_{a7}),
\end{align}
where we have used the relation \ref{gelambda}. The 
corresponding charge density for all subgroups
is given by:
\begin{align}\label{rhocoulombrho}
\rho (\vec{r}) &= \frac{\lambda}{\sin(\theta)} \delta(\vec{r}) - \frac{\lambda}{4\pi r^3 \sin^3(\theta)} + \frac{2g \sign(\lambda)}{r^3_0 \sin (\theta)} \left( \frac{r_0}{r} \right)^4 \left(\gamma P_1(\cos(\theta)+ \epsilon Q_1(\cos(\theta))\right)^2.
\end{align}
It presents a strong anisotropy
that is  a  non-Abelian effect from  the vector potential.
Similarly to the  case of $l=0$ (and also to the 
magnetic field of the previous spherically symmetric Coulomb potential)
 the presence of the terms with $1/(\sin^n \theta)$ - odd n - may  be an indication of 
a dipolar type configuration with the unusual presence of analytical 
cuts for $\theta=0, 2 \pi$.
Similarly to other situations addressed in the present work,
this color-charge density could be associated to a classical gluon cloud around
a punctual quark in the origin similarly
 to  the solution of the previous section for 
a spherical symmetric Coloumb potential (\ref{SectionCoulomb}).

\begin{figure}[H]
    \centering
    \begin{minipage}{0.49\textwidth}
        \centering
        \includegraphics[width=0.9\textwidth]{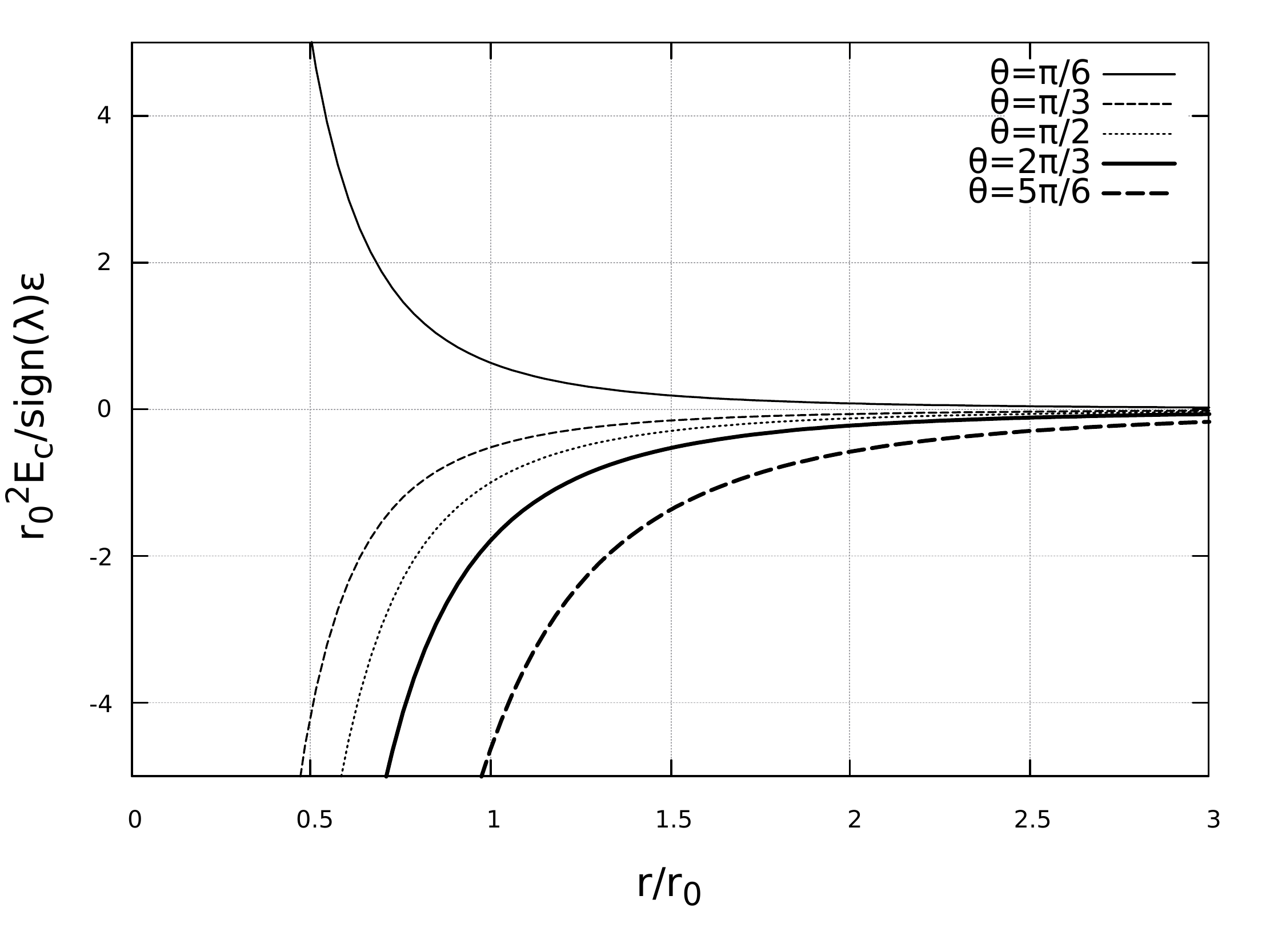}
\caption{$\vec{E}_c \cdot \hat{\phi}$, Eq. (\ref{ECoulombrho}),
 as a function of $r/r_0$ for different $\theta$, and $\gamma=0$.
 }
        \label{fig:Ecoulombrho}
    \end{minipage}\hfill
    \begin{minipage}{0.49\textwidth}
        \centering
        \includegraphics[width=0.9\textwidth]{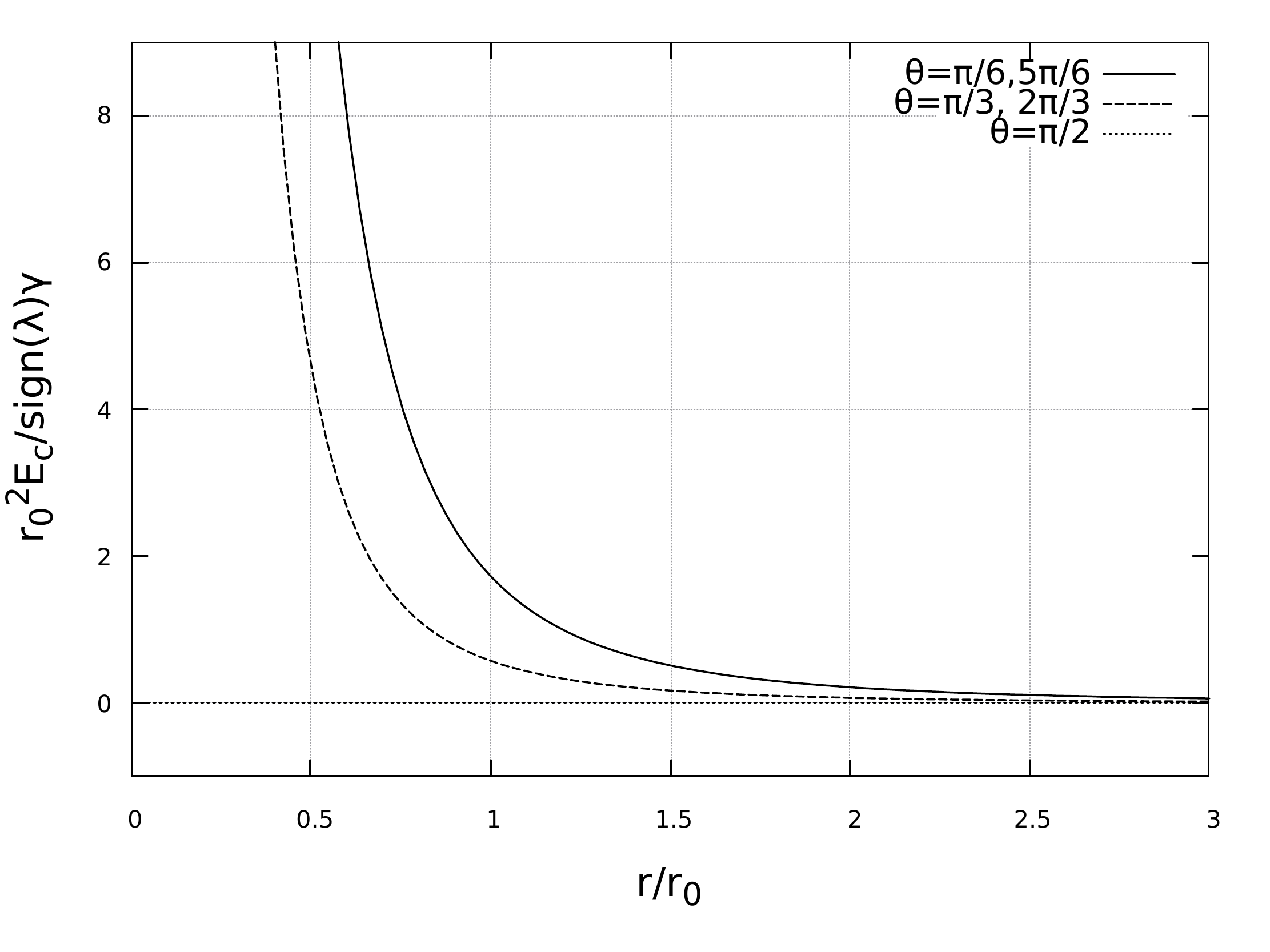}
\caption{$\vec{E}_c \cdot \hat{r} = E_{c,r}$, Eq. (\ref{ECoulombrho})
 as a function of $r/r_0$ for different $\theta$, and $\epsilon=0$.
 }
        \label{fig:Ercoulombrho}
    \end{minipage}
\end{figure}

\begin{figure}[H]
    \centering
    \begin{minipage}{0.49\textwidth}
        \centering
        \includegraphics[width=0.9\textwidth]{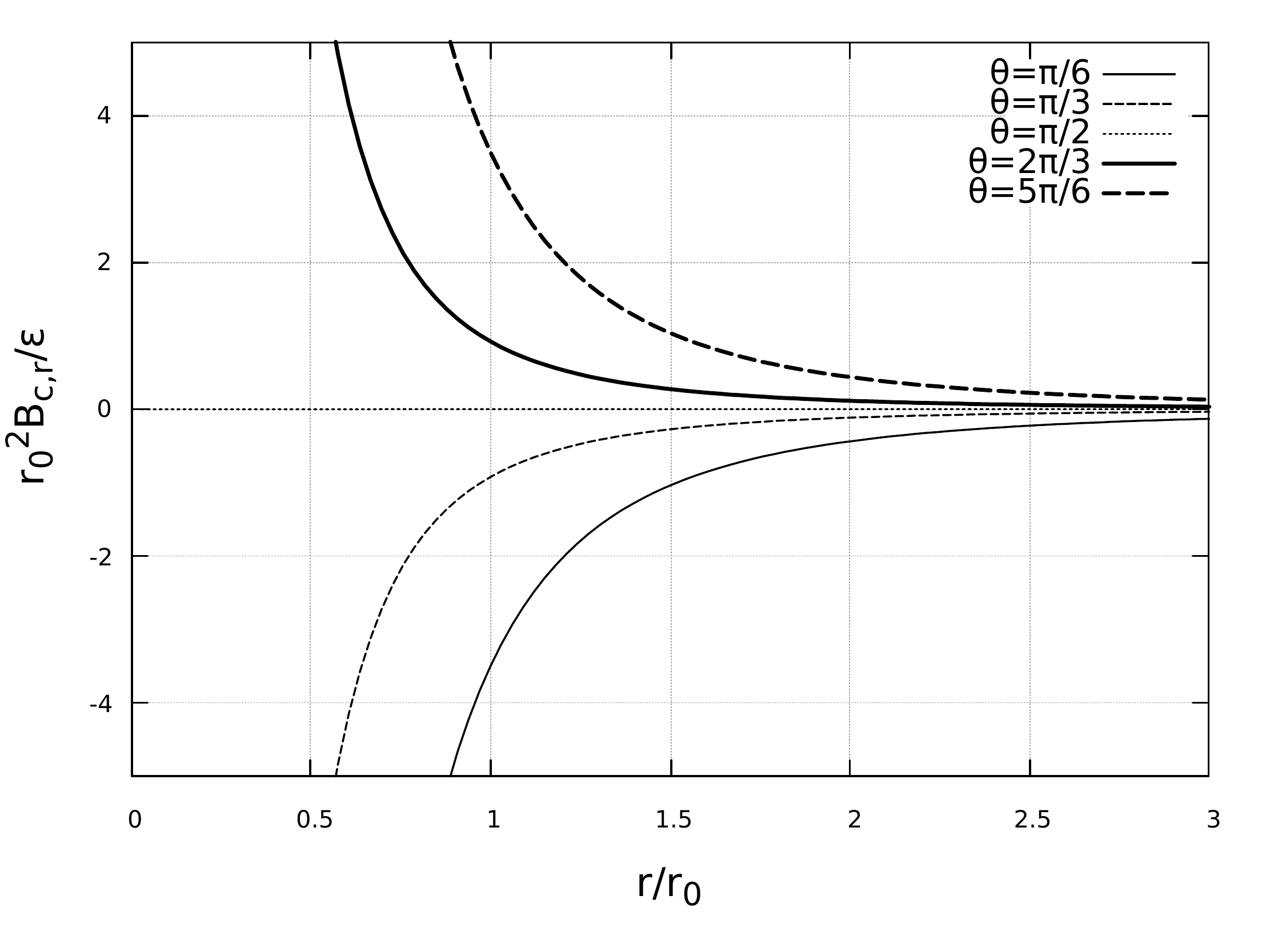}
\caption{$\vec{B}_c \cdot \hat{r} \equiv B_{c,r}$, equation \ref{BCoulombrho}, as a function of $r/r_0$ for different $\theta$, and $\gamma=0$.}
         \label{fig:Brcoulombrho-eps}
    \end{minipage}\hfill
    \begin{minipage}{0.49\textwidth}
        \centering
        \includegraphics[width=0.9\textwidth]{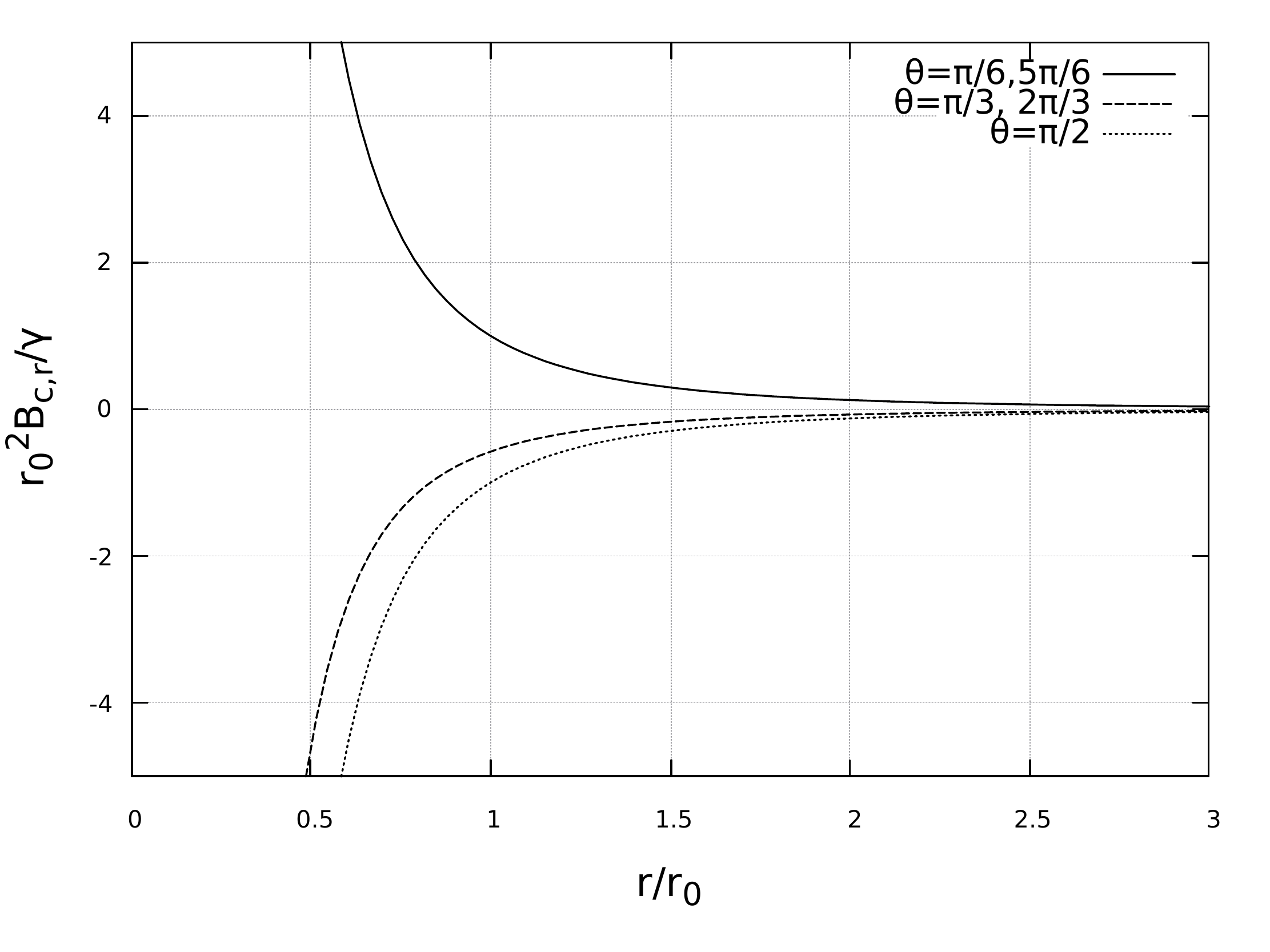}
\caption{$\vec{B}_c \cdot \hat{r} \equiv B_{c,r}$, equation \ref{BCoulombrho}, as a function of $r/r_0$ for different $\theta$, and $\epsilon=0$.}
    \label{fig:Brcoulombrho-gamma}
    \end{minipage}
\end{figure}

\begin{figure}[H]
    \centering
    \begin{minipage}{0.49\textwidth}
        \centering
        \includegraphics[width=0.9\textwidth]{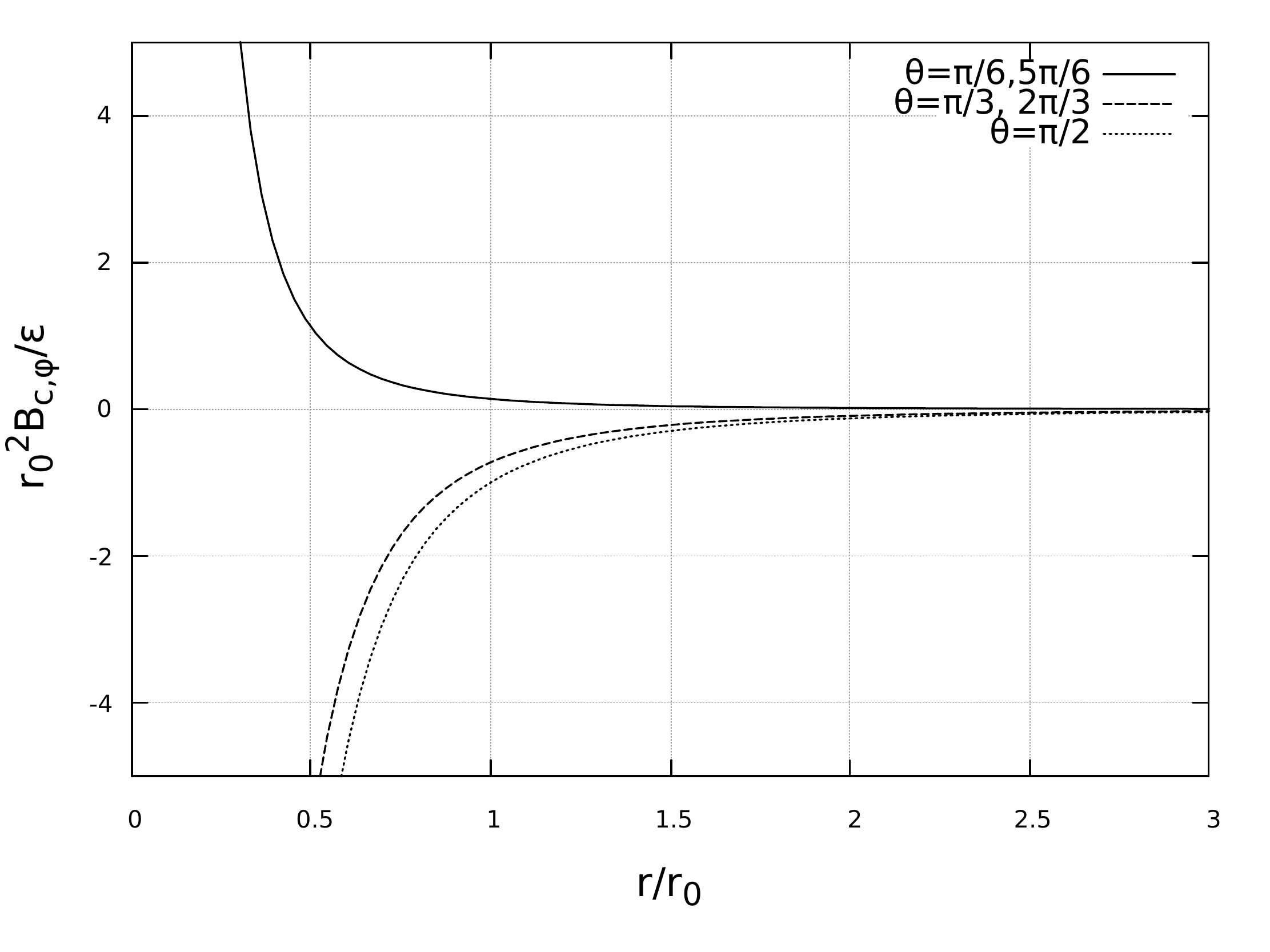}
\caption{$\vec{B}_c \cdot \hat{\phi} \equiv B_{c,\phi}$, equation \ref{BCoulombrho}, as a function of $r/r_0$ for different $\theta$, and $\gamma=0$.}
        \label{fig:Bthetacoulombrho-eps} 
    \end{minipage}\hfill
    \begin{minipage}{0.49\textwidth}
        \centering
        \includegraphics[width=0.9\textwidth]{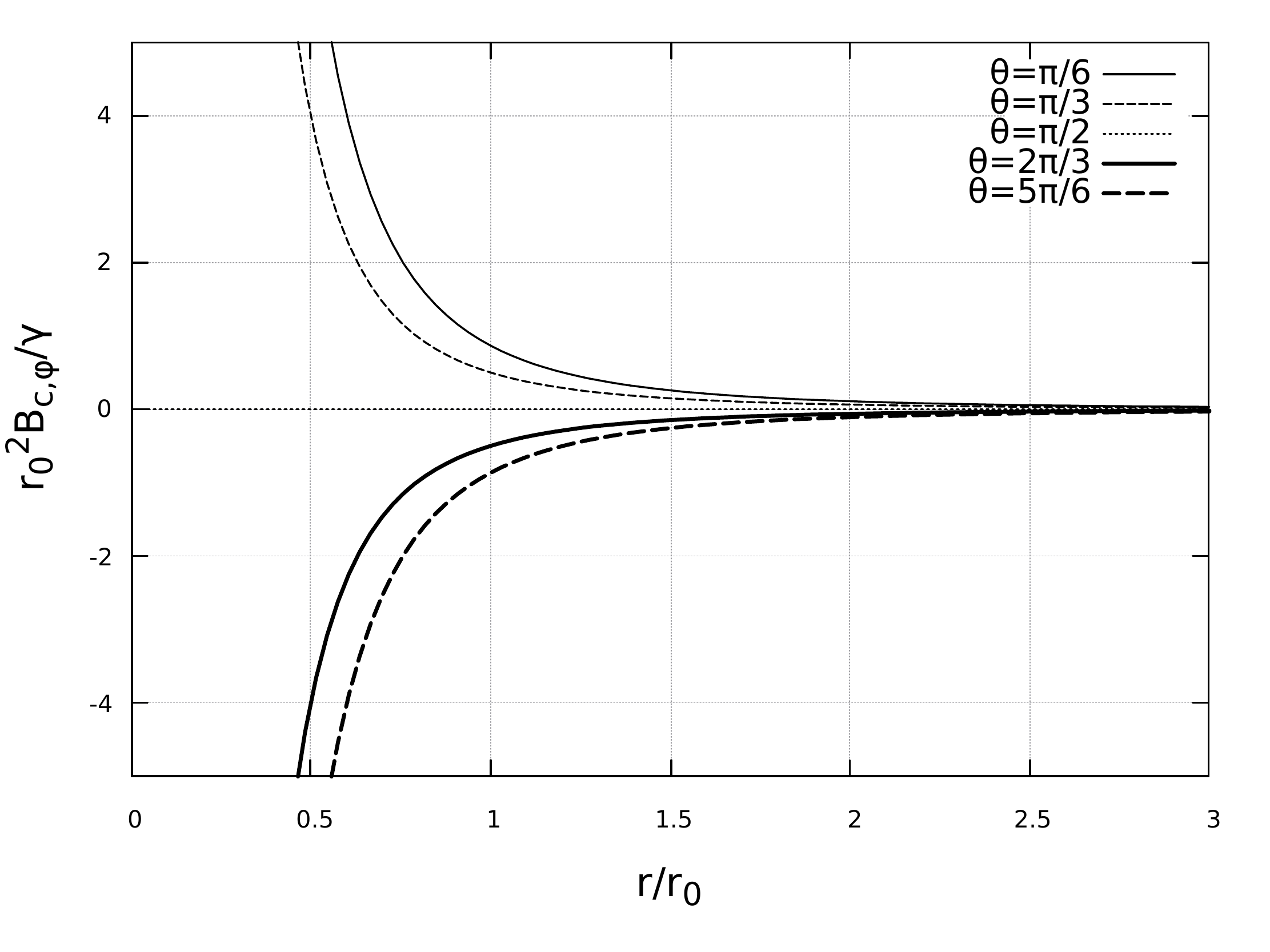}
\caption{$\vec{B}_c \cdot \hat{\phi} \equiv B_{c,\phi}$, equation \ref{BCoulombrho}, as a function of $r/r_0$ for different $\theta$, and $\epsilon=0$.}
    \label{fig:Bthetacoulombrho-gam}
    \end{minipage}
\end{figure}

 In Figs. (\ref{fig:Ecoulombrho}) and (\ref{fig:Ercoulombrho})
 the (normalized, dimensionless)
 strength of the component  $E_{\phi}$  of the c-electric field
are shown respectively for $\gamma=0$ and $\epsilon = 0$
in Eq. (\ref{ECoulombrho})
 for  different angles $\theta$.
For  $\theta=0, \pi$
the Ec component is singular.
This non-Abelian component of the Ec field has a singularity of $1/r^3$.
Similarly to the case of $l=0$  there is a component of the c-electric field
that does not change sign and another that changes sign depending on the 
octant they are placed.
Similarly the radial and angular (normalized and dimensionless) 
components of $B_c$, $B_{c,r}$ and $B_{c,\theta}$, are shown
separately for components with $\gamma$ and $\epsilon$ 
in
Figs. (\ref{fig:Brcoulombrho-eps},\ref{fig:Brcoulombrho-gamma})
- for $B_{c,r}$ - and 
Figs. (\ref{fig:Bthetacoulombrho-eps},\ref{fig:Bthetacoulombrho-gam})
- for $B_{c,\phi}$ - 
for some angles $\theta$.
They have similar behavior as the Ec field, with singularities
 for $\theta=0, \pi$, except that
they are zero for different angles.
 These color-electric and magnetic fields have a similar 
shape as the fields to 
 the (1) lower component $l=0$ shown in section (\ref{sec:asymcoulombl0})
and  to the  (2) 
spherical symmetric Coulomb potential of the previous 
section, as it can be seen in
Figs. (\ref{figfig1Ec},\ref{figfig2Bcr}) and (\ref{figfig2Btheta})
although the specific angular dependencies are different.

-
-

\begin{figure}[H]
\centering
\includegraphics[width=0.441\textwidth]{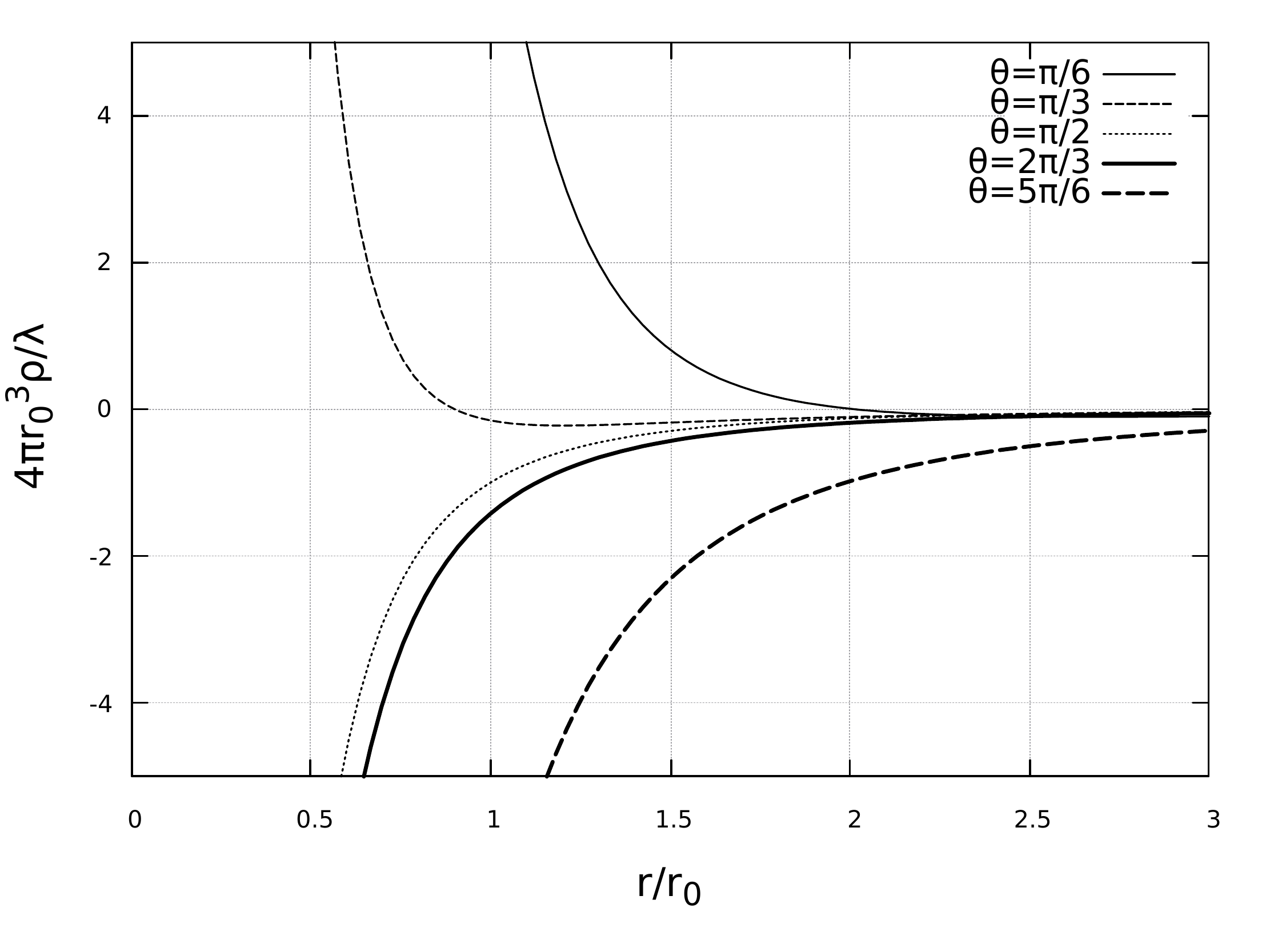}
\caption{Charge density, equation \ref{rhocoulombrho}, as a function of $r/r_0$ for different $\theta$, with $g=1$ and $\gamma=\epsilon=1$. }
    \label{fig:rhocoulombrho}
\end{figure}

The color-charge density, Eq. (\ref{rhocoulombrho}),
is shown in Fig. (\ref{fig:rhocoulombrho}) for the same angles considered for the previous figures.
Again the change of sign of the color-charge density 
occurs for a particular angle  $\theta$ whose determination is not trivial for the 
case of $l=1$.
 This configuration corresponds to a somewhat dipole type color-charge distribution
for which, however, a cutoff in the angle $\theta \neq 0, \pi$ is required to avoid 
the analytical cuts.
It is interesting to note that the curves that are positive for small $r$ ($\pi/6$ and $\pi/3$)
 cross the horizontal axis in the points in which  $\rho(\vec{r})=0$ according to 
Eq. (\ref{rhocoulombrho}).
These points of zero color charge do not define a closed surface containing  the origin 
(where the color charge responsible for the (modified) Coulomb potential lies)
 because of the 
analytical cuts.

\section{Linear potential}
\label{sec:linearpot}

In this section, we'll impose  the scalar potential to be given by a linearly rising potential
without imposing a charge distribution from the beginning.
Therefore we go back to the Eqs. (\ref{SEMISYM1}-\ref{SEMISYM3}).
Consider the scalar potential to be given by:
\begin{eqnarray} \label{kappar}
\varphi  (\vec{r}) = \kappa r,
\end{eqnarray}
where $\kappa$ is a constant that carries the color charge from Eq. (\ref{qeta}).
The unknown  quark density $\rho(\vec{r})$ will be considered to be the one that 
makes eq. (\ref{kappar}) a solution of the eq. (\ref{SEMISYM1}).
The vector potential will be considered to have only the angular components. 
Eq.  \ref{SEMISYM1} can be seen therefore as a constraint, in which case reads:
\begin{eqnarray} \label{linear-rho}
\rho (\vec{r}) = 2 g^2 \left( A_\theta^2 (\vec{r}) + A_\phi^2 (\vec{r})
 \right) \kappa r - \frac{2 \kappa}{r},
\;\;\;\;\;
\mbox{ for } \;\;
r \leq R_{max},
\end{eqnarray}
that might be seen as a constraint for the angular components of the vector potential.
 Also, we notice that in order for $\rho$ to be {\it naturally} contained in some region, 
the vector potential must 
go to zero faster than the linear potential increases as $r \rightarrow \infty$. 
Otherwise, an external boundary may be imposed, maybe by quantum fluctuations.
In any case, the decreasing part, $2\kappa/r$, must be confined in a region for which a closed
 wall such as
a sphere can be imposed. 
As a consequence, all the solutions will be valid inside such a closed region
in which there is an extended color-charge distribution.

From \ref{SEMISYM2} and the prescription for the linear $\varphi$ potential,
 we may obtain a solution by re-scaling $R_\theta$ by:
\begin{align}
R_\theta(r) = \frac{e^{-i\frac{g\kappa}{2}r^2}M\left(\sqrt{ig\kappa}r \equiv x\right)}{r} ,
\end{align}
where the resulting equation for  $M(x= \sqrt{i g \kappa r} )$ is the following:
\begin{align}
\frac{d^2M}{dx^2}-2x\frac{dM}{dx}-M=0,
\end{align}
wich has the solutions:
\begin{align}
M = aH_{-1/2}(x) + b e^{x^2/2} \sqrt{x} I_{-1/4} \left( \frac{x^2}{2} \right),
\end{align}
where $H_\alpha(x)$ and $I_\alpha(x)$ are the generalized Hermite and Bessel (1st kind) functions, respectively. 
Then:
\begin{align}
R_\theta = \frac{e^{-i\frac{g\kappa}{2}r^2}}{r} \left( aH_{-1/2}(\sqrt{ig\kappa}r) + b e^{x^2/2} \sqrt{x} I_{-1/4} \left( \frac{ig\kappa r^2}{2} \right) \right).
\end{align}
For the phi-component, that is separable, it can be written:
\begin{align}
A_\phi(\vec{r}) = R_\phi(r) \Theta_\phi(\theta),
\end{align}
and the following equations ($x \equiv \cos(\theta)$) are obtained:
\begin{align}
r^2 \frac{d^2R_\phi}{dr^2} + 2r \frac{dR_\phi}{dr} + \left( (g\kappa)^2 r^4 - l(l+1) \right) R_\phi &=0;\\
(1-x^2)\frac{d^2\Theta_\phi}{dx^2} - 2x \frac{d\Theta_\phi}{dx} + 
\left(l(l+1)-\frac{1}{1-x^2} \right) \Theta_\phi &=0.
\end{align}
The equation for $\Theta_\phi$ is the Legendre associated equation of degree $l$ and order $\pm 1$ \cite{butkov}. 
The equation for $R_\phi$ is solved by the re-scaling:
\begin{align}
R_\phi(r) = \frac{S\left(\frac{g\kappa r^2}{2} \equiv v\right)}{\sqrt{r}}.
\end{align}
This change of variable leads to a Bessel's equation \cite{butkov}, whose solution can be written as:
\begin{align}
R_\phi(r) = \frac{c_l J_{\frac{2l+1}{2}} \left(\frac{g\kappa r^2}{2} \right) + d_l Y_{\frac{2l+1}{2}} \left(\frac{g\kappa r^2}{2} \right)}{\sqrt{r}}.
\end{align}
Then, the complete solution for the vector potential is:
\begin{align}
\vec{A}(\vec{r}) = \frac{e^{-i\frac{g\kappa}{2}r^2}}{r \sin(\theta)} \left( aH_{-1/2}(\sqrt{ig\kappa}r) + b e^{x^2/2} \sqrt{x} I_{-1/4} \left( \frac{ig\kappa r^2}{2} \right) \right) \hat{\theta} + \sum_{l=1}^\infty \frac{c_l J_{l+\frac{1}{2}} \left(\frac{g\kappa r^2}{2} \right) + d_l Y_{l+\frac{1}{2}} \left(\frac{g\kappa r^2}{2} \right)}{\sqrt{r}} P^1_l(\cos(\theta)) \hat{\phi}.
\end{align}

Now,  by requiring only real-valued functions
such that $\vec{A}$ is real,
 it follows that $a=b=0$.
 Consider now the simplest case,  $l=1$, for which it can be written:
\begin{align} \label{Avec}
\vec{A}(\vec{r}) = -\frac{c_1 J_{3/2} \left(\frac{g\kappa r^2}{2}\right) + d_1 Y_{3/2} \left(\frac{g\kappa r^2}{2}\right)}{\sqrt{r}} \sin(\theta) \hat{\phi},
\end{align}
where:
\begin{align}
J_{3/2}(x) &= \sum_{k=0}^{\infty} \frac{(-1)^k (x/2)^{2k+3/2}}{k! \Gamma(k+5/2)},
\\
Y_{3/2}(x) &=  \sum_{k=0}^{\infty} \frac{(-1)^k (x/2)^{2k-3/2}}{k! \Gamma(k-1/2)} .
\end{align}
Note that $R_\theta$ has a dimensionless constant of integration 
 and $R_\phi$  has its constant with dimension  length$^{-\frac{1}{2}}$.

Consider the case where the singularities at the origin in the vector potential is  limited to
 the $1/\sqrt{r}$ that appears dividing the $J$ and $Y$ in Eq. (\ref{Avec}).,
and  $d_1=0$. By re-defining the constant the $c_1$ as:
\begin{align} \label{c1-r1}
c_1 &= \sign(c_1)\frac{1}{\sqrt{r_1}},
\end{align}
where:
\begin{align}
\sign(x) = \begin{cases}
+1, \mbox{ if } x>0,\\
-1, \mbox{ if } x<0,
\end{cases}
\end{align}
the solution becomes:
\begin{align}
\vec{A}(\vec{r}) = - \sign(c_1) \sqrt{\frac{r_1}{r}} J_{3/2} \left(\frac{g\kappa r^2}{2}\right) \frac{\sin(\theta)}{r_1} \hat{\phi}.
\end{align}

\begin{figure}[H]
    \centering
    \begin{minipage}{0.49\textwidth}
        \centering
        \includegraphics[width=0.9\textwidth]{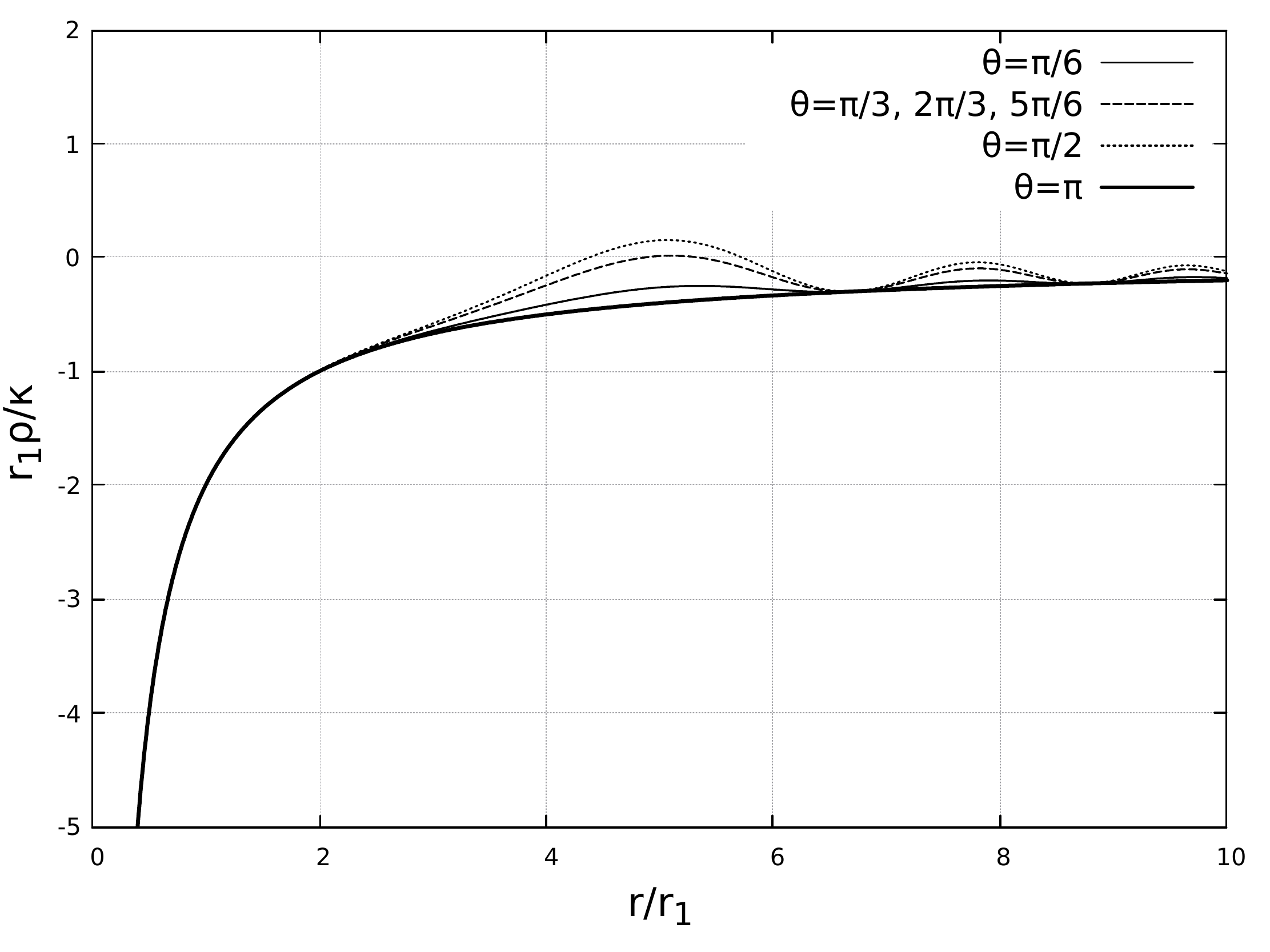}
\caption{ The (normalized and dimensionless) 
color-charge distribution
$\rho(r)$, Eq. (\ref{Linearrho} as a function of $r/r_1$, for $g=1$
for $\gamma= 0.1$.}
        \label{fig:Linearrho}
    \end{minipage}\hfill
    \begin{minipage}{0.49\textwidth}
        \centering
        \includegraphics[width=0.9\textwidth]{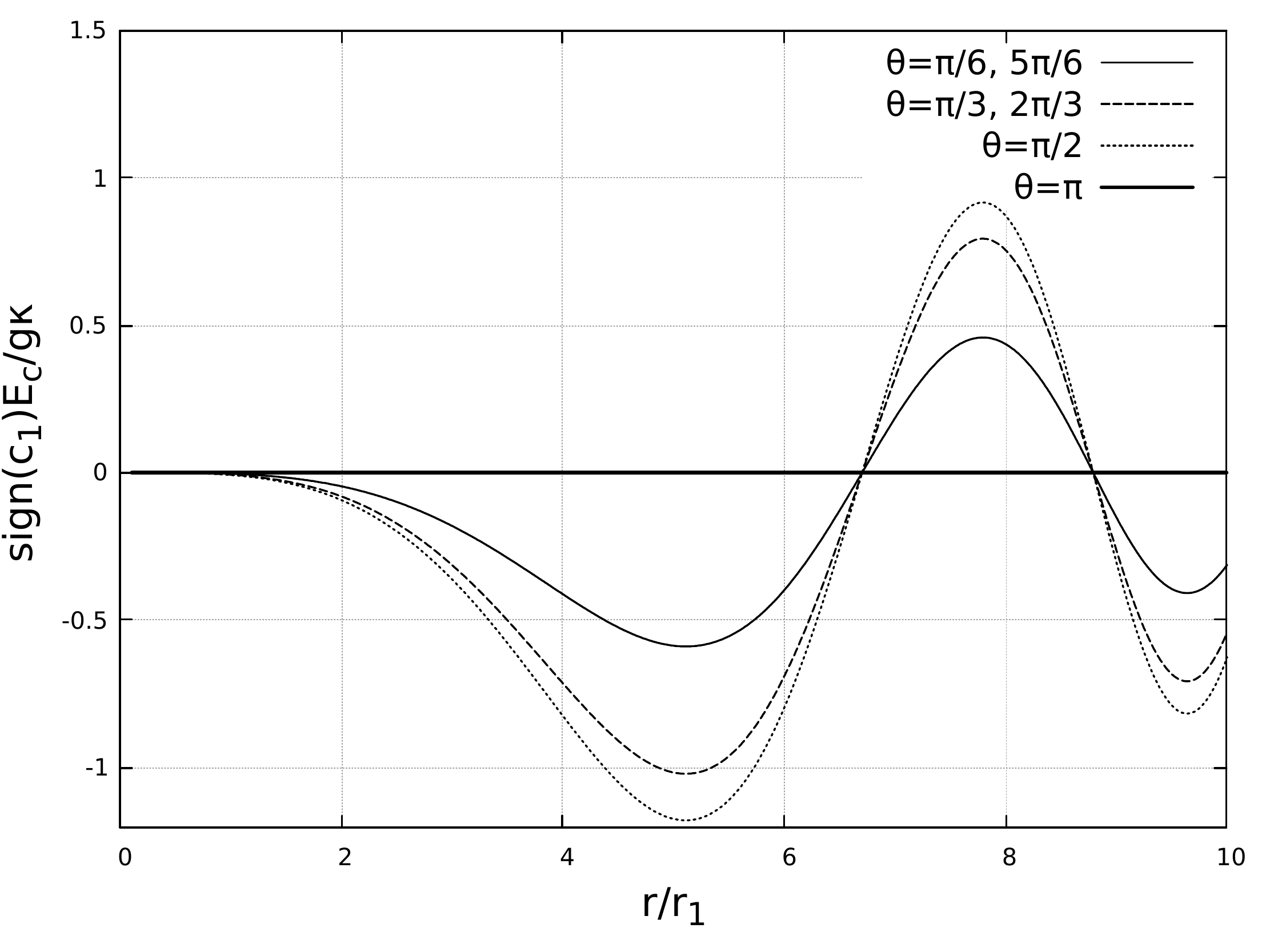}
\caption{ The normalized and dimensionless $E_c$,
Eq. (\ref{LinearEc}),  as a function of $r/r_1$
for $\gamma= 0.1$.}
        \label{fig:LinearE}
    \end{minipage}
\end{figure}

With this, the charge density, C-Electric and C-Magnetic fields for each of the 
SU(2) subgroups   with the condition \ref{sigma0}) are, respectively:
\begin{align}\nonumber
\mbox{\textbf{I-spin:}}\\
\label{LinearEc}
\vec{E}_a(\vec{r}) &=\kappa \hat{r} \delta_{a3} -g\kappa \sign(c_1) \sqrt{\frac{r}{r_1}} J_{3/2} \left(\frac{g\kappa r^2}{2}\right) \sin(\theta) \hat{\phi} (\delta_{a1}-\delta_{a2});
\\ \nonumber
\vec{B}_a(\vec{r}) &= -\sign(c_1) \frac{1}{r_1^2} \left[ \hat{r} \left( \frac{r_1}{r} \right)^{3/2} 2J_{3/2} \left(\frac{g\kappa r^2}{2}\right) \cos(\theta) \right.\\ \label{LinearBc}
&\left. - \hat{\theta} \sin(\theta) \left( \left( \frac{r_1}{r} \right)^{3/2} 2 J_{3/2} \left(\frac{g\kappa r^2}{2}\right) - g\kappa r_1^2 J_{5/2} \left(\frac{g\kappa r^2}{2}\right) \right)   \right](\delta_{a1}+\delta_{a2});\\ \nonumber
\mbox{\textbf{V-spin:}}\\
\vec{E}_a(\vec{r}) &=\kappa \hat{r} \left( \frac{1}{2}\delta_{a3} + \frac{\sqrt{3}}{2} \delta_{a8} \right) -g\kappa \sign(c_1) \sqrt{\frac{r}{r_1}} J_{3/2} \left(\frac{g\kappa r^2}{2}\right) \sin(\theta) \hat{\phi} (\delta_{a4}-\delta_{a5});
\\ \nonumber
\vec{B}_a(\vec{r}) &= -\sign(c_1) \frac{1}{r_1^2} \left[ \hat{r} \left( \frac{r_1}{r} \right)^{3/2} 2J_{3/2} \left(\frac{g\kappa r^2}{2}\right) \cos(\theta) \right. \\
&\left. - \hat{\theta} \sin(\theta) \left( \left( \frac{r_1}{r} \right)^{3/2} 2 J_{3/2} \left(\frac{g\kappa r^2}{2}\right) - g\kappa r_1^2 J_{5/2} \left(\frac{g\kappa r^2}{2}\right) \right)   \right](\delta_{a4}+\delta_{a5});\\ \nonumber
\mbox{\textbf{U-spin:}}\\
\vec{E}_a(\vec{r}) &=\kappa \hat{r} \left( \delta_{a3} - \frac{\sqrt{3}}{3} \delta_{a8} \right) -g\kappa \sign(c_1) \sqrt{\frac{r}{r_1}} J_{3/2} \left(\frac{g\kappa r^2}{2}\right) \sin(\theta) \hat{\phi} (\delta_{a6}-\delta_{a7});
\\ \nonumber
\vec{B}_a(\vec{r}) &= -\sign(c_1) \frac{1}{r_1^2} \left[ \hat{r} \left( \frac{r_1}{r} \right)^{3/2} 2J_{3/2} \left(\frac{g\kappa r^2}{2}\right) \cos(\theta) \right. \\
&\left. - \hat{\theta} \sin(\theta) \left( \left( \frac{r_1}{r} \right)^{3/2} 2 J_{3/2} \left(\frac{g\kappa r^2}{2}\right) - g\kappa r_1^2 J_{5/2} \left(\frac{g\kappa r^2}{2}\right) \right)   \right](\delta_{a6}+\delta_{a7}),
\end{align}
The corresponding charge density will be given by:
\begin{align}\label{Linearrho}
\rho (\vec{r}) &= \frac{\kappa}{ r_1} \left( 2g^2 \sin^2(\theta) 
 \left[ J_{3/2} \left(\gamma \left(\frac{r}{r_1}\right)^2\right) \right]^2 - 2\frac{r_1}{r} \right).
\end{align}
This color-charge density is shown in Fig. (\ref{fig:Linearrho}) for different angles 
as a function of the (normalized) radial coordinate $r/r_1$ for 
$\gamma = 0.1$ and $g=1$.
Note that for a small solid angle around $\pi/2$ the color charge density reaches zero(s)
for quite shorter distances as compared to its possible reaching distances.
These values were also adopted for the next  Figs. 
(\ref{fig:LinearE}),(\ref{fig:LinearBr}) and (\ref{fig:LinearBtheta})
for the c-electric and c-magnetic fields.
For larger values of  $\gamma$, the c-electric and the c-magnetic  fields oscillate progressively 
more at short distances, 
so that for smaller values of $\gamma$ the oscillations still happens although they take place for larger 
$r/r_1$.
The strongly anisotropic behavior is present in the chromo-electric and chromo-magnetic fields
that are shown in Figs. (\ref{fig:LinearE},\ref{fig:LinearBr}) and (\ref{fig:LinearBtheta})
for different angles $\theta$.
The field $E_c$ and the component $B_{c,r}$ have their zeros at the same point and 
the component $B_{c,\theta}$ has zero in smaller values of $r/r_1$.
Only the radial component $B_{c,r}$ changes sign for specific directions,
specifically for $-\pi/2 < \theta < \pi/2$.
 Note that, again the Ec and Bc fluxes are highly anisotropic, similarly to the 
effects discussed above for the strict Coulomb  potential.

\begin{figure}[H]
\centering
\includegraphics[width=0.441\textwidth]{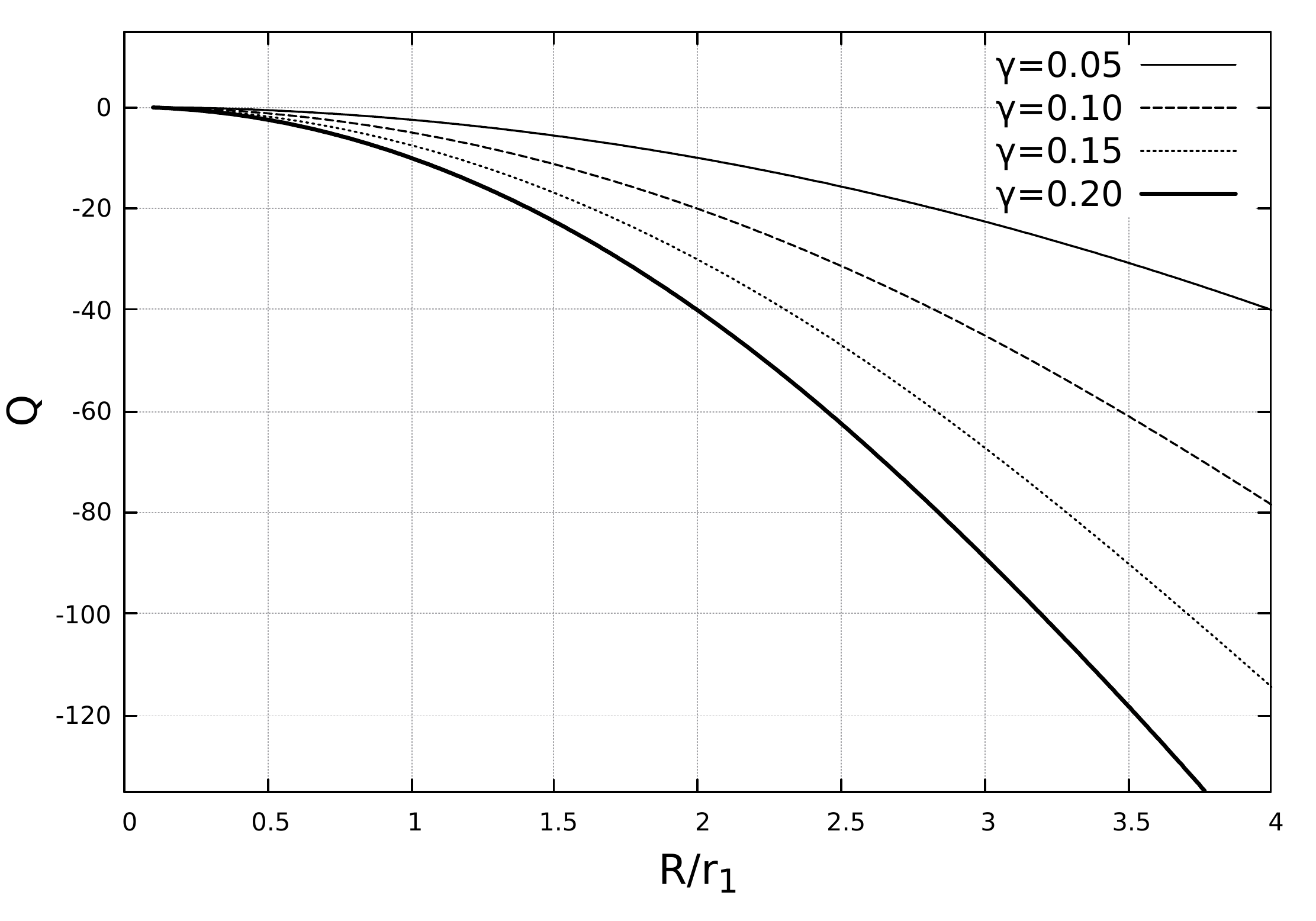}
\caption{ Total color-charge contained in a sphere of radius $R$
as a function of $R/r_1$,
for different values of $\gamma$. }
    \label{fig:Qtotlinear}
\end{figure}

The parameter $r_1=1/c_1^2$ is related to the normalization of 
gauge vector potential, $\vec{A} \cdot \hat{\phi}$, 
that, together with the dimensionless parameter
 $\gamma=g r^2 \kappa/2$, settle possible zeros for the color-charge density
$\rho(r,\theta)$, that is highly anisotropic. 
Analogously to the case of anisotropic Coulomb potential
a surface of zero color-charge density can also be defined by:
\begin{align}\label{relacaolinear}
\frac{r_1}{r} = g^2 \sin^2(\theta) \left[ J_{3/2} \left( \gamma \left( \frac{r}{r_1} \right)^2 \right) \right]^2.
\end{align}
It can be seen, however, that for $\theta=0,\pi$,
there are no zeros of the color-charge density.
 The non-Abelian contribution is intrinsically 
non-spherically symmetric.
Therefore, the shape of the relation \ref{relacaolinear} gives us information about the
 shape of the localized charge distribution although it is not limited to a closed region in space.
The color charge density might also change its sign as one increases the radial coordinate $r$
and in this case, a total zero color charge might be found for an extended color distribution.
In some sense,  this is remarkably similar to a bag model for a "non-spheric bag" 
 and  possibly, a different mechanism or 
external restriction of its spatial distribution should be imposed.
These solutions are more well behaved than 
the solutions found for the extension of the Coulomb potential
of Sec. (\ref{sec:nonsphericalcoulomb}).
We did not find a straightforward way to fix the constant $\kappa$, as it would be desirable,
although this is possible in a {\it precarious way}
 for a particular value of $\gamma$
and a particular radius $r=R_0$ the following definition appeared above:
\begin{eqnarray}
\kappa = \frac{2 \gamma_0}{ g R_0^2}
\end{eqnarray}
where $\gamma_0 = \gamma (R_0^2/r_1^2)$.
The linear potential found in lattice QCD is rather expected to be leading
at larger distances in comparison to the shorter distances in which the Coulomb potential, and 
others must be more important.
In the specific   radial coordinate  $R_0$ is related to the 
characteristic  distance $r_1$ above, it is associated to the vector potential normalization 
from Eq. (\ref{c1-r1}).
For $\gamma = 0.1$ the color-charge density reaches zero, $\rho(\vec{r})=0$,
 for a solid angle around $\theta= \pi/2$
 for $R_0/r_1 \sim 4$  from Fig. (\ref{fig:Linearrho}).
By adopting $R_0/r_1 = x$,  and the 
same normalization for both the scalar and vector potentials
 $ |\frac{1}{r_1}| \sim |\kappa R_0|$,
one has:
$\kappa = g/(2 x^4 \gamma^2 R_0^2)$.
By considering $x \sim 4$,  $g\simeq 1$, $R_0\sim 0.5$ fm. 
 and $\gamma \sim 0.1$  
it  yields  $\kappa \sim 0.03$ GeV$^2$
that is somewhat smaller  than $\sigma_{latt} \sim 0.19$ GeV$^2$
 \cite{SM,bali}.
The total color-charge contained in a sphere of radius $R$
as a function of the normalized radius $R/r_1$ is presented in Fig. (\ref{fig:Qtotlinear}).

\hspace{1cm}
\begin{figure}[H]
    \centering
    \begin{minipage}{0.49\textwidth}
        \centering
        \includegraphics[width=0.9\textwidth]{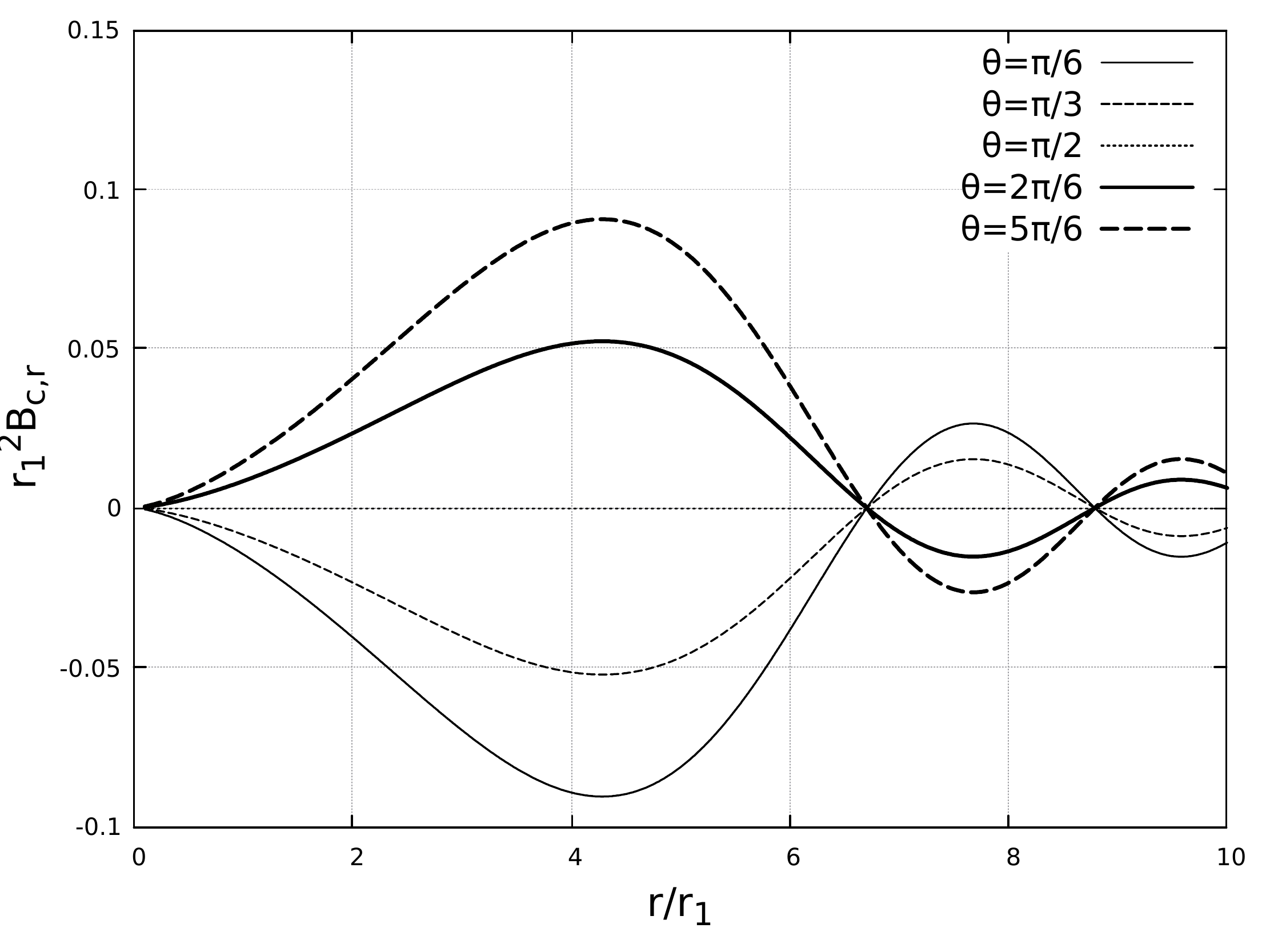}
\caption{  The normalized and dimensionless 
$\vec{B}_c \cdot \hat{r} \equiv B_{c,r}$, Eq. (\ref{LinearBc}), 
 as a function of $r/r_1$
for $\gamma= 0.1$.
}
        \label{fig:LinearBr}
    \end{minipage}\hfill
    \begin{minipage}{0.49\textwidth}
        \centering
        \includegraphics[width=0.9\textwidth]{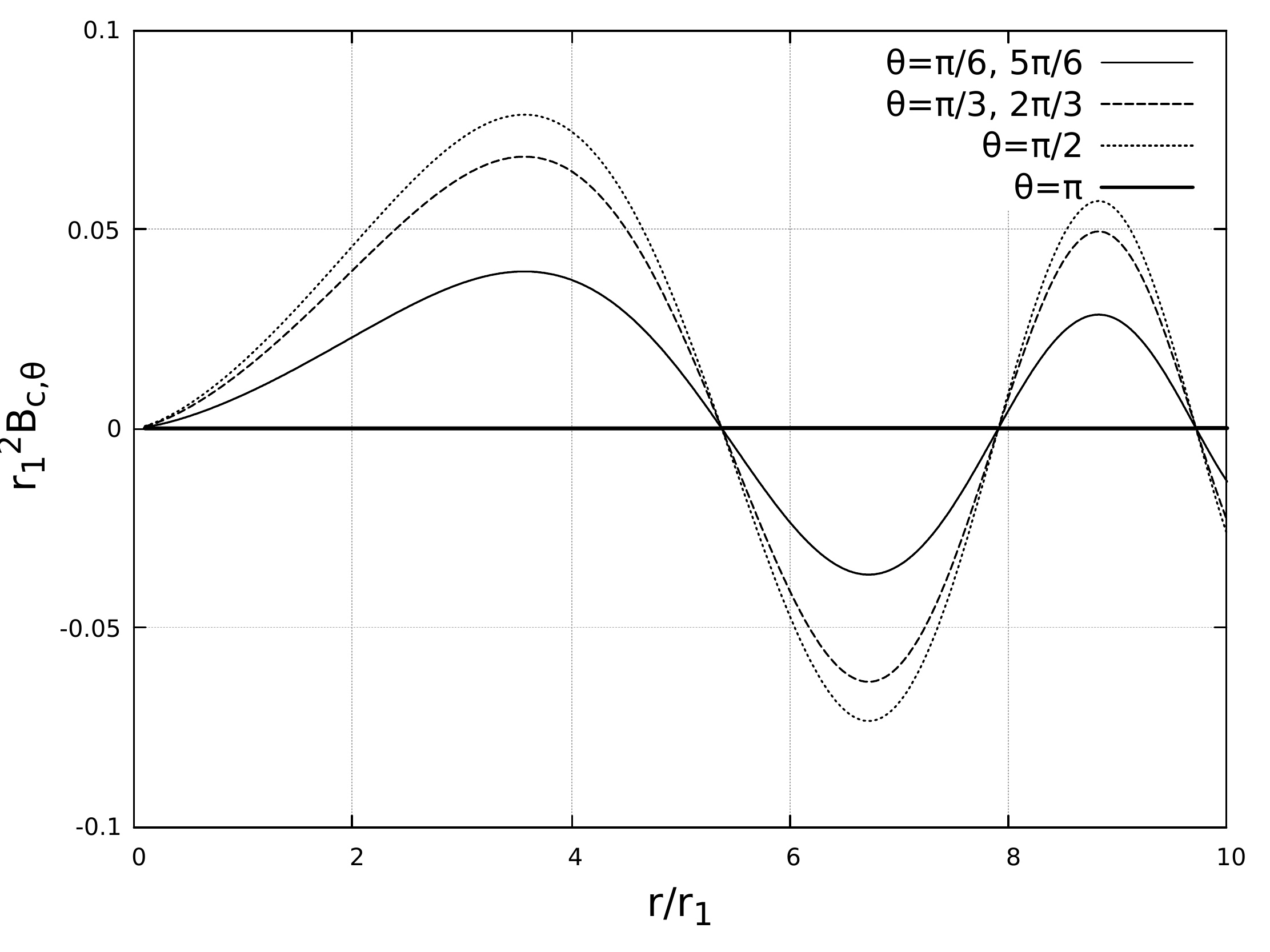}
\caption{
 The normalized and dimensionless 
$\vec{B}_c \cdot \hat{\theta} \equiv B_{c,\theta}$,
Eq. (\ref{LinearBc}),
 as a function of $r/r_1$
for $\gamma= 0.1$.}
        \label{fig:LinearBtheta}
    \end{minipage}
\end{figure}

\section{   Yukawa potential }
\label{sec:yukawa}

In this section the following spherical symmetric Yukawa type potential will be considered:
\begin{align}
\varphi (\vec{r}) = - C  \frac{e^{- \frac{r}{r_0} } }{r},
\end{align}
where $r_0$ is a parameter to be determined.
The equations \ref{SEMISYM1} - \ref{SEMISYM3} 
can be solved by separation of variables
($A_\phi  (\vec{r})= R_\phi(r) \Theta_\phi(\cos(\theta) \equiv x)$)
 whose resulting equations can be written as:
\begin{align}
\label{YukawaRtheta}
r^2 \frac{d^2R_\theta}{dr^2} + 2r\frac{dR_\theta}{dr} + (g  C)^2 e^{-2r/r_0} R_\theta &= 0,
\\ \label{YukawaRphi}
r^2 \frac{d^2R_\phi}{dr^2} +2r\frac{dR_\phi}{dr} + 
\left( (g  C)^2 e^{-2r/r_0} - l(l+1) \right) R_\phi &= 0,
\\ \label{Yukawathetax}
(1-x^2) \frac{d^2\Theta_\phi}{dx^2} - 2x \frac{d\Theta_\phi}{dx} + 
\left( l(l+1) - \frac{1}{1-x^2} \right)
 \Theta_\phi &=0.
\end{align}
Equations (\ref{YukawaRtheta}) and (\ref{YukawaRphi}) do not depend on the normalization of 
$R_\theta$ and $R_\phi$ and therefore the more general solution may  be 
obtained by multiplied them by a constant $a$, i.e $a R_\theta$ and $a R_\phi$ 
that will be searched numerically.
Again, the solution for the angular part, equation 
(\ref{Yukawathetax}), is given by Legendre's associated 
functions of order $\pm 1$. It's interesting to see that the equation for 
$R_\theta$ is a particular case of the one for $R_\phi$, namely, $l=0$. 
The resulting color-charge density that yield the solutions above is given by
\begin{eqnarray}
\label{Yukawarho}
\rho(\vec{r}) = \frac{C e^{- \frac{r}{r_0}}}{r} \left( \frac{1}{r_0^2} - 2g^2 A^2(\vec{r}) \right) .
\end{eqnarray} 
To total charge $Q$ given by integration of \ref{Yukawarho} diverges if the functions 
$R_\phi,R_\theta,\Theta_{\theta}$ are given by the associated Legendre's functions of the second kind. Therefore, if we impose that the only acceptable solutions are given by the $P^m_l(\cos(\theta))$ (implying that $R_\theta = 0$), we have:
\begin{align}
Q = 4\pi C \left(1-g^2 \frac{2(l+m)!}{(2l+1)(l-m)!} \int_0^\infty e^{-r/r_0}r (R_\phi)^2 dr \right)
\end{align}

The equation for $\Theta_\phi$, \ref{Yukawathetax}, can be solved for $x$, arriving at:
\begin{align}
\Theta_\phi(\theta)= c \cosh \left( \ln  \sqrt{\frac{1-x}{1+x}}  \right) + b \sinh \left( \ln \sqrt{\frac{1-x}{1+x}}  \right),
\end{align}
where $c$ and $b$ are the integration constants. Then, the full solution for $\vec{A}$ is given by:
\begin{align}\label{YukawaA}
\vec{A}(\vec{r}) = \frac{R_\theta (r)}{\sin(\theta)} \hat{\theta} +  R_\phi (r)
 \left( c \cosh \left( \ln  \sqrt{\frac{1-\cos(\theta)}{1+\cos(\theta)}}  \right) + b \sinh \left( \ln \sqrt{\frac{1-\cos(\theta)}{1+\cos(\theta)}}  \right) \right)\hat{\phi},
\end{align}
where the normalization of each of the components are implicit and they are obtained numerically.

Analytical solutions for $R_\theta$ and $R_\phi$ have not been found,
and 
numerical solutions were searched 
for the following boundary conditions:
\begin{align}
\label{Yukawabc1}
\mbox{\textbf{BC1:}}
\;\;\;
  \lim_{r = \delta r\rightarrow0} R_{\theta,\phi} &\sim - \frac{1}{\delta r} \to -\infty;
\nonumber
\\
\lim_{r=\delta r\rightarrow0} \frac{dR_{\theta,\phi}}{dr} & 
\sim  \frac{1}{(\delta r)^2} \to \infty,
\end{align}
and:
\begin{align}
\label{Yukawabc2}
\mbox{\textbf{BC2:}}
\;\;\;
\lim_{r\rightarrow0} R_{\theta,\phi} &\sim  \frac{1}{\delta r} \to \infty;
\nonumber
\\
\lim_{r\rightarrow0} \frac{dR_{\theta,\phi}}{dr} & \sim -  \frac{1}{(\delta r)^2} \to  -\infty.
\end{align}
The resulting profiles for the lowest $l \ge 1$ exhibit somewhat
unphysical oscillatory or divergent behavior that are not exhibited.
So, below we show solutions for $l=0$ in which case
we have the same equations for $R_\theta$ and $R_\phi$.

\begin{figure}[H]
    \begin{minipage}{0.49\textwidth}
        \includegraphics[width=1\textwidth]{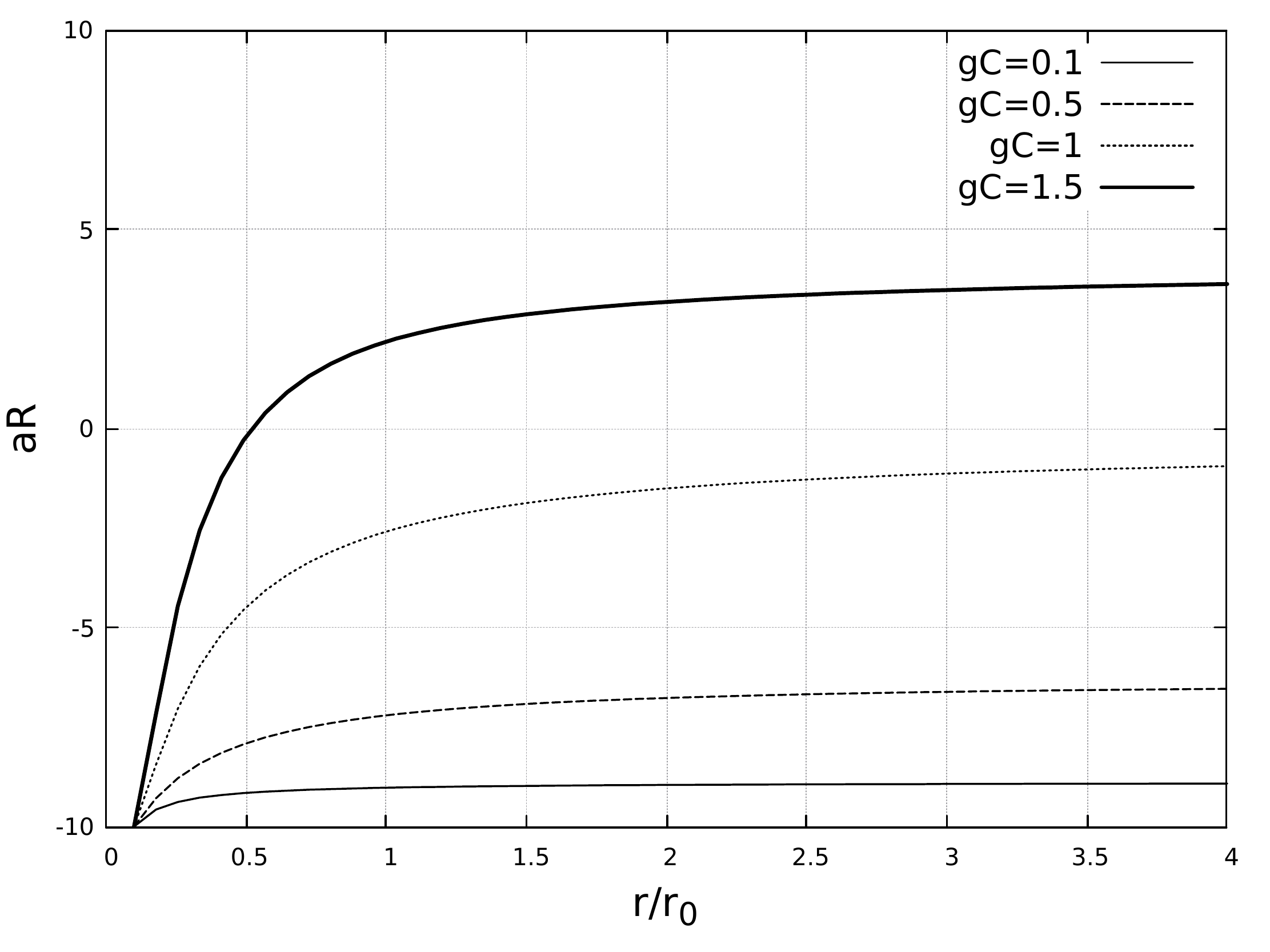}
\caption{Numerical solutions for Eq. (\ref{YukawaRphi}) for the boundary condition at the origin (\ref{Yukawabc1})
and  for different values of $gC$, and the integration constant $a$ made explicit.}
    \label{fig:Yukawa1}
    \end{minipage}\hfill
    \centering
    \begin{minipage}{0.49\textwidth}
        \centering
        \includegraphics[width=1\textwidth]{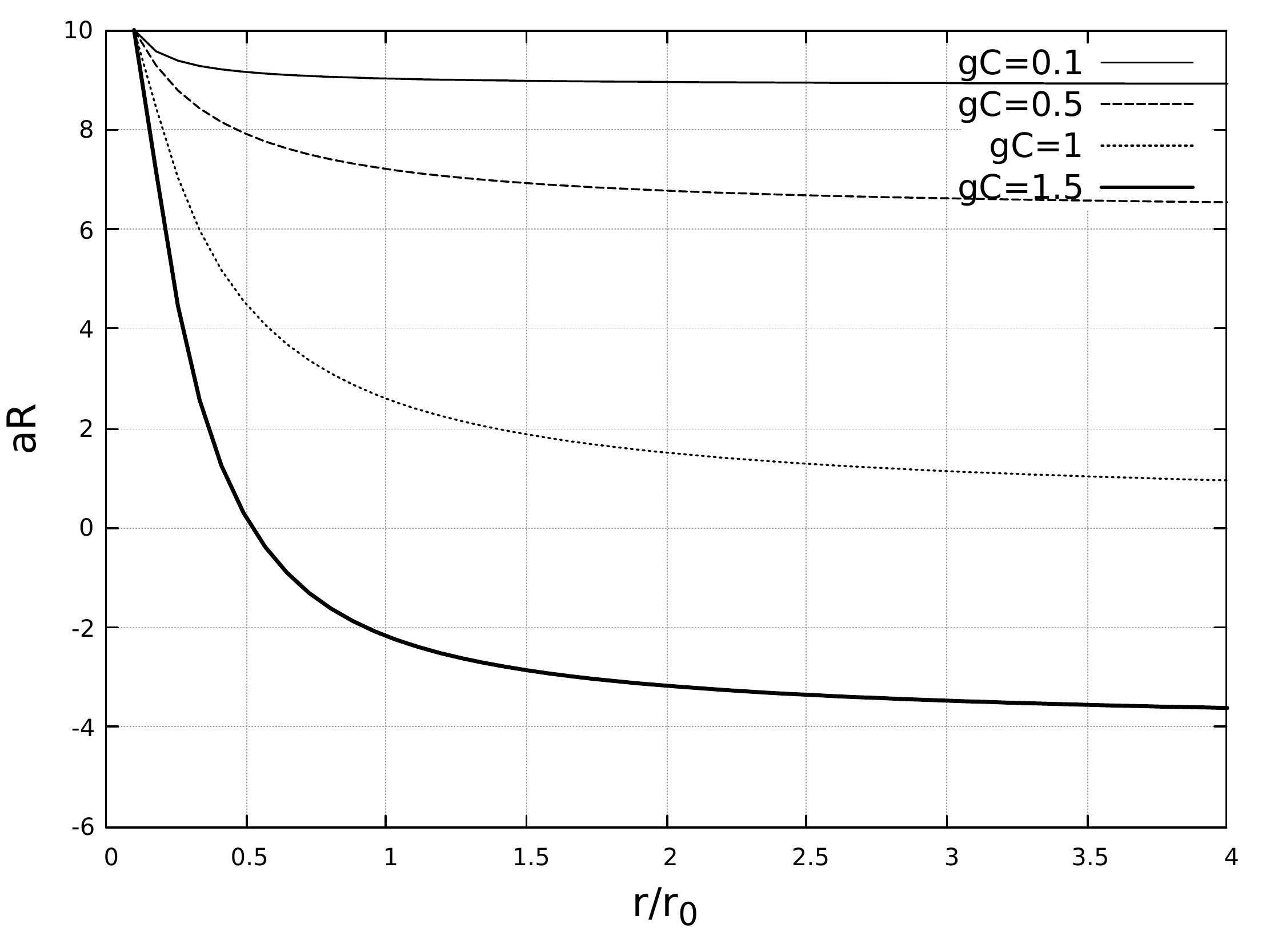}
\caption{Numerical solutions for Eq. (\ref{YukawaRtheta}) for the boundary condition at the origin 
(\ref{Yukawabc2})
and  for different values of $gC$, and the integration constant $a$ made explicit.}
    \label{fig:Yukawa2}
    \end{minipage}
\end{figure}

In Fig. (\ref{fig:Yukawa1}),
 solutions for the normalized/dimensionless $C R=C R_\theta,R_\phi$ with $l=0$ 
for the boundary conditions at the origin given by (\ref{Yukawabc1})
are shown. 
By increasing the value of $gC$ the asymptotic value 
$R (r\to \infty)$ becomes more negative.
In Fig. 
(\ref{fig:Yukawa2}),
 solutions for $C R=C R_\theta,R_\phi$ with $l=0$ 
for the boundary conditions at the origin given by (\ref{Yukawabc2})
are shown. 
By increasing the value of $gC$ the asymptotic value $R (r\to \infty)$ becomes more positive.
In both cases, however, for $(gC)^2 \approx2.33$
(i.e. $gC \simeq 1.53$)
  the asymptotic values go to zero 
$R(r\to \infty) = 0$.
 
\begin{figure}[H]
\centering
    \begin{minipage}{0.49\textwidth}
  \centering
\includegraphics[width=1\textwidth]{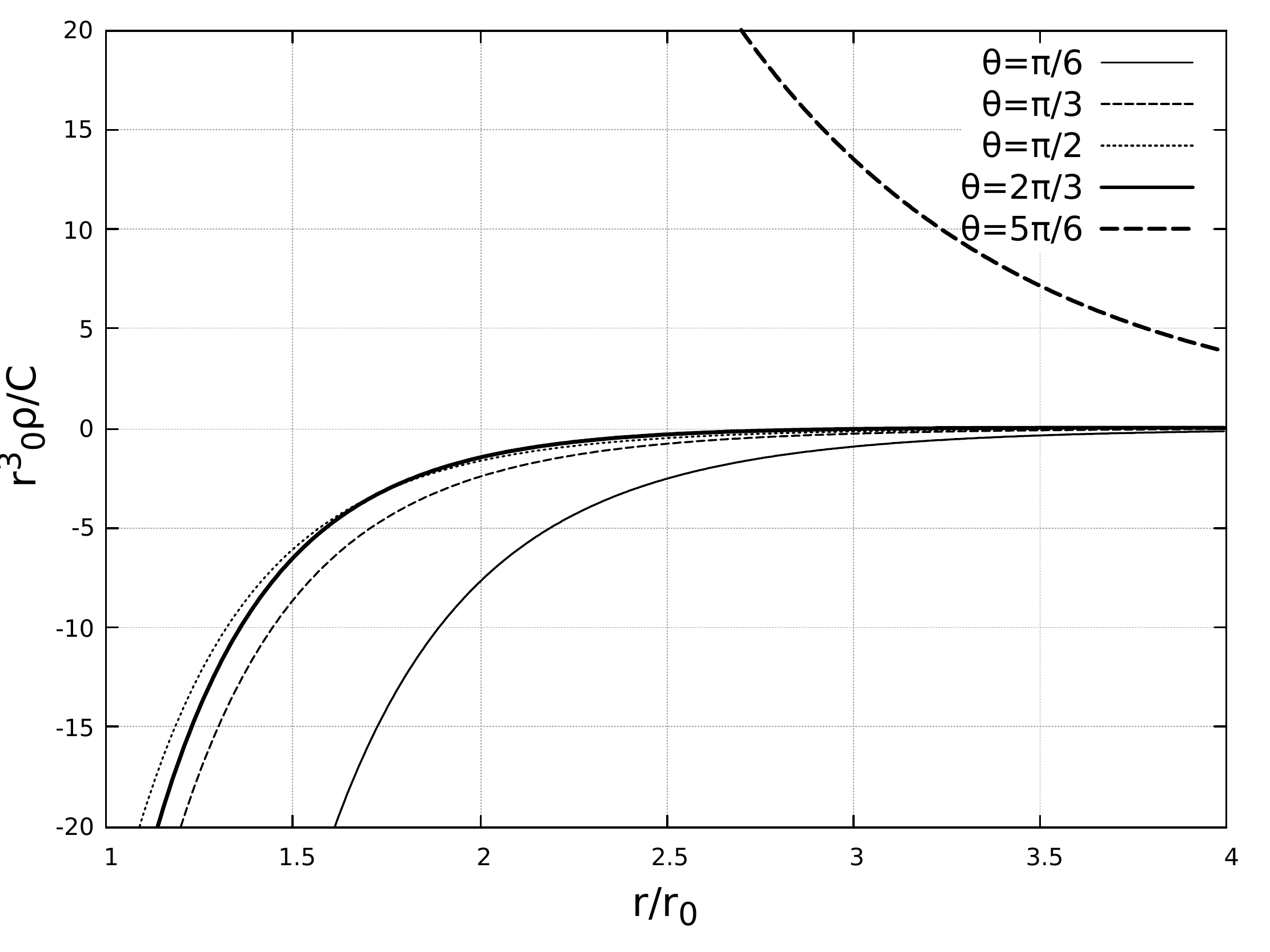}
\caption{Normalized charge density in equation \ref{Yukawarho}, as a function of $r/r_0$ for different $\theta$. Here, $g=1$ was used and $C^2=2.33$.}
\label{fig:YukawaRho}
\end{minipage}\hfill
\centering
\begin{minipage}{0.49\textwidth}
\centering
\includegraphics[width=1\textwidth]{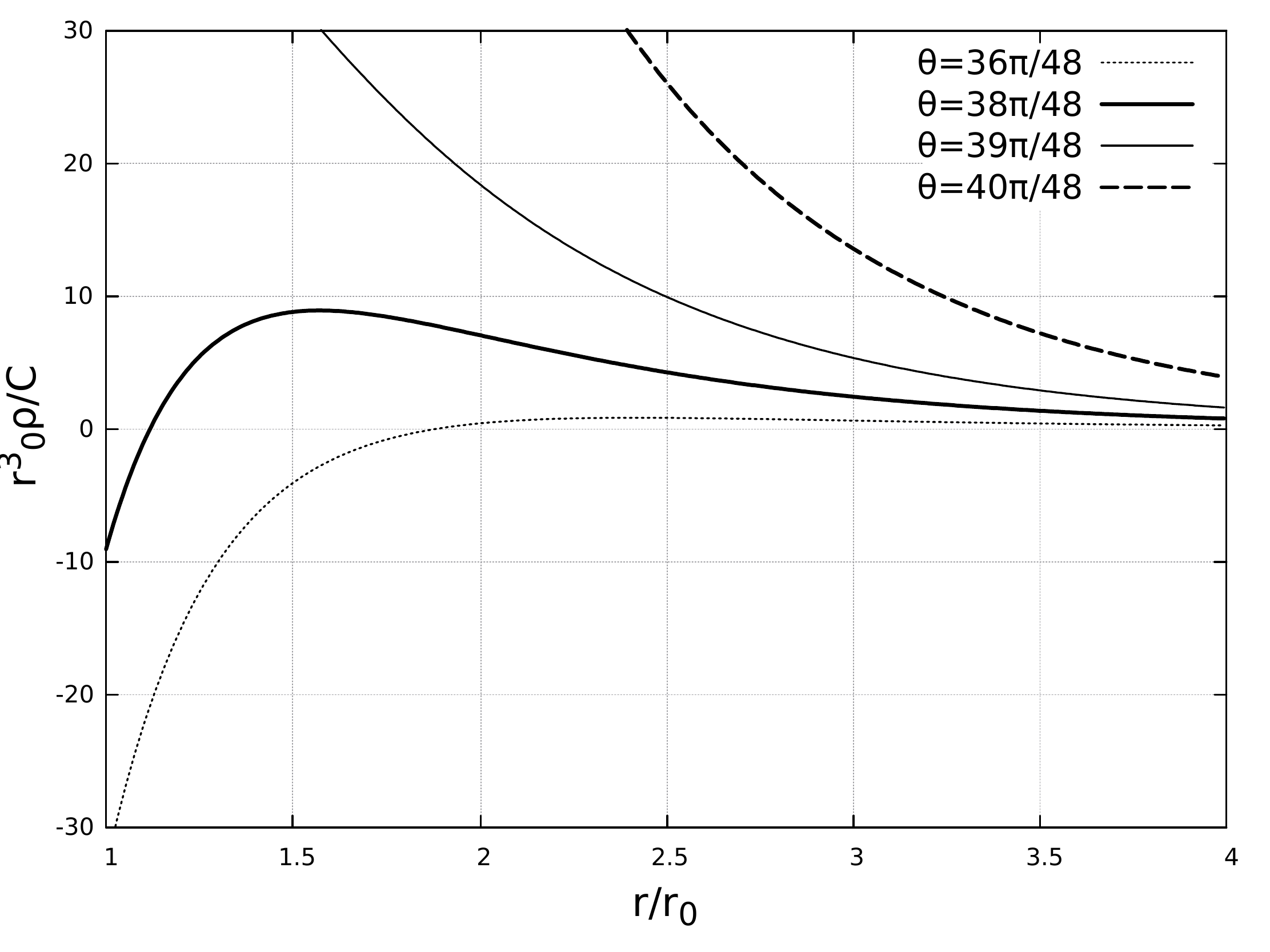}
\caption{Normalized charge density in equation \ref{Yukawarho}, as a function of $r/r_0$ for different $\theta$. Here, $g=1$ was used and $C^2=2.33$.}
\label{fig:YukawaRho2}
    \end{minipage}
\end{figure}

The charge density associated with the solution in equation \ref{YukawaA}, for $(gC)^2 = 2.33$, given by equation (\ref{Yukawarho}), is shown in figures  (\ref{fig:YukawaRho}) and (\ref{fig:YukawaRho2}).
In these Figures, the  numerical solutions
are normalized  by  
$a=r_0$ by considering $c=d=1$.
In Fig.  (\ref{fig:YukawaRho}) the overall behavior
for the  (normalized-dimensionless) color-charge distribution
 is shown for the same angles 
considered in the previous figures.
In Fig.  (\ref{fig:YukawaRho2}) the same quantity is exhibited
for  specific angles that 
help to show  how the color-charge distribution changes sign around $\theta \sim 19\pi/24$
($\theta \sim 0.79 \pi$),
and correspondingly the inverse change of sign close to $\theta \sim 29\pi/24$
($\theta \sim 1.21 \pi$).
The dipole-type configuration is therefore 
deformed such that
the negative color-charge lies within a smaller solid angle.

\section{ Constant scalar potential in a dipole-like spatial distribution  }
\label{sec:Vconstant}

Let us consider a finite  region of 3-dimensional space in which the classical scalar potential 
is constant, i.e.
\begin{align}
\varphi = V = const. 
\end{align}
Equation (\ref{EquaçãoGeneralizada3}) for the vector potential,
a sort of non-Abelian generalization of the Amp\`ere's Law in the  Coulomb Gauge \ref{CoulombGauge},
 is reduced to the Helmholtz equation, \cite{butkov}. 
 By neglecting  the solutions that diverge in the origin
we have the following solution:
\begin{align}\nonumber
\vec{A} (\vec{r}) =& \sum_{l=0}^{\infty} \sum_{m=-l}^{l} j_l(gVr) Y^m_l (\theta,\phi) \left( \hat{r} \left( a_l^m \sin(\theta) \cos(\phi) + b^l_m \sin(\theta) \sin(\phi) + c^{l}_m \cos(\theta) \right) \right.\\
&+ 
\left.\hat{\theta} \left( a^l_m \cos(\theta) \cos(\phi) + b^l_m \cos(\theta) \sin(\phi) - c^l_m \sin(\theta) \right) + \hat{\phi} \left( -a^l_m \sin(\phi) + b^l_m \cos(\phi) \right) \right),
\end{align}
where $a^l_m, b^l_m$ and $c^l_m$ are the integration constants, $j_l(x)$ are the spherical 
Bessel functions and $Y^m_l(\theta,\phi)$ are the spherical harmonics. 
 Eq.
(\ref{EquaçãoGeneralizada2}), in this case, reduces to:
\begin{align}\label{AeRho}
\vec{A}^2   (\vec{r}) \equiv A^{2}  (\vec{r}) = \frac{\rho (\vec{r})}{2g^2V}  .
\end{align}
Therefore the constant scalar potential solution is possible provided the 
color charge density has a very particular shape.

As we can see, since $A^2$ is positive, if $V<0$ ($V>0$), then $\rho < 0$ ($\rho>0$). 
With that in mind, we choose to use this solution to a model 
for a   color-anticolor dipole configuration restricted to 
two semi-spheres with  color charges with opposite sign
 of radius $R_0$, by means of   the  following prescription:
\begin{eqnarray} \label{varphi-dipole}
\varphi (\vec{r}) = V \left( H(-r+R_0,\theta+\frac{\pi}{2}) - H(-r+R_0, \theta - \frac{\pi}{2}) \right),
\;\;  V >0,
\end{eqnarray}
where $H(x)$ is a two dimensional Heaviside function
for the coordinates $r$ and $\theta$. 
Note that, according to \ref{qeta}, this is a model for two pairs of dipoles, i.e.
 four color charges that create a color-neutral system.
This discontinuity may be thought by  considering a very  tiny region without
color charge that separates the two different regions, for $\varphi = + V$ and $\varphi = - V$.
Eventual chromo-electromagnetic fluxes in this tiny transition region will be neglected.
This sort of configuration would be possible in electromagnetic system for two conducting 
semi-spheres with superficial charge distribution, being this picture however 
not suitable for 
the present Yang Mills system because of strong non-Abelian effects:
in spite of being equipotential volumes the chromo-electric fields are not zero.
 From eq. 
(\ref{AeRho}) it corresponds to a dipolar configuration
being that in both regions the vector potential $A^2(\vec{r})$ is the same.
 It's important to notice that $\vec{A}$ can be non-zero only in the region where $\varphi \neq 0$. 
We can achieve this continuously by imposing that for $r=R_0$ one has  $\vec{A}=0$. 
By picking up a single component $l$, this implies:
\begin{align}
gVR_0 = Z^l_n,
\end{align}
where $Z^l_n$ is the $n-$th zero of the spherical Bessel function of order $l$. 
Some of these zeros are given in table \ref{tabela3}.

\begin{table}[H]
\centering
\resizebox{0.3\columnwidth}{!}{
\begin{tabular}{|c|c|c|c|}
\hline
$l$ & $Z^l_1$ & $Z^l_2$ & $Z^l_3$ \\
\hline
1 & 4.493 & 7.725 & 10.904\\
\hline
2 & 5.764 & 9.095 & 12.323\\
\hline
3 & 6.988 & 10.417 & 13.698\\
\hline
\end{tabular}}
\caption{Zeros for spherical Bessel function, of different lowest  orders of $l$.}
\label{tabela3}
\end{table}

As an example, consider the lowest solution in $l$ ($l=1$) that 
 is given by:
\begin{eqnarray} \nonumber
\vec{A} (\vec{r}) &=&
  \left( \frac{\sin(gVr)}{(gVr)^2} - \frac{\cos(gVr)}{gVr} \right) 
\cos(\theta) \left( \hat{r} \left( a^1 \sin(\theta) \cos(\phi) 
+ b^1 \sin(\theta) \sin(\phi) + c^1 \cos(\theta) \right) 
\right.\\
&+& 
\left.\hat{\theta} \left( a^1\cos(\theta) \cos(\phi) + b^1 \cos(\theta) \sin(\phi) - c^1 \sin(\theta) \right) + \hat{\phi} \left( -a^1 \sin(\phi) + b^1 \cos(\phi) \right) \right),
\;\;\;\; r \leq R_0.
\end{eqnarray}
By redefining the integration constant:
\begin{eqnarray} \label{abc-r0}
\sqrt{(a^1)^2 + (b^1)^2 + (c^1)^2} = \frac{1}{r_0},
\end{eqnarray}
the resulting charge density, c-electric and c-magnetic fields  
 - in the region $r<R_0$ - 
are given by:
\begin{align} \label{rhoVconsdip}
\rho (\vec{r}) &=  I_\rho
\left( \frac{2  \; g^2 V}{r_0^2} \right) \frac{1+((gVr)^2-1)\cos^2(gVr)-2gVr 
\sin(gVr)\cos(gVr)}{(gVr)^4} \cos^2(\theta) 
,\\  \label{EcVconsdip}
\vec{E}_a   (\vec{r}) 
&= \frac{gV}{r_0} \left( \frac{\sin(gVr)}{(gVr)^2} - 
\frac{\cos(gVr)}{gVr} \right) \cos(\theta) \hat{k}(\delta_{ac}-\delta_{ad}),
\\  \label{BcVconsdip}
\vec{B}_a  (\vec{r})
 &= \frac{gV}{r_0} \cos(\theta) \sin(\theta) 
\left( \frac{\sin(gVr)(3-(gVr)^2) -3gVr \cos(gVr)}{(gVr)^3} \right)\hat{\phi}(\delta_{ac}+\delta_{ad}),
\end{align}
where the effect of the discontinuity in $\varphi(\vec{r})$, that 
yields a Dirac-delta function,  was omitted.
In the above equation for  the color-charge density equation it was used:
\begin{eqnarray}
I_\rho =  \begin{cases}
1, \mbox{ for } \theta <\pi/2,
\\
-1, \mbox{ for } \theta > \pi/2.
\end{cases}
\end{eqnarray}
Note that all the cases \textit{I,V,U}-spin have the same result.

\begin{figure}[H]
    \begin{minipage}{0.49\textwidth}   
       \includegraphics[width=0.9\textwidth]{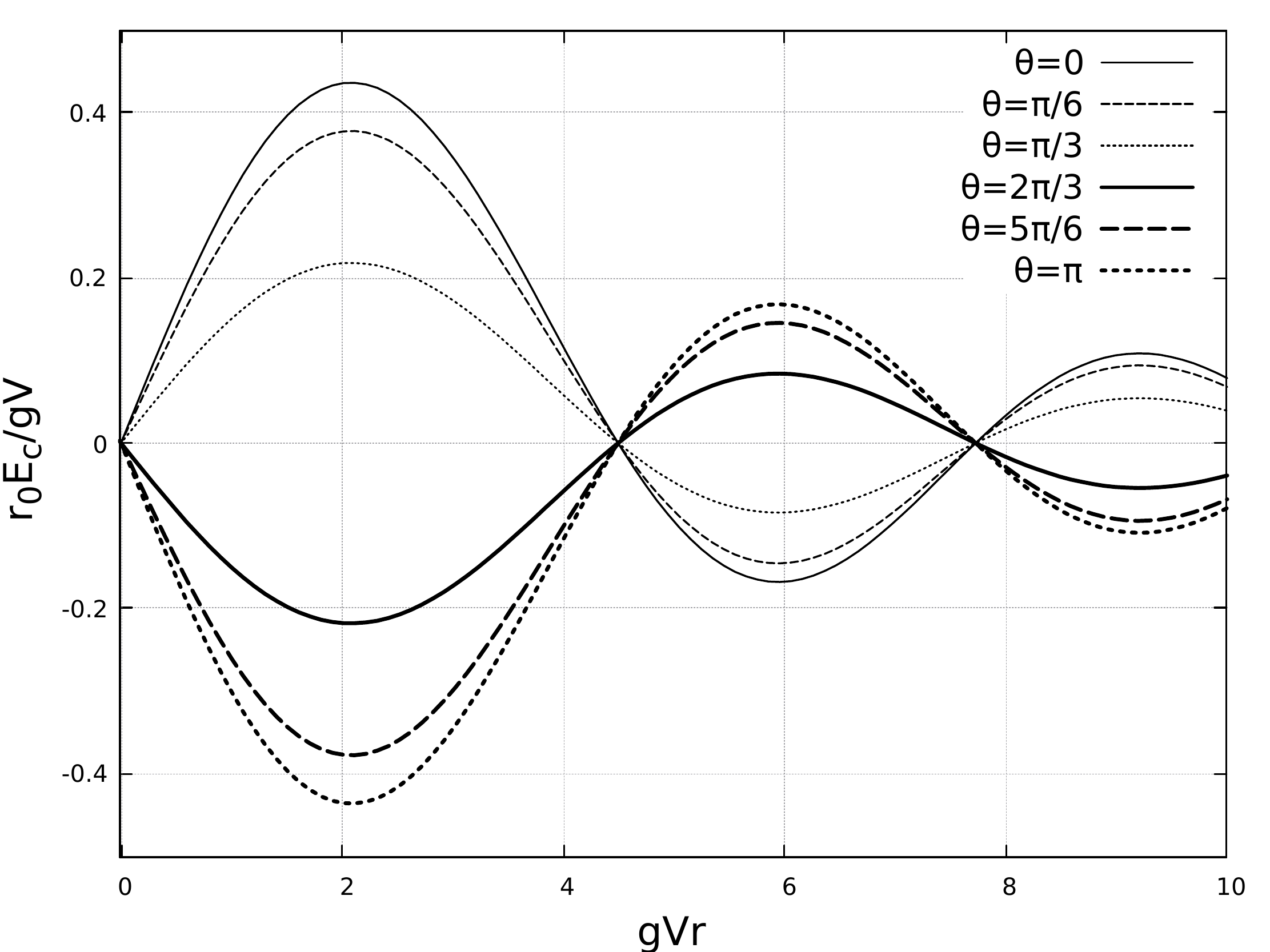}
\caption{$|\vec{E}_c| = E_c$ field, Eq. (\ref{EcVconsdip}), 
as a function of $gVr$ for different angles $\theta$.}
  \label{VconsdipEc}   
    \end{minipage}\hfill
\begin{minipage}{0.49\textwidth}
\includegraphics[width=0.9\textwidth]{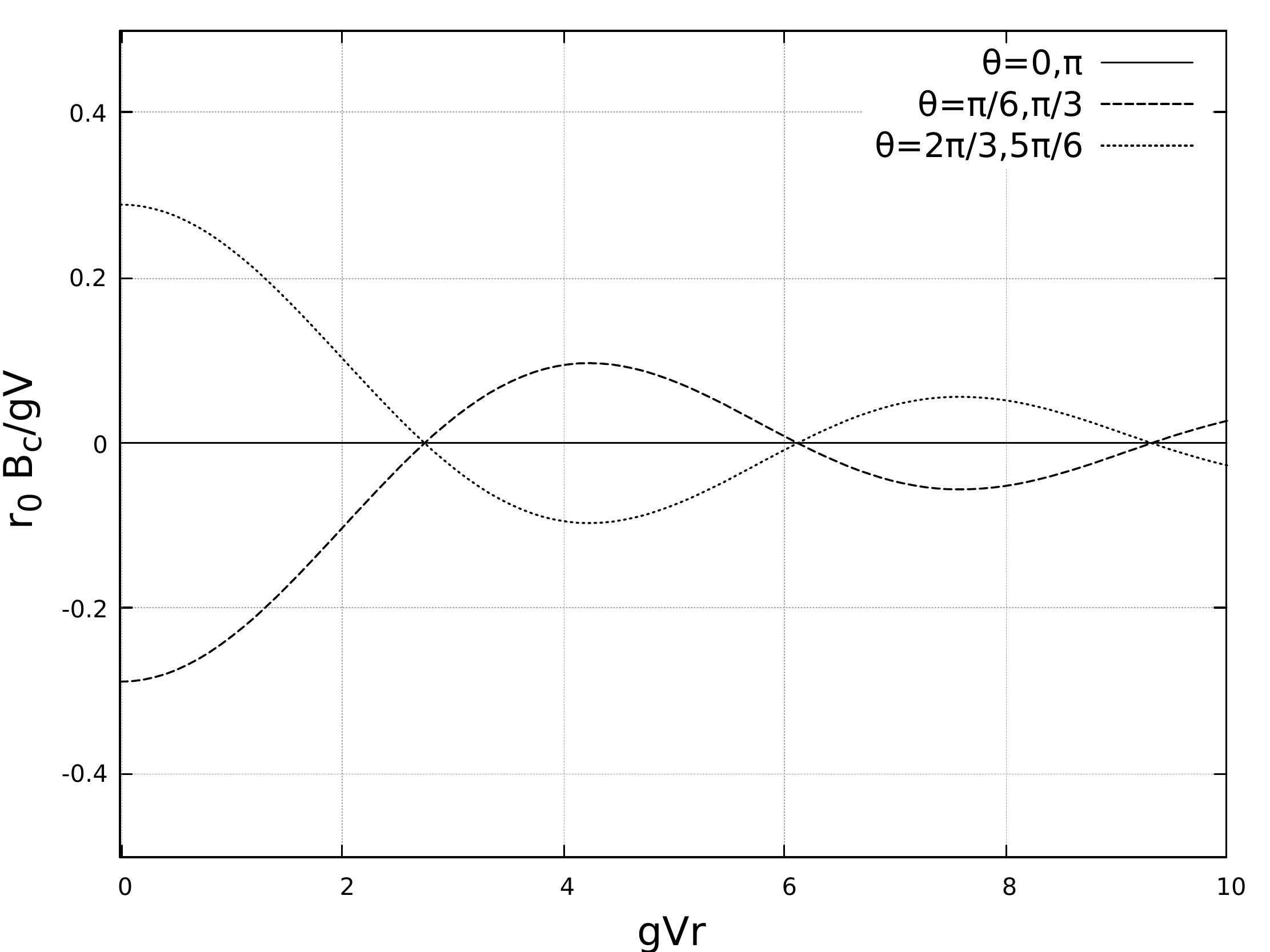}
\caption{$|\vec{B}_c| = B_c$ field, eq. (\ref{BcVconsdip}),
 as a function of $gVr$ for different angles $\theta$ .
 }
  \label{VconsdipBc}
\end{minipage}
\end{figure}

In Figs.  (\ref{VconsdipEc}) and (\ref{VconsdipBc})  
the
normalized/dimensionless
 c-electric field, Eq.  (\ref{EcVconsdip}),  and c-magnetic field, Eq. (\ref{BcVconsdip})
for the lowest $l=1$, are exhibited 
as a function of the dimensionless variable $g V r$
for different angles $\theta$.
The oscillatory behavior must be an indication of the sharp transition between the inner and outer parts of the sphere defined by 
Eq. (\ref{varphi-dipole})
that define the dipole-like configuration.
The radius of the sphere may be identified with the zero of the color-charge density that
is the same as the first zero of the c-electric field.
It can be identified with the lowest zero of the Bessel function:
\begin{eqnarray} \label{R0-constantV}
R_0 \simeq \frac{ 4.493}{ g V}.
\end{eqnarray}
The c-magnetic field, however, has a finite value at $r=R_0$.

\begin{figure}[H]
    \centering
    \begin{minipage}{0.5\textwidth}
        \centering
        \includegraphics[width=0.9\textwidth]{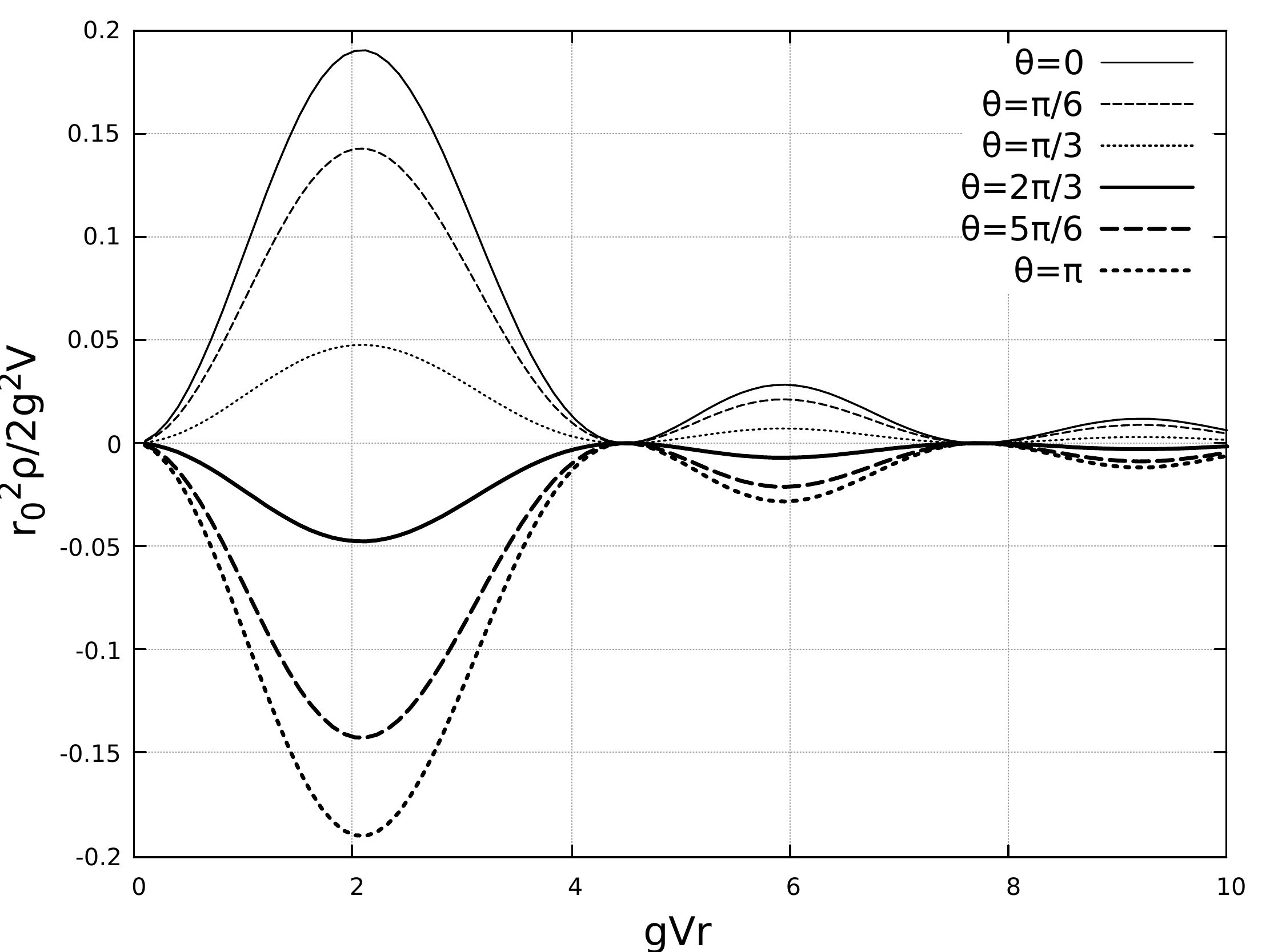}
\caption{Charge density, eq. (\ref{rhoVconsdip}), as a function of $gVr$ for different angles $\theta$.}
  \label{VconsdipRHO}
    \end{minipage}\hfill
\end{figure}

In Fig.
 (\ref{VconsdipRHO})
the corresponding normalized and dimensionless color-charge density is shown as a function
of the variable $g V r$.
The zero is, of course, at the same point as the first zero of the c-electric field
defined above.

\FloatBarrier

\subsection{ Classical model for a meson}
\label{sec:classicalmeson}

Consider this dipole-type
 configuration may be associated to a classical model for a  meson configuration
of radius $R_0$
to some extent inspired in the MIT bag model \cite{bagmodel}.
Eq. (\ref{R0-constantV}) suggests that, for $R_0 \simeq 0.9$ fm and  $g=1$,
one might have $V 
\sim 1$  GeV
that is approximately  a typical energy scale of hadron physics.

The corresponding 
total energy inside the sphere of radius $R_0$ for $l=1$
is obtained by integrating eq. (\ref{EMTensor00}).
It yields the following:
\begin{align}
U(l=1) = \frac{\pi}{ gVr_0^2} \left( \frac{gVR_0}{3} + \frac{\sin(2gVR_0)}{6} 
- 2\frac{\sin^2(gVR_0)}{3gVR_0} + \frac{4}{15}  I(gVR_0)  \right),
\end{align}
where:
\begin{align}
I_1(gVR_0) &= \int_0^{gVR_0}  \frac{\left( (3-x^2)\sin(x)-3x\cos(x) \right)^2}{x^4}  dx.
\end{align}
The total charge contained on the top hemisphere is given by:
\begin{align}
Q= 2\pi \int_{0}^{\pi/2} \int_0^{R_0} \rho r^2 \sin(\theta) dr d\theta= \frac{\pi}{3g(r_0V)^2}I_2(gVR_0),
\end{align}
where:
\begin{align}
I_2(gVR_0)= \int_0^{gVR_0} \frac{1+(x^2-1)\cos^2(x)-2x 
\sin(x)\cos(x)}{x^2} dx.
\end{align}
And it's clear that for the bottom hemisphere the total charge is given by the same values, times $-1$.

For the values in table (\ref{tabela3}), the resulting lowest energies and charges 
are  exhibited in Table (\ref{tabela4}).
\begin{table}[H]
\centering
\resizebox{0.4\columnwidth}{!}{
\begin{tabular}{|c|c|c|c|}
\hline
 & $Z^1_1$ & $Z^1_2$ & $Z^1_3$ \\
\hline
$\frac{gVr_0^2U({l=1})}{\pi}$ & 1.998 & 3.546 & 5.046\\
\hline
$\frac{3g(r_0V)^2Q}{\pi}$ & 2.141 & 3.7989 & 5.407 \\
\hline
\end{tabular}}
\caption{Values of the (normalized) energy and charge (top hemisphere) for different zeros of the spherical Bessel function.}
\label{tabela4}
\end{table}

\begin{figure}[H]
\centering
\includegraphics[width=0.45\textwidth]{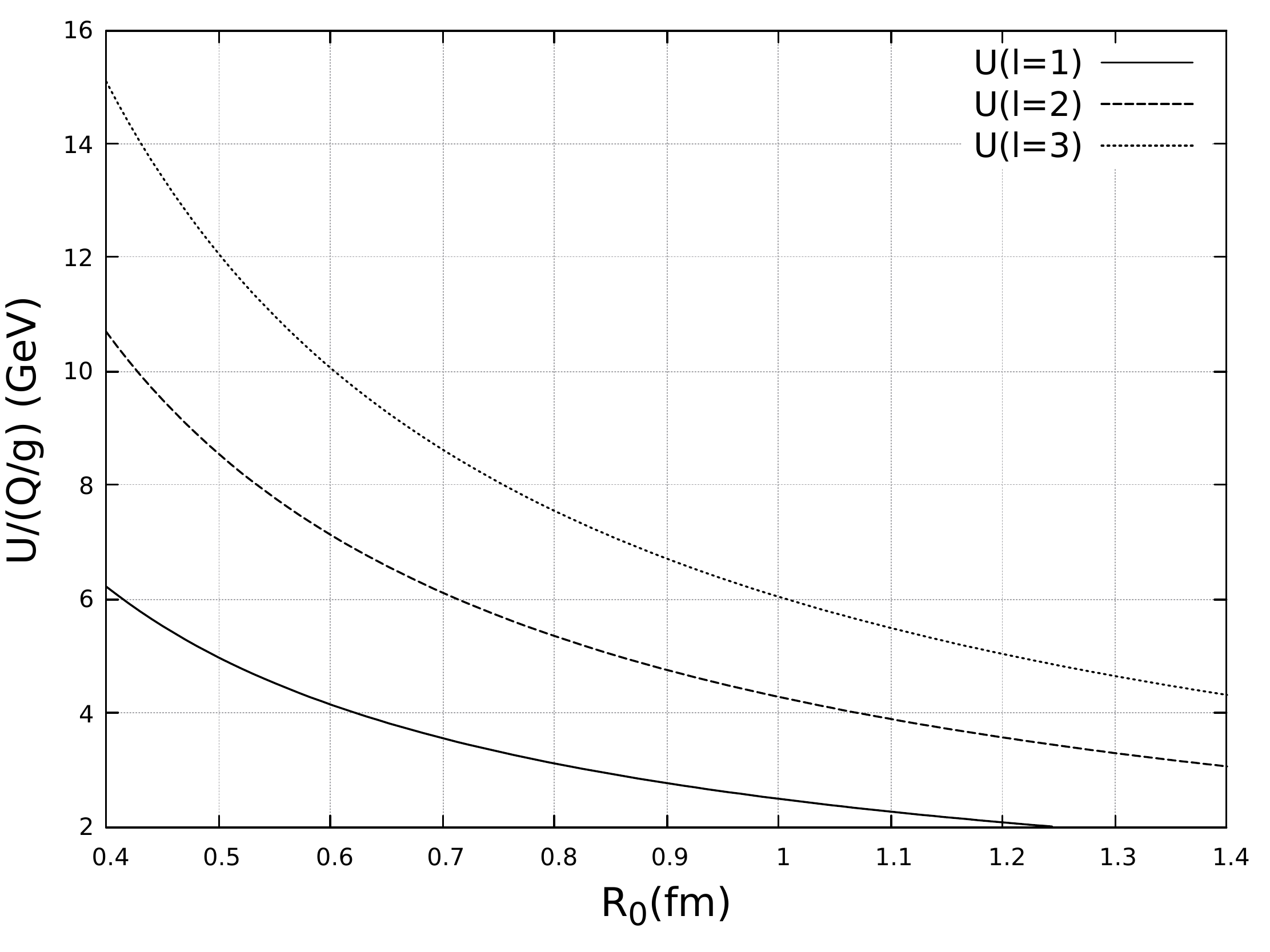}
\caption{Values of the energy $U/(Q/g)$, equation 
\ref{DipoleLikeEnergyFinal} as a function of $R_0$.}
\label{fig:U-Q-R0}
\end{figure}

Note that the constant of  integration is to be associated 
with a normalization for the vector potential - in Eq. (\ref{abc-r0}), while the 
normalization for the scalar is given by the constant $V$. 
So, rewriting $r_0$ in therms of $V$,
Eq. (\ref{R0-constantV}) as:
\begin{align}
r_0 V = \epsilon, 
\end{align}
where the new constant $\epsilon$  represents 
an eventual
 deviation of the vector potential normalization from the scalar one.
 The following  three relations arise:
\begin{align}\label{DipoleRadi}
R_0 &= \frac{Z^l}{gV};\\ \label{DipolerEnergy}
U &= \frac{V \pi}{g \epsilon^2} F(Z^l);\\\label{DipoleCharge}
Q &= \frac{\pi}{3g \epsilon^2} G(Z^l),
\end{align}
where the parameters   $F$ and $G$ are the ones in table \ref{tabela4}
for a particular zero of the Bessel function $Z^l$.
 The total  energy of the configuration 
 may be written as:
\begin{align}\label{DipoleLikeEnergyFinal}
U = 3QV \frac{F(Z^l)}{G(Z^l)} = \frac{3 Q}{gR_0} \frac{Z^l F(Z^l)}{G(Z^l)}.
\end{align}
The value of the charge $Q$ is associated with the combinations in \ref{qeta}, 
which are normalized. Then, we may have $Q=1$.
The total  energy of the classical dipole-configuration,
Eq. \ref{DipoleLikeEnergyFinal},  
 may be associated to a classical model for tetraquark's states,
 two color and two anti-color with total color charge zero.
 The corresponding energies $U$ for particular  $l=1,2,3$
as functions of the radius $R_0$  are shown in Fig. (\ref{fig:U-Q-R0}).

The relation of radius and total energy exhibit values somewhat similar to the ones found 
the phenomenology of heavy tetraquarks as discussed in Ref \cite{Htetraquarks}
and references quoted therein.
Masses of mesons candidates to be
heavy tetraquarks with at least two charm/anti-charm quarks, $c$ or $\bar{c}$, were
 measured to be at least 3900 MeV 
and 
masses of candidates to 
 heavy tetraquarks, with at least one bottom  or anti-bottom quark $b$ or $\bar{b}$,  
have been found to be larger than at least 10 GeV.
Tetraquarks with two quarks 
$b$, $\bar{b}$ or $\bar{b}b$ 
to be larger than 11 GeV or with three quarks $b$ of the order of 14- 15 GeV.
For charmed meson with three $c$-quarks the typical size of a compact tetraquark is 
$0.1$ fm whereas a molecule (mesons bound state) has a typical scale of the order of 
1 fm.

\section{ Summary and conclusions}

In this work, some aspects of the $N_c=3$ classical Yang-Mills equations,
in the Coulomb gauge,  were discussed in 
the presence of color sources in the spherical coordinate system
by means of 
the equations of the $SU(2)$ sub-groups.
Although the equations are  non-Abelian they contain  different limits in which
they behave effectively as if they were Abelian, 
independently of the 
value of the coupling constant.
By assuming the vector and scalar potentials to be fully  spherically symmetric, i.e. 
dependent only on the radial coordinate, the only possible non-trivial solution 
was found to be given by the Abelian limit.
Conversely, by imposing the scalar potential to be spherically symmetric
the vector potentials were found to develop strong anisotropies.

Different  specific solutions for the scalar potential were considered separately and 
the consequences were investigated.
The charge distributions were left, to some extent,
free such that these scalar  gauge potential  solutions 
  remain valid.
The following scalar potentials were considered:
the Coulomb potential and an anisotropic extension, a  linear potential 
and a Yukawa potential.
Also a sort of dipole-type configuration of constant scalar potential inside 
a sphere of radius R, being a semi-sphere of positive constant scalar potential $V$ 
and a semi-sphere of negative $V$.
In some situations, the existence of a color charge distribution was associated to 
a non zero vector potential such as in the case of the Yukawa potential, being that
the range of the potential is associated to the typical range of the color-charge distribution.
For these cases,
outside the region in which color-charge distribution is displaced there are no non-Abelian effects.
In some cases therefore, 
 there is a need to impose a maximum radius R where the color-charge distribution remains
 and for that, a model have been  adopted to give further meaning for the resulting configuration.
For that, the reasoning of the 
 MIT bag model was considered and it turns out that results might contribute 
for the dynamical emergence of a bag size.
Furthermore, most of the configurations analyzed
 might contribute to the formation of a classical picture of a 
constituent quark in the sense of a quark that is dressed by non-Abelian effects
being highly anisotropic.

By enforcing the scalar potential to be a Coulomb-potential in the presence of 
a punctual color-charge and an ad hoc three-dimensional color-charge distribution,
and by enforcing the c-electric field Gauss Law to be satisfied,
it generates a constraint between the color-charge distribution and the 
components of the vector potential.
Solutions for $\vec{A} (\vec{r})$ were found analytically.
A condition for  the coupling constant and color-charge normalization was found to be:
$ 0 \leq \frac{ q g }{4 \pi} = \sqrt{ l(l+1) - \beta } \leq \frac{1}{2}$
where $l$ is the angular momentum
and $\beta$ is a constant of integration.
The resulting charge distribution and c-electric and c-magnetic fields
contain singularities in the origin being
strongly anisotropic. 
For $\theta=\pi/2$   they reduce to the 
usual punctual charge and  Coulomb potential,
being that the magnetic field
has a behavior like $1/r^3$
 in the direction $\hat{\theta}$ - as
 a purely non-Abelian effect. 
The c-magnetic field in general, however,
presents  analytical cuts
for $\theta=0, \pi$, besides the
singular behavior at $r=0$.
The chromo-electric and chromo-magnetic fluxes,
and the energy density,
 are therefore strongly anisotropic.
The total charge $Q$ contained inside 
 a sphere of radius $R$ is given by:
\begin{align} \label{Q-coulomb2}
Q = q \left( 1 + \frac{g^2}{T-1} \left( \frac{r_0}{R} \right)^{2(T-1)} \right).
\end{align}
Since $T > 1$, for a fixed $R$ the quark color-charge is increased due to the 
gauge potentials.

A classes of
anisotropic extensions of the Coulomb potential were searched to be of the form: 
$\varphi (r, \theta) = f(\theta)/r$, where $f(\theta)$ was envisaged to have specific shapes.
Solutions with $f(\theta) = \frac{\lambda \cos^n( \theta)}{\sin (\theta)}$ were 
found for
$n=0$.
In this case, analytical solutions were found for the vector potential in the $\hat{\phi}$ direction
for the lower angular momenta $l=0,1$ with usual
 singularity in the origin.
However, similarly to the strict Coulomb potential,  there appears non-analytical behavior 
for $\theta = 0,  \pi$.
The  corresponding 
color-charge density  presents singularities both at the origin, for  $1/r^n$ (n=3,4),
and cuts in the directions $\theta = 0, \pi$.
Components of  c-electric and c-magnetic fields may 
change sign at different  angles, depending on the angular momenta $l=0,1$
for example for $\theta = \pi/2$ or $\theta=\pi$,
and this suggests a (non trivial or asymmetric) dipole-like color-charge configuration
that can be naturally associated to the anisotropic Coulomb potential shape.
The analytical cuts happen as  if a current were passing along these directions
$\theta=0,\pi$, but it is 
rather a static non-spherical  non-Abelian effect that may mimic the effect of a 
color-electric current in terms of the set of  Abelian equations of motion.
This can also be understood for the other anisotropic solutions with analytical cuts,

By enforcing the gauge scalar potential to have a linearly increasing  solution, 
$\varphi = \kappa r$, by adopting a similar  procedure to the one 
for the Coulomb potential,
without the punctual charge,
analytical solutions for the vector potential were also found with the 
corresponding c-electric and c-magnetic fields.
They are given by Bessel functions and are strongly oscillatory with 
the anisotropic behavior.
The oscillatory behavior suggests the need of limiting the solution to be valid 
inside a finite region at the example of bag-like models.
The vector potential, c-electric field, and the components of the c-magnetic field
disappear not necessarily in the same  radial coordinate and the 
color charge distribution may have zero's in particular directions, but not all of them,
for particular values of the parameters such as $\kappa$ and 
the field normalization. 
There are directions in which the color-charge distribution 
does not go to zero, except in infinite.
A {\it precarious} dynamical way to define $\kappa$ was proposed by 
imposing the solution to be valid inside a finite spatial region.
By considering a typical distance for which 
color-charge distribution go to zero 
for $\gamma \sim 0.1$  
it  yields  $\kappa \sim 0.03$ GeV$^2$
that is  smaller than  lattice results $\sigma_{latt} \sim 0.19$ GeV$^2$
 \cite{SM,bali}.
The resulting configuration is therefore strongly anisotropic
and the c-electric and c-magnetic fluxes are also strongly anisotropic.

The case of a Yukawa potential for the scalar potential, $\varphi = -  \frac{C}{r} e^{r/r_0}$, 
was also addressed
being that the length scale $r_0$ was dynamically associated to
 the color-charge distribution.
By considering separation of variables for the vector potential, 
analytical cuts were found in the  $\theta-$dependence for angles $\theta=0,\pi$.
Only numerical solutions for the radial dependence of the vector potential
 could be found in this case.
For that, singular boundary conditions in the origin  were considered and 
solutions were found as functions of $r/r_0$.
A determination of $r_0$, the range of the Yukawa potential, was not possible
although it is traced back to the color-charge distribution.
The color-charge distribution is highly anisotropic being that 
a more drastic behavior happens  around the analytical cut at
$\theta=\pi$.

Finally, one configuration of positive and negative 
constant scalar potential, inside 
two semi-spheres, were defined  leading to a sort of 
dipole type configuration.
The oscillatory vector potential presents zeros, from  the spherical Bessel function of order l,
being that the resulting c-electric and c-magnetic fields 
fluxes are strongly anisotropic.
Whereas the c-electric field has the same zeros of the color-charge distribution
the c-magnetic field has its zero at a smaller radial coordinate.
By imposing continuity of the solutions, some boundary conditions were chosen.
This dipole type configuration inspired a sort of classical 
 bag-model for (heavy) tetraquarks since
the dipole configuration for $\varphi$ is directly defined for 
a combination of two color charges according to 
the choices in Table (\ref{tabtabela1}) for V-spin and U-spin.
The masses of the configuration, defined as the total energy of the c-electric and c-magnetic fields, 
can
 present the order of magnitude
of those measured or  predicted for heavy tetraquarks.

Several of the solutions found above for the vector potentials and color-charge
distributions
were  found to be  strongly anisotropic  with 
 analytical cuts.
The chromo-electric and magnetic fluxes in 
hadrons, mesons and baryons, are necessarily anisotropic
and this comparison might suggest that emergence of these 
anisotropic fluxes of c-electric and c-magnetic lines
may be part of the confinement mechanism.
As an outcome of this work, it can be stated that, at the classical Yang-Mills level,
the gauge scalar potentials that were considered would be more suitably
used  
as long as  some further  corresponding
anisotropic
 vector potentials
 and  color-charge distributions, are considered. 
These issues might have relevant effects in the hadron spectroscopy when
calculated with Schrodinger or Dirac equation 
type or Bethe-Salpeter equations modeled with gauge potentials.
There are available estimations of the anisotropy's effects on the gluon propagator
\cite{gluonpro-latt-anisotropic} however 
the eventual role of these classical solutions for the full quantum problem is to be understood in the 
future.
Further consequences of the breaking of  rotational invariance were not 
investigated in this work.

\section*{ Acknowledgements}

The authors thank short discussions with 
 P. Sikivie, J. Greensite and G. Bali.
F.L.B.
(CNPq-312072/2018-0 , 
CNPq-421480/2018-1 and 
CNPq-312750/2021-8) and I. de M.F. thank  support from  CNPq.

 \appendix

\section{ Appendix A: The equations for  $SU(2)$} 
\label{EquaçõesSU2}

\setcounter{equation}{0}
 
\renewcommand{\theequation}{A.\arabic{equation}}

As pointed out in the previous section, we may reduce the problem from $SU(3)_c$ 
color group to a subgroup sector, given by an $SU(2)_{I,V,U}$. 
The prescriptions \ref{aligment1} and \ref{alignment2} are now generalized for containing all potentials for the $SU(2)$ case, which is given by the set of equations (for equivalent non-abelian directions, with arguments similar to past section):
\begin{align}
\nabla^2 \sigma &= - j_\sigma\\
\nabla^2 \vec{\sigma} &= 0;\\
\nabla^2 \varphi - 2g^2 \left( A^2 \varphi - A^0 \vec{A} \cdot \vec{\varphi} \right) &= - \rho;\\
\nabla^2 \vec{\varphi} + 
 2g^2 \left( A^0 \varphi \vec{A} - (A^0)^2 \vec{\varphi} + \vec{A} \times \left( \vec{A} \times \vec{\varphi} \right) \right) &= 0;\\
\nabla^2 A^0 -g^2\left( \vec{\varphi} \cdot \left( \vec{A} \varphi - A^0 \vec{\varphi} \right) \right) &= 0;\\
\nabla^2 \vec{A} - g \left( \frac{1}{2} \vec{\nabla} \times \left( \vec{\varphi} \times \vec{A} \right) - \vec{\varphi} \times \vec{\nabla} \times \vec{A} + \varphi \vec{\nabla} A^0 \right.
\nonumber
\\
\left. -  A^0 \vec{\nabla} \varphi  \right) - g^2 \left(\frac{1}{2}  \vec{\varphi} \times \left( \vec{\varphi} \times \vec{A} \right)  + \varphi \left( A^0 \vec{\varphi} - \vec{A} \varphi \right) \right) &=0,
\end{align}
where $A^0$ stands for the scalar potentials in the non-abelian directions, $\vec{A}$ the non-abelian vector potentials and $\vec{\varphi},\vec{\sigma}$ are the vector correspondents of the scalars found in table \ref{tabtabela1}.

The version of Eqs. (\ref{eom1},\ref{jacobi1}), 
with time dependence and in Lorentz gauge, is (for U-spin, since the others are completely equivalent):
\begin{align} \label{equations4}
\nabla^2\xi &= - \sigma;\\
\frac{\partial}{\partial t} \vec{\nabla} \xi &= \vec{\sigma};\\
\nabla^2\varphi+ 4g (\vec{A}_6 \cdot \frac{\partial}{\partial t}\vec{A}_7 - \vec{A}_7 \cdot \frac{\partial}{\partial t}\vec{A}_6) - 4g^2(A_6^2 + A_7^2)  \varphi &= -\rho;\\
\frac{\partial}{\partial t} \vec{\nabla} \varphi +4g \left( \vec{\nabla} \times \vec{A}_6 \times \vec{A}_7 + \vec{A}_6 \times \vec{\nabla} \times \vec{A}_7 \right) &= \vec{\rho};\\
\frac{\partial}{\partial t} \vec{\nabla} \cdot \vec{A}_6 + 2g \vec{\nabla}\varphi \cdot \vec{A}_7 + g \varphi \vec{\nabla}\cdot \vec{A}_7 &= j^0_6;\\
\frac{\partial}{\partial t} \vec{\nabla} \cdot \vec{A}_7 - 2g \vec{\nabla}\varphi \cdot \vec{A}_6 - g \varphi \vec{\nabla}\cdot \vec{A}_6 &= j^0_7;\\
\vec{\nabla}(\vec{\nabla} \cdot \vec{A}_{6}) - \vec{\nabla}^2 \vec{A}_{6} + \frac{\partial^2}{\partial t^2} \vec{A}_{6} + g \left( 2\varphi \frac{\partial \vec{A}_7}{\partial t} + \frac{\partial \varphi}{\partial t} \vec{A}_7\right) - g^2 \varphi^2 \vec{A}_{6} + 4 g^2 \vec{A}_7 \times \left( \vec{A}_6 \times \vec{A}_7 \right) &= \vec{j}_{6};\\
\vec{\nabla}(\vec{\nabla} \cdot \vec{A}_{7}) - \vec{\nabla}^2 \vec{A}_{7} + \frac{\partial^2}{\partial t^2} \vec{A}_{7} + g \left( 2\varphi \frac{\partial \vec{A}_6}{\partial t} + \frac{\partial \varphi}{\partial t} \vec{A}_6\right) - g^2 \phi^2 \vec{A}_{7} - 4 g^2 \vec{A}_6 \times \left( \vec{A}_6 \times \vec{A}_7 \right) &= \vec{j}_{7}.
\end{align}

\section*{Appendix B: Chromo-electric and chromo-magnetic fields}
\label{appEcBc}

\setcounter{equation}{0}
\renewcommand{\theequation}{B.\arabic{equation}}

Analogous to the electromagnetic fields, one can define:
\begin{eqnarray}
F_{i0} = E_i &\rightarrow& \vec{E}_a = - \frac{\partial \vec{A}_a }{\partial t}
- \nabla A^0_a  + g f_a^{bc} A_b^0 \vec{A}_c,;\\
F_{ij} = \epsilon_{ijk} B_k &\rightarrow& 
\vec{B}_a = \nabla \times \vec{A}_a + g f_a^{bc} (\vec{A}_b \times \vec{A}_c ),
\end{eqnarray}
Besides the Euler-Lagrange equations given in the text, a non-Abelian generalization of the 
Faraday's law, 
by choosing 
$\mu=t, \nu = i, \lambda=j$ in Eq. (\ref{jacobi1}),
is obtained from the topological constraint of the SU(3) Lie algebra.
It  yields:
by multiplying by $\epsilon_{ijk}$ and by using $\epsilon_{ijk} \epsilon_{ijl} = 2 \delta_{kl}$
\begin{eqnarray}
2 \partial_t B_k + 2 (\nabla \times E)_k 
- 
i g \left( 
2 [ A_0, B_k] + 2  (A_j \times E_i )_k   - 2 (E_j \times A_i )_k \right).
\end{eqnarray}

The set of four equations can be written as:
\begin{eqnarray}
\nabla \cdot \vec{E}_a + f_{abc} \vec{E}_b \cdot \vec{A}_c &=&
g \bpsi \xlam_a \gamma^0 \psi,
\\
\nabla \times \vec{E}_a + f_{abc} ( A_{0b} \vec{B}_b - \vec{A}_c \times \vec{E}_b )
 &=&
- \frac{\partial \vec{B}_a }{\partial t}  - g f_{abc} A_b^0 \vec{B}_c,
\\
\nabla \cdot \vec{B}_a + f_{abc} \vec{A}_c \cdot \vec{B}_b = 0 ,
\\
\nabla \times \vec{B}_a + f_{abc} (\vec{A}_c \times \vec{B}_b - A_{0c} \vec{E}_b )
&=& \frac{ \partial \vec{E}_a }{\partial t} +   g \bpsi \xlam_a \vec{\gamma} \psi
+ g f_{abc} A_b^0 \vec{E}_c,
\end{eqnarray}
It is not possible, however, to write these equations solely in terms of
$E_c$ and $B_c$, as it is well known, being needed to account 
for scalar and vector potentials as well, except in very particular cases.

For the prescriptions adopted in the work 
for the different components of the  gauge scalar and vector potentials, 
Eq. (\ref{aligment1},\ref{alignment2}),
 C-Electric and C-Magnetic fields can be written as:
\begin{align}
\vec{E}_a =& \left(-\vec{\nabla} A^0_3 \right) \delta_{a3} + \left( -\vec{\nabla} A^0_8 \right) \delta_{a8} + \left( -\vec{\nabla} A^0 - g\left( \vec{\varphi} A^0 - \varphi \vec{A} \right) \right) \delta_{ac} + \left( - \vec{\nabla} A^0 + g\left(\vec{\varphi} A^0 - \varphi \vec{A} \right) \right) \delta_{ad},\\
\vec{B}_a =& \left( \vec{\nabla} \times \vec{A}_3  \right) \delta_{a3} + \left( \vec{\nabla} \times \vec{A}_8 \right) \delta_{a8} + \left( \vec{\nabla} \times \vec{A} + g \vec{\varphi} \times \vec{A} \right) \delta_{ac} + \left( \vec{\nabla} \times \vec{A}  - g \vec{\varphi} \times \vec{A} \right) \delta_{ad}.
\end{align}

\subsection*{ Energy-momentum tensor and Continuity equations }
\label{appemtensor}

The gauge-invariant traceless energy-momentum tensor
by omitting quarks contributions to avoid the problem of the energy of
a single quark-source, or its size,
can be written as:
%
\begin{eqnarray}
\theta^{\mu\nu} = \frac{1}{2} \mbox{Tr} ( -F^{\mu\rho} F_\rho^\nu + \frac{1}{4} g^{\mu\nu} F^{\rho\sigma}
F_{\rho\sigma} ) .
\end{eqnarray}

%

The most relevant two components are the energy density and pressure that can be written in general 
in terms  of the   chromo-electric and chromo-magnetic fields, $\vec{E}_c$ and $\vec{B}_c$,
defined as usual in the Appendix (\ref{appEcBc}).
By adding the quark current it can be written that:
\begin{eqnarray} \label{EMTensor00}
\theta^{00} &=& \frac{1}{2} ( \vec{E}_a \cdot \vec{E}_a + \vec{B}_a \cdot \vec{B}_a )
+ j^i_a A^a_i
\\ \label{EMTensor01}
\theta^{0i} &=& 
(\vec{E}_a \times \vec{B}_a )_i,
\\
\theta_{ij} &=& 
- E_i^a E_j^a - B_i^a B_j^a + 
\frac{\delta_{ij}}{2} ( \vec{E}_a \cdot \vec{E}_a + \vec{B}_a \cdot \vec{B}_a ).
\end{eqnarray}



%
%


%
%
From the Noether Current defined in \ref{noethercurr}, we may write the continuity equations as:
\begin{align}\label{ContinuityEq}
\frac{\partial j^0_a}{\partial t} + \vec{\nabla} \cdot \vec{j}_a + gf_{abc} \left( A^0_b j^0_c - \vec{A}_b \cdot \vec{j}_c \right)=0.
\end{align}


\begin{thebibliography}{99}




 \bibitem{YM}
C.N. Yang, R.L. Mills, 
Conservation of Isotopic Spin and Isotopic Gauge Invariance,
Phys. Rev. 96, 191 (1954).


\bibitem{SM}
W.N. Cottingham, D.A. Greenwood,
An Introduction to the Standard Model of Particles,
 2nd Ed. Cambridge (2007).

\bibitem{QCD}
T.P. Cheng, L.F. Li,  Gauge Theory of Elementary Particles, 
Oxford (1984).


\bibitem{gaugetheory}
L. O'Raifeartaigh, N. Straumann,
Gauge theory: Historical origins and some modern developments,
Rev. of Mod. Phys. 72, 1 (2000).



 
\bibitem{monopole-rossi}
P. Rossi,  Exact Results in the Theory of non-Abelian magnetic monopoles,
Phys. Rept.  86, 317 (1982).
 

\bibitem{actor-review}
A. Actor, 
Classical solutions of SU(2) 
Yang-Mills theories, Rev. of Mod. Phys. 51, 461 (1979).


\bibitem{wu-yang-rosen}
T.T. Wu,  and C. N. Yang, 1968, in Properties of Matter
Under Unusual Conditions, edited by H. Mark and S. Fernbach
(Interscience, New York).
Rosen, G., 1972, J. Math. Phys. 13, 595.


\bibitem{instantons1}
G. 't Hooft,  Magnetic monopoles in unified gauge theories. Nuclear Physics B 79, 
 276 (1974). 
A. M. Polyakov, Particle spectrum in the quantum field theory. JETP Letters 20,
 194 (1974).
G. 't Hooft, F. Bruckmann,
Monopoles, Instantons and Confinement, arXiv:hep-th/0010225.

\bibitem{instantons-meron}
E. Elizalde, Phys. Lett. B77,  73 (1978).



\bibitem{bali}
G. Bali, QCD forces and heavy quark bound states,
Phys.Rept.343, 1  (2001).


\bibitem{review1}
N. Brambilla, {\it et al}, QCD and strongly coupled gauge theories: challenges and
perspectives,
Eur. Phys. J. C 74, 2981 (2014).

\bibitem{griffiths}
D. Griffiths, Introduction to Elementary Particles,
Wiley, (1987).



\bibitem{sikivie-weiss-prd}
P. Sikivie, N. Weiss,
Classical Yang-Mills theory in the presence of external sources
Phys. Rev. D 18,  3809
(1978).


\bibitem{sikivie-weiss-prl}
P. Sikivie, N. Weiss,
Screening Solutions to Classical Yang-Mills Theory,
Phys. Rev. Lett. 40, 1411  (1978).

 
\bibitem{mandula} 
J.E. Mandula, Classical Yang-Mills potentials,
Phys. Rev. D14, 3497  (1976)


\bibitem{jackiw} R. Jackiw, L. Jacobs, C. Rebbi, 
Static Yang-Mills fields with sources, 
Phys. Rev. D20,  474 (1979).



\bibitem{cilinder}
S. M. Mahajan, P.M. Valanju, 
Finite-energy classical solutions to Yang-Mills theories, Phys. Rev. D35, 2543
(1987)


\bibitem{plane-color} 
R. E. Crandall, D. J. Griffiths, N. A. Wheeler, R. A. Mayer
Fields of an infinite plane of color,
Phys. Rev. D25, 1143 (1982)




\bibitem{current-sources}
L. Mathelitsch, H. Mitter, F.  Widder, 
Stationary Yang-Mills fields with current sources,
Phys. Rev. D 25, 1123 (1982)




\bibitem{passarino}
G. Passarino,
Yang-Mills theories in the presence of classical plane-wave fields:
stability properties,
Phys. Lett. B 176, 135 (1986)


\bibitem{gausslaw-greek}
A. Tsapalis, E. P. Politis, X. N. Maintas, and F. K. Diakonos,
Gauss’ law and nonlinear plane waves for Yang-Mills theory
Phys. Rev. D93, 085003 (2016)


\bibitem{ikeda-miyachi-1962}
M. Ikeda, Y. Miyachi,
On the Static and Spherically Symmetric Solutions
of the Yang-Mills Field,
Progress of Theoretical Physics 27,  474 (1962).


\bibitem{constant-gauge}
T.N. Tudron, 
Instability of constant Yang-Mills fields generated by constant gauge potentials,
Phys. Rev. D22, 2566 (1980)



\bibitem{huang-levi}
S. Huang and A. R. Levi, Phys. Rev. D49, 6849 (1994); A. R. Levi, subtleties and
fancies in gauge theory nontrivial vacuum, hep-lat/9409002.


\bibitem{constant-Bc1}
G.K. Savvidy, Phys. Lett. B71, 133 (1977).
S.G. Matinyan and G.K. Savvidy, Nucl. Phys. B134, 539  (1978)
 N.K. Nielsen and P. Olesen, Nucl. Phys. B144, 376  (1978).
J. Ambjorn, N.K. Nielsen and P. Olesen, Nucl. Phys.
B152, 75  (1979).
J. Ambjorn, P. Olesen, Nucl. Phys. BI70, 60 (1980).


\bibitem{wetterich-reuter-1994}
M. Reuter, C. Wetterich,
Search for the QCD ground state, Phys. Lett. B 334, 412 (1994).

\bibitem{coimbra-quarkonia}
R.A. Coimbra, O. Oliveira, 
Heavy quarkonia from classical SU(3) Yang Mills configurations,
Eur. Phys. J. A 31, 718 (2007)


\bibitem{initial-cond1}
A. Kovner, L. McLerran, H. Weigert,
Gluon production from non-Abelian Weizsacker-Williams fields
in nucleus-nucleus collisions,  Phys. Rev. D52, 6231 (1995).

\bibitem{stability}
S. Bazak, S. Mrowczynski,
Stability of Classical Chromodynamic Fields, Phys. Rev. D 105, 034023 (2022).



\bibitem{rhic3}
B. Schenke, S. Schlichting, R. Venugopalan,
Azimuthal anisotropies in p+Pb collisions from classical Yang-Mills dynamics,
Phys. Lett. B747, 76 (2015).


\bibitem{cgc-class-YM}
O. Philipsen, B. Wagenbach, S. Zafeiropoulos,
From the colour glass condensate to filamentation:
systematics of classical Yang–Mills theory,
Eur. Phys. J. C  79, 286  (2019).


\bibitem{shear-YM}
H. Matsuda, et al,
Shear viscosity of a classical Yang-Mills field,
Phys. Rev. D102, 114503 (2020).


\bibitem{3dim-YM}
M. Frasca, 
Confinement in a three-dimensional Yang–Mills theory,
Eur. Phys. J. C 77:255, (2017)

\bibitem{greensite} J. Greensite, An Introduction to the Confinement Problem,
Springer, Lectures notes in Physics 821, (2011).



\bibitem{fluxtube-qqb}
J.W. Flower, S.W. Otto, 
The field distribution in SU(3) lattice gauge theory,
Phys. Lett 160B, 128  (1985)
 R. Sommer, Nucl. Phys. B291 (1986) 673.
 G. S. Bali, C. Schlichter and K. Schilling,
Phys. Rev. D51 (1995) 5165.


\bibitem{fluxtube-qqb2}
 R. W. Haymaker, V. Singh and Y. Peng,
Phys. Rev. D53 (1996) 389.
M. Luscher, G. Munster, P. Weisz, 
How thick are chromo-electric flux tubes?,
Nucl. Phys. B180,  1 (1981).


\bibitem{fluxtube-baryon}
F. Okihary, R. Woloshyn,
A study of colour field distributions in the baryon,
Nucl. Phys. B 129-130(Proc.Suppl.), 745 (2004).
 J. Flower, CALT-68-1378 (1986).
 H. Ichie, V. Bornyakov, T. Streuer and G.
Schierholz, Nucl. Phys. B (Proc. Suppl.) 119
 (2003) 751, hep-lat/0212036.




\bibitem{instabilities}
C. Cardona, T. Vachaspati,
Instability of a uniform electric field in pure non-Abelian Yang-Mills theory,
Phys. Rev. D 104, 045009  (2021).

\bibitem{instability-sikivie}
P. Sikivie, 
 Instability of Abelian field configurations in Yang-Mills theory,
Phys.Rev. D20, 877 (1979).


\bibitem{constant-gauge}
S. J. Chang,  N. Weiss,
 Instability of constant Yang-Mills fields 
 Phys. Rev. D 20, 869 (1979).


\bibitem{greenberg}
O.W. Greenberg,
 Quarks,
 Ann. Rev. Nucl. Part. Sci. 28, 327 (1978).


\bibitem{laplacianoMoon} P. Moon, D.E. Spencer,
The meaning of the vector Laplacian,
Journal of the Franklin Institute 256, 6 (1953).



\bibitem{hadicke-pohle-spher-sym}
A. Hadicke, H.-J. Pohle, 
Solutions of the SU(3) gauge field equations with spherical symmetry,
Phys. Lett. B137,  193  (1984)


\bibitem{Loos-1965}
H.G. Loss, The range of gauge fields, Nucl. Phys. 72, 677 (1965).

\bibitem{butkov} E. Butkov,  Mathematical Physics,  (1983).

\bibitem{alpha-qcd}
 D. J. Gross and F. Wilczek, Ultraviolet behavior of nonabelian gauge theories, Phys.
Rev. Lett. 30, 1343 (1973); H. D. Politzer, Reliable perturbative results for strong
interactions?, Phys. Rev. Lett. 30, 1346 (1973).

\bibitem{lattice-g}
A. Deur, S. J. Brodsky, G. F. de Teramond,
he QCD Running Coupling,
Prog. Part. Nuc. Phys. 90 1 (2016).


\bibitem{CYM-book}
V. Rubakov,
Classical Theory of Gauge Fields, translated S.S. Wilson,
Princeton (2002).


\bibitem{bagmodel}
A. Chodos,
R. L. Jaffe, K. Johnson, C. B. Thorn, V. F. Weisskopf,
 New extended model of hadrons, Phys. Rev. D 9, 3471 (1974).
A.  Chodos, 
R. L. Jaffe, K. Johnson,  C. B. Thorn, 
Baryon structure in the bag theory, Phys. Rev. D 10, 2599 (1974).



\bibitem{Htetraquarks}
X.-Z. Weng,  W.-Z. Deng,  S.-L. Zhu,
Triply heavy tetraquark states,
Phys. Rev. D 105, 034026 (2022).




\bibitem{plane-color} 
R. E. Crandall, D. J. Griffiths, N. A. Wheeler, R. A. Mayer
Fields of an infinite plane of color,
Phys. Rev. D25, 1143 (1982)



\bibitem{cilinder}
S. M. Mahajan, P.M. Valanju, 
Finite-energy classical solutions to Yang-Mills theories, Phys. Rev. D35, 2543
(1987)


\bibitem{current-sources}
L. Mathelitsch, H. Mitter, F.  Widder, 
Stationary Yang-Mills fields with current sources,
Phys. Rev. D 25, 1123 (1982)




\bibitem{gluonpro-latt-anisotropic}
Y. Nakagawa, A. Nakamura, T. Saito, H. Toki,
Scaling study of the gluon propagator in 
Coulomb gauge QCD on isotropic and anisotropic lattices,
Phys. Rev. 83, 114503 (2011).












\end{thebibliography}
\end{document}